\documentclass[aps,pra,twocolumn,amsmath,superscriptaddress,longbibliography]{revtex4-2}

\newcommand{\bea}{\begin{eqnarray}}
\newcommand{\eea}{\end{eqnarray}}
\newcommand{\beq}{\begin{equation}}
\newcommand{\eeq}{\end{equation}}

\newcommand{\Si}{\Sigma}

\newcommand{\black}[1]{\textcolor[rgb]{0.00,0.00,0.00}{#1}}
\usepackage{amsmath}
\usepackage[urlcolor=blue,colorlinks=true,citecolor=blue,linkcolor=blue,pdfstartview={FitH},bookmarks=false]{hyperref}
\usepackage{appendix}
\usepackage{graphicx}
\usepackage{longtable}
\usepackage{epsfig}
\usepackage{dcolumn}
\usepackage{bm}
\usepackage{bbm}
\usepackage{amssymb}
\usepackage{multirow}
\usepackage{times,color}
\usepackage{hyperref}
\usepackage{amsmath}
\usepackage{color}
\usepackage{subfigure}
\usepackage{float}
\usepackage{epstopdf}
\usepackage{ulem}

\begin{document}
\title{Exploring quantum criticality and ergodicity-breaking dynamics in spin-1 Kitaev chains via single-ion anisotropies}

\author{Wen-Yi Zhang}
\affiliation{College of Physics, Nanjing University of Aeronautics and Astronautics, Nanjing 211106, China}
\affiliation{Key Laboratory of Aerospace Information Materials and Physics (NUAA), MIIT, Nanjing 211106, China}

\author{Qing-Min Hu}
\affiliation{College of Physics, Nanjing University of Aeronautics and Astronautics, Nanjing, 211106, China}
\affiliation{Key Laboratory of Aerospace Information Materials and Physics (NUAA), MIIT, Nanjing 211106, China}

\author{Jie Ren}
\affiliation{Department of Physics, Changshu Institute of Technology, Changshu 215500, China}

\author{Liangsheng Li}
\affiliation{National Key Laboratory of Scattering and Radiation, Beijing 100854, China}
             
\author{Wen-Long You}
\email{wlyou@nuaa.edu.cn}
\affiliation{College of Physics, Nanjing University of Aeronautics and Astronautics, Nanjing, 211106, China}
\affiliation{Key Laboratory of Aerospace Information Materials and Physics (NUAA), MIIT, Nanjing 211106, China}

\begin{abstract}
We investigate topological gauge-theory terms and quantum criticality in a spin-1 Kitaev chain with general single-ion anisotropies (SIAs). The ground-state phase diagram, including the Kitaev spin liquid (KSL) and gapless dimer phases, is determined by the infinite time evolving block decimation (iTEBD) method. A quantum phase transition between the KSL and dimer phases occurs by varying uniaxial SIA, analogous to the confinement-deconfinement transition in the lattice Schwinger model with a topological $\theta$ angle of $\pi$. Introducing rhombic SIA shifts this angle from $\pi$, resulting in $y$- and $x$-ferroquadrupole phases. The transition between these phases can occur through a crossover in the KSL phase or a genuine phase transition along a deconfined line. We map the spin-1 Hamiltonian to an effective spin-1/2 PXP Hamiltonian, with uniaxial SIA corresponding to uniform detuning and rhombic SIA to staggered detuning. We explore the hierarchical fragmentation of the Hilbert space, revealing that quantum many-body scars (QMBSs) emerge under weak uniform detuning, while slow dynamics under large staggered detuning is accurately captured by a second-order effective Hamiltonian via the Schrieffer-Wolff transformation. Our work establishes a framework for simulating topological $\theta$ angles and ergodicity-breaking dynamics, bridging higher-spin generalizations of scarred models with lattice gauge theories, potentially realizable using state-of-the-art cold-atom quantum simulators.
\end{abstract}

\date{\today}

\maketitle
%\pacs{05.70.Jk,64.70.Tg ,71.10.Fd,71.45.Gm}

\section{Introduction}
\label{intro}

Recent advancements in synthetic quantum matter have propelled the exploration of non-equilibrium dynamics in isolated quantum systems.
The eigenstate thermalization hypothesis (ETH) that sets the foundation of statistical physics, still emerges as a central framework in quantum physics~\cite{Srednicki1999,Rigol2014,Rigol2016,Deutsch2018}. Rydberg atoms, with their high principal quantum numbers, exhibit strong repulsive interactions when close together, preventing two such atoms from occupying nearby space~\cite{Browaeys2020}. Quantum simulators utilizing Rydberg atoms are capable of emulating quantum many-body dynamics and probing unprecedented  ground-state quantum phases~\cite{Ho2019,Giudici2022,Yao2022}. A seminal experiment involving a Rydberg atom chain, initiated from a density wave state, showed persistent quantum revivals~\cite{Bernien2017}. This observation elucidated that embedded within the predominantly thermal spectrum, a minority of eigenstates significantly contribute \black{to} the atypical dynamics. These zero-measure non-thermal eigenstates delineate as the quantum many-body scar (QMBS). Recognized as a paradigm of weak ergodicity breaking in disorder-free nonintegrable models, QMBSs have \black{attracted} widespread attention and have become a \black{significant area} of active research. A kinetically constrained PXP model was derived to %effectively 
elucidate the mechanisms responsible for the observed phenomena in Rydberg atom experiments~\cite{Turner2018Nature}. Subsequently, the oscillatory dynamics for certain initial states was also observed across various experimental platforms~\cite{Pan2023,Zhang2023}. Meanwhile, multiple approaches have been proposed for constructing Hamiltonians with small invariant subspaces that  weakly violate ergodicity.
Analytic expressions for QMBS states have also been identified in the spin-1 Affleck-Kennedy-Lieb-Tasaki (AKLT) model~\cite{Moudgalya2018}. Other systems that display QMBSs include the transverse field Ising ladder~\cite{Kareljan2020}, the two-dimensional PXP model~\cite{Serbyn2020,Lin2020}, the spin-1 XY model~\cite{Schecter2019,Chattopadhyay2020,Gotta2023}, and non-Hermitian many-body systems~\cite{Chen2023}. \black{Scar states in specific models can be generated using various approaches, such as %kinetic constraint models~\cite{Turner2018Nature,TurnerPRB2018}, 
projector embedding~\cite{Omiya2023,John2024,Wang2024}, the spectrum-generating algebra (SGA)~\cite{Andrei2020,Sanada2023,Deng2023,Pakrouski2020,ODea2020,Pakrouski2021,Shibata2020,Moudgalyaprx2022}, the Krylov restricted thermalization~\cite{Sanjay2022,Francica2023,Andreadakis2023} and the Floquet scars~\cite{Mizuta2020,Machado2020}. }
%group theory~\cite{Pakrouski2020,ODea2020,Pakrouski2021}, commutant algebra~\cite{Moudgalyaprx2022}, 
%quasi-integrable systems~\cite{Shibata2020,Wang2024},
 %In our work, the \blue{kinetic constraints} introduced by the $\mathbb{Z}_2$ local gauge charges in the spin-1 Kitaev model provide a hybrid mechanism that contributes to weak ergodicity breaking and the emergence of ergodicity-breaking dynamic states.}

%The approaches for constructing Hamiltonians 
% are predominantly divided into \sout{three}\black{some} categories\black{~\cite{Sanjay2022}}: the projector embedding method~\cite{Omiya2023,Markus2023,John2024}, the spectrum generating algebra technique~\cite{Andrei2020,Sanada2023,Deng2023}, and the Krylov restricted thermalization strategy~\cite{Francica2023,Andreadakis2023}. 
  
While researchers have explored QMBSs across diverse models, the seminal PXP model, where these states were first identified,  has never faded from view.    \black{The mechanisms underlying scar states in the PXP model have been examined from various perspectives, including the SGA~\cite{Dmitry2019}, Krylov restricted thermalization~\cite{Bull2020}, and hidden projector embedding~\cite{Markus2023}. Despite these efforts, a complete understanding of this formally simple model  remains elusive. } An explicit description of certain QMBS states in the PXP model has been provided through matrix product states~\cite{Lin2019}. By introducing perturbations to the PXP model~\cite{Mark2020}, 
one can either engineer perfect revivals~\cite{Dmitry2019} or achieve integrability~\cite{Anushya2019}. Recent studies have demonstrated that the PXP model is \black{exactly} embedded in the spin-1 Kitaev model~\cite{You2020,You2022}, in which the uniaxial single-ion anisotropy (SIA) \black{$\propto \sum_j (S_j^z)^2$ plays a role equivalent to that of the static uniform detuning~\cite{Zhangwy2023}}.  Specifically, in the context of achieving targeted Rydberg excitations, the detuning can be meticulously adjusted at any designated site within an atom array using external Raman lasers~\cite{Ebadi2021,Lukin2021_2}. Most prior research has assumed uniform detuning in Rydberg atoms~\cite{Anushya2019,Ghosh2018,Whitsitt2018,Keesling2019,Bull2019,Mukherjee2020_2}, but latest studies have increasingly focused on the effects of staggered detuning~\cite{Surace2020_2}. Recent advancements in experiments with Rydberg atom arrays have demonstrated that constrained Hamiltonian dynamics can be connected to U(1) lattice gauge theory (LGT)~\cite{Zache2019,Zache2022}, creating a mathematical equivalence between the PXP model and the lattice Schwinger model, which incorporates a topological angle $\theta\!= \!\pi$~\cite{Desaules2023,Pan2023_2}, and the deviation $\theta\!-\! \pi$ of the lattice Schwinger model is equivalent to an additional staggered field applied to the Rydberg system.
Notably, Pan {\it et al.} have successfully implemented a tunable $\theta$ angle in U(1) LGT with an atom-number-resolved optical lattice quantum simulator, allowing for the direct observation of the confinement-deconfinement transition~\cite{zhang2023observation}. 

%\black{In the paper}, we propose a method for realizing topological gauge-theory terms and quantum criticality in spin-1 Kitaev chain with generic SIAs. \black{ The enlarged parameter space renders the simulation of a tunable $\theta$-angle. The kinetic constraints introduced by the $\mathbb{Z}_2$​ local gauge charges in the spin-1 Kitaev model   }  We illustrate that the staggered detuning present in the extended PXP model can be effectively mimicked by the incorporation of rhombic SIAs \black{$\propto \sum_j [(S_j^x)^2-(S_j^y)^2 ]$} in spin-1 Kitaev model. Note that both the staggered terms in the spin-1/2 PXP model and the rhombic SIA terms in spin-1 Kitaev model can be equivalently mapped to the topological $\theta$-angle term in gauge theory. Through the demonstrations, the intriguing correspondence between these different models is established. 
  
\black{In this paper, we explore the relationship between the quantum criticality of the spin-1 Kitaev model, driven by generic single-ion anisotropies (SIAs), and the non-ergodic behavior observed in the PXP model with staggered detuning. By expanding the parameter space of the spin-1 Kitaev model, we develop a comprehensive phase diagram that encompasses both quantum critical and non-ergodic regimes. Our analysis reveals that different forms of SIA lead to distinct physical effects, which become particularly transparent when the system is mapped onto an effective spin-1/2 PXP model. This mapping establishes a clear connection between the quantum critical points of the Kitaev model and the non-ergodic dynamics in the PXP model. The results not only clarify the interplay between quantum criticality and non-ergodicity but also provide experimentally testable insights for realizing topological gauge-theory terms in platforms such as Rydberg atom arrays.}

The structure of this paper is as follows: In Sec. \ref{sec:KA}, we introduce a one-dimensional spin-1 Kitaev model with tunable three-component SIAs and employ both the exact diagonalization (ED) method and the infinite time-evolving block decimation (iTEBD) algorithm~\cite{Vidal2007} to determine the ground-state phase diagram. In Sec. \ref{sec:KED}, we scrutinize the quantum criticality in the spin-1 Kitaev model by varying both uniaxial and rhombic SIAs. Furthermore, we utilize a Schrieffer-Wolff transformation in Sec. \ref{sec:Schrieffer-Wolff transformation} to develop a perturbative description of the spin-1 Kitaev chain in the large limit of rhombic SIA. In Sec. \ref{sec:spin-1/2}, we demonstrate the spin-1 Hamiltonians can be exactly mapped to the spin-1/2 extended PXP Hamiltonians in the ground-state manifold characterize by all positive $\mathbb{Z}_2$ gauge charges. In addition to the gauge constraint, the Hilbert space exhibits a further fragmentation in the strong-confinement limit. We compare the quench dynamics for various values of rhombic SIAs. The glassy dynamics can be well understood by the effective Schrieffer-Wolff Hamiltonian from the perspective of Hilbert space fragmentation. Finally, a summary is provided in Sec. \ref{Summary and conclusions}.

\section{Spin-1 Kitaev model with three-component single-ion anisotropies}
\label{sec:KA}

\begin{figure*}[tb]
\centering
\includegraphics[width=2\columnwidth]{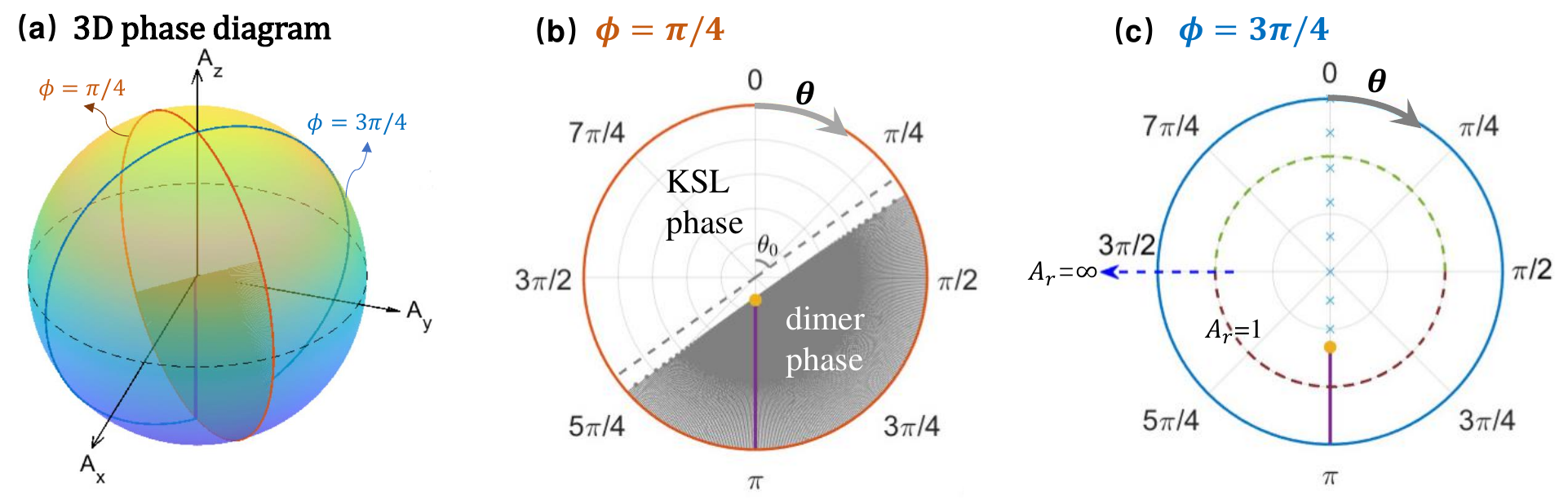}
\caption{ Phase diagram of Hamiltonian (\ref{eq:HKA}) with $A_x=A_r \sin \theta \cos \phi$, $A_y=A_r \sin \theta \sin \phi$, and $A_z=A_r \cos \theta$, calculated by the iTEBD method with bond dimension $\chi=120$. (a) Representation of the phase diagram in three-dimensional spherical coordinates. The orange (blue) line denotes the cross-section for $\phi = \pi/4$ ($\phi = 3\pi/4$), which will be detailed in (b) [(c)]. (b) Two-dimensional cross-section for $\phi = \pi/4$, i.e., $A_x = A_y$ and $A_r \in [0,5]$. The grey semi-infinite shaded region indicates the gapless dimer phase, with the grey dashed line corresponding to the phase boundaries as $A_r \to \infty$, where $\theta_0 = 0.3041\pi$ denotes the angle with the $z$-axis. (c) Two-dimensional cross-section for $\phi = 3\pi/4$, i.e., $A_x = -A_y$, and $A_r \in [0,1.5]$. In this case, Eq. (\ref{eq:HKA}) reduces to Hamiltonian (\ref{eq:HKED}). The dimer phase degrades into a ray line, represented by the purple line. Blue arrow indicates where the strength of the rhombic SIA $A_{x} \to \infty$ and the uniaxial SIA $A_{z}=0$. Dashed lines represent $A_r = 1$, and light blue crosses indicate that the system undergoes a crossover rather than a phase transition as the lowest excitation gap never closes. See more details in the main text.
}
\label{fig:phase_diagram3D}
\end{figure*}

We start with the one-dimensional spin-1 Kitaev models with tunable three-component 
 SIAs. The Hamiltonian is given by
\begin{eqnarray}
\label{eq:HKA}
\hat{H}_{\rm KA}\!= \!\sum_{j=1}^{N/2}K_j \!(S_{2j-1}^x S_{2j}^x\! +\!S_{2j}^y S_{2j+1}^y )\!+\!\sum_{\alpha,j=1}^N \! A_{\alpha}\! (S_j^\alpha)^2.  
\end{eqnarray}
Here, $A_{\alpha}~(\alpha=x,y,z)$ denotes the amplitude of the $\alpha$-component SIA, and $K_j$ parameterizes the strength of the bond-dependent Kitaev exchange coupling. $S_j^{\alpha}$ ($\alpha = x,y,z$) represents the $\alpha$-component of the spin-1 operator at the $j$th site among total $N$ sites, obeying the SU(2) algebra as $[S_i^{\alpha},S_j^{\beta}] = i \delta_{ij} \epsilon_{\alpha \beta \gamma} S_j^\gamma$, 
where  $\epsilon_{\alpha \beta \gamma}$ is the  
antisymmetric tensor. Hereafter, we assume the uniform coupling with $K_j=1$ unless stated otherwise.
We adopt a specific spin-1 representation defined as follows:
\begin{eqnarray}
\label{xyzbases}
\vert x\rangle\!&=&\!\frac{1}{\sqrt{2}}(\vert -1\rangle\!-\!\vert1\rangle),
\vert y\rangle\!=\!\frac{i}{\sqrt{2}}(\vert -1\rangle\!+\!\vert1\rangle),
\vert z\rangle\!=\!\vert 0\rangle, \quad
\end{eqnarray}
where $\vert m\rangle$ represents the eigenstate of the spin operator $S^z$ with eigenvalues $m = -1$, $0$, $1$. In this representation, we have \black{$(S^\alpha)_{ \beta \gamma} = -i\epsilon_{\alpha \beta \gamma}$}.
To parameterize the amplitudes ${A_\alpha}$ for the $\alpha$ component of SIA, we employ spherical coordinates:
\begin{eqnarray}
\label{Threecomponent_sphere}
A_x= A_r \sin \theta \cos \phi,  
A_y= A_r \sin \theta \sin \phi,  
A_z= A_r \cos \theta. \quad
\end{eqnarray}
For convenience, we here define the parameterization of the Hamiltonian in Eq. (\ref{eq:HKA}) using the polar angle $\theta \in [0, 2\pi)$, the azimuthal angle $\phi \in [0, \pi)$, and the radius $A_r$.  \black{The constraint $(\boldsymbol{S}_j)^2 = S(S+1) = 2$ inherently reduces the degrees of freedom for spin-1 operators. For a fixed $\phi$, the parameters $\theta$ and $A_r$ provide two independent degrees of freedom to define the direction of anisotropy. To illustrate, we select $\phi = \pi/4$ and $\phi = 3\pi/4$. At $\phi = \pi/4$, the $x$ and $y$ components are equal, i.e., $A_x = A_y$. In contrast, $\phi = 3\pi/4$ corresponds to a direction where the $x$ and $y$ components have opposite signs, i.e., $A_x = -A_y$.}
%Considering the phase transition from the gapless dimer phase to the Kitaev spin liquid (KSL) phase occurs  at $A_{zc}=-0.655$ for $\theta=\pi$~\cite{Zhangwy2023}, a more general  expression for the critical points is established as follows:
%\begin{eqnarray}
% A_{rc}\!=\!- \!\frac{A_{zc} \sin \theta_0}{  \sin(\theta-\theta_0)  }  \textrm{ and }  \theta_0\!<\!\theta\!<\!\pi\!+\!\theta_0, \textrm{ for } \phi=\pi/4, 
% \end{eqnarray}
%where $\theta_0 \equiv \arctan(\sin^{-1} \phi)=0.3041\pi$. 

To effectively delineate the phase structure, we present the comprehensive phase diagram of $\hat{H}_{\rm KA}$ in Eq. (\ref{eq:HKA}) using three-dimensional $A_\alpha$ coordinates, as shown in Fig. \ref{fig:phase_diagram3D}(a).  The orange and blue lines represent the phase planes for $\phi=\pi/4$ and $\phi=3\pi/4$, respectively, with detailed behaviors explored through cross-sectional views.
%At $\theta=\pi$, a phase transition is triggered by increasing $A_r$ from the Kitaev spin liquid (KSL)  phase to the gapless dimer phase  Considering the phase transition from the gapless dimer phase to the Kitaev spin liquid (KSL) phase occurs at $A_{zc}=-0.655$~\cite{Zhangwy2023}, a more general  expression for the critical points is established as follows:
\black{In both cases, for $\theta=\pi$, a phase transition is induced by increasing $A_r$, shifting from the Kitaev spin liquid (KSL) phase to the gapless dimer phase. Considering that the KSL-dimer transition occurs at $A_{zc}=-0.655$~\cite{Zhangwy2023}, a more general expression for the critical points can be established as follows:
 \begin{eqnarray}
 A_{rc}\!=\!- \!\frac{A_{zc} \sin \theta_0}{  \sin(\theta-\theta_0)  }  \textrm{ and }  \theta_0\!<\!\theta\!<\!\pi\!+\!\theta_0, \textrm{ for } \phi=\pi/4, 
 \end{eqnarray}
 where $\theta_0 \equiv \arctan(\sin^{-1} \phi)=0.3041\pi$.}  For $\phi=\pi/4$ in Fig. \ref{fig:phase_diagram3D}(b), the emergence of the dimer phase is distinctly marked by the grey shaded region, where the grey dashed line that indicates $\theta_0$ as the pivotal angle in relation to the $z$-axis delineates the phase boundary in the limiting case, i.e., $A_r \to \infty$.  Figure \ref{fig:phase_diagram3D}(c) shows the phase diagram of the Hamiltonian in Eq. (\ref{eq:HKED}) for $\phi=3\pi/4$. The dimer phase is represented along the purple line, while the KSL phase encompasses the region beyond this boundary. As discussed in Sec. \ref{sec:KED}, within the KSL phase, the region with $\theta$ ranging from $0$ to $\pi$ is identified as the $x$-ferroquadrupole phase, while for $\theta$ spanning $(\pi, 2\pi)$, it corresponds to the $y$-ferroquadrupole phase. The light blue crosses denote crossover rather than phase transitions, as indicated by the absence of gap closure in the lowest excitation spectrum and the continuity observed in the quadrupole correlations. In fact, the phase diagram is entirely consistent with that of the $U(1)$ quantum link model with a topological $\theta$ angle~\cite{Halimeh2022}.
 The KSL-dimer transition \black{corresponds to} the confinement-deconfinement transition~\cite{Buyens2016}.  The yellow dot marks the 2D $\mathbb{Z}_2$ tricritical point, also known as Coleman's quantum phase transition~\cite{COLEMAN1976}. Since the phases on either side of this critical point are identical, the transition can be bypassed by following specific paths within the parameter space.  \black{In this sense, this tricritical point can be described as an unnecessary quantum critical point~\cite{Senthil2019,Prakash2023}.}
%Figure \ref{fig:phase_diagram3D}(c) depicts the phase diagram of the Hamiltonian given in Eq. (\ref{eq:HKED}) for $\phi=3\pi/4$.  Interestingly, the KSL-dimer transition is equivalent to the confinement-deconfinement transition~\cite{Buyens2016}. Moreover, the yellow dot highlights the 2D $\mathbb{Z}_2$ tricritical point, known as Coleman’s quantum phase transition point~\cite{COLEMAN1976}. Notably, the yellow dot is also dubbed as unnecessary quantum critical point~\cite{Senthil2019,Prakash2023}. Since the phases on either side of this critical point are identical, the transition can be circumvented by following certain symmetric paths within the entire parameter space. 
%In fact, the phase diagram is completely consistent with that of the $U(1)$ QLM with topological $\theta$-angle~\cite{Halimeh2022}. 

\begin{figure}[ht!]
\centering
\includegraphics[width=0.85\columnwidth]{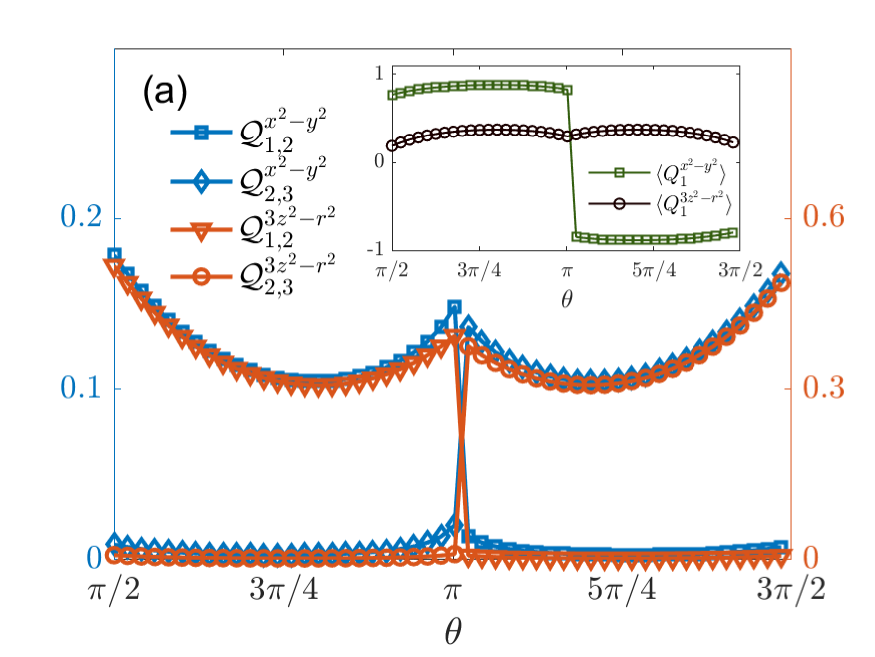}
\includegraphics[width=0.85\columnwidth]{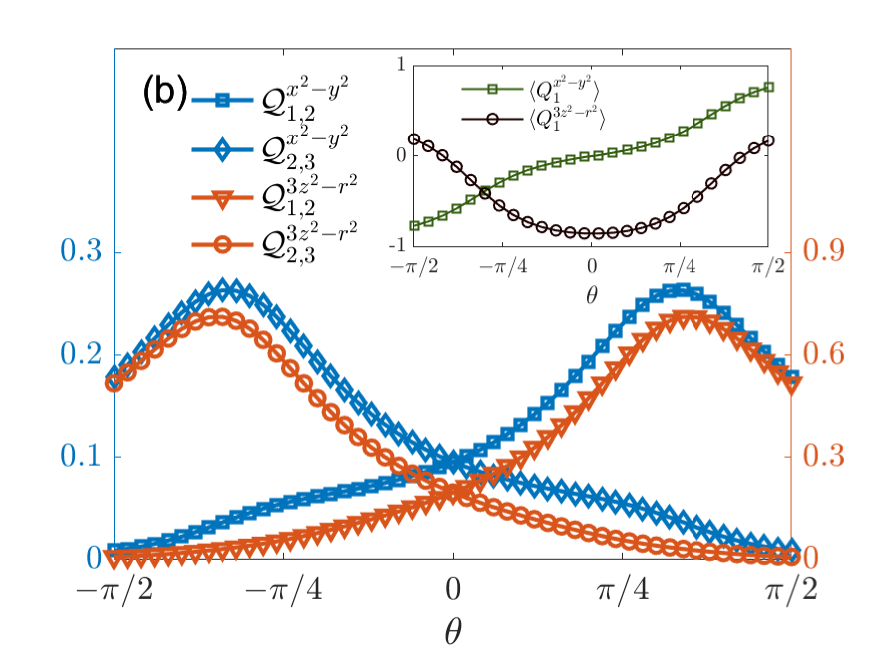}
\includegraphics[width=0.85\columnwidth]{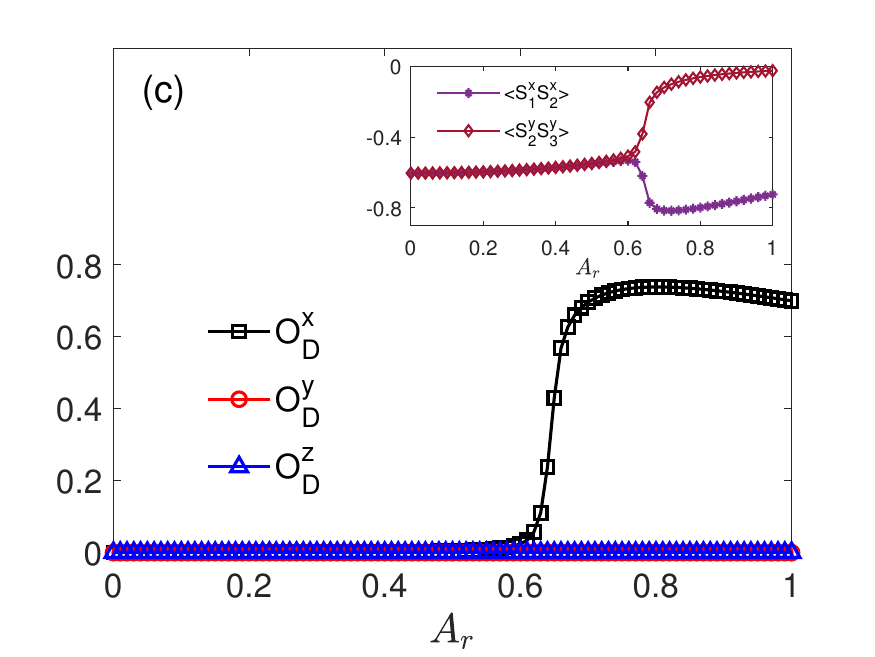}
 \caption{ %Order parameters 
 \black{Ground-state expectation values of various quantities  for the Hamiltonian in Eq. (\ref{eq:HKED}).} (a) Quadrupole correlation functions $\mathcal{Q}_{1,2}^{x^2-y^2}$,  $\mathcal{Q}_{2,3}^{x^2-y^2}$, $\mathcal{Q}_{1,2}^{3z^2-r^2}$ and $\mathcal{Q}_{2,3}^{3z^2-r^2}$ at $A_r=1$ [see Fig. \ref{fig:phase_diagram3D}(c) dark red dashed line] as a function of $\theta$. The inset shows quadrupole moment components $\langle Q_1^{x^2-y^2} \rangle$ and $\langle Q_1^{3z^2-r^2}\rangle$ at $A_r=1$. %[see Fig. \ref{fig:phase_diagram3D}(c) dark red dashed line] as a function of $\theta$.
  (b) Quadrupole correlation functions at $A_r=1$ [see Fig. \ref{fig:phase_diagram3D}(c) green dashed line] as a function of $\theta$. The inset shows quadrupole moment components at $A_r=1$.  %[see Fig. \ref{fig:phase_diagram3D}(c) green dashed line] as a function of $\theta$. 
  (c) The three components of dimer order parameter $O_D$ as a function of $A_r$. The transition from the KSL phase to the dimer phase occurs at $\theta=\pi, A_{rc}=0.655$. \black{The inset shows nonzero two-spin correlation functions. Calculations are performed using the iTEBD method with a bond dimension of  $\chi=120$. } }
  %Here we use the iTEBD method and the bond dimension is set as $\chi=120$.}
 \label{fig:quadrupole}
\end{figure} 

\section{Effect of tunable single-ion anisotropy for $\phi = 3\pi/4$ }
\label{sec:KED}

%Next, we focus on the scenario where $\phi = 3\pi/4$ (i.e., $A_x = -A_y$), the model simplifies to a spin-1 Kitaev chain with both tunable rhombic and uniaxial single-ion anisotropies.
\black{We now focus on the scenario  where $\phi = 3\pi/4$ (i.e., $A_x = -A_y$) and provide a detailed analysis of the results shown in Fig. \ref{fig:phase_diagram3D}(c). In this scenario, the model reduces to a spin-1 Kitaev chain with tunable rhombic and uniaxial single-ion anisotropies.} 
The Hamiltonian in such case is given by
\begin{eqnarray}
\label{eq:HKED}
\hat{H}&=&  \sum_{j=1}^{N/2} (S_{2j-1}^x S_{2j}^x +S_{2j}^y S_{2j+1}^y ) \nonumber \\
&-&\frac{A_{r} \sin \theta}{\sqrt{2}} \sum_{j=1}^N  \left[(S_j^x)^2 - (S_j^y)^2\right] +  A_{r} \cos \theta  \sum_{j=1}^N (S_j^z)^2, \quad \;
\end{eqnarray}
where the second term represents the strength of rhombic SIA and the third term denotes the amplitude of uniaxial SIA.  

%In our investigation of quantum phases in one-dimensional spin-1 Kitaev models endowed with tunable three-component SIAs, we firstly examine the ground state local magnetizations and find $\langle S_j^{\alpha} \rangle=0$ $(\alpha=x, y, z)$, across the model's parameter space. 
In our investigation of quantum phases in one-dimensional spin-1 Kitaev models with tunable three-component SIAs, we first examine the local magnetizations of the ground state and find $\langle S_j^{\alpha} \rangle = 0$ $(\alpha = x, y, z)$ across the entire parameter space. This absence of local magnetization indicates the preservation of time-reversal symmetry.  We then calculate the two-point spin-spin correlations $\mathcal{C}^{\alpha \beta}_{i,j}=\langle S^{\alpha}_i S^{\beta}_j \rangle$, where $\alpha,\beta=x,y,z$. \black{Notably, the Hamiltonian in Eq. (\ref{eq:HKA})  is invariant under the operation $C_{4z} T_1$ that combines a $\pi/2$-rotation about the $z$-axis $(C_{4z})$ and a single-site translation $(T_1)$.} This screw symmetry imposes equalities in the finite-size system expressed as
\begin{eqnarray}
\label{jointsymmetry}
&&\mathcal{C}^{xx}_{1,2} = \mathcal{C}^{yy}_{2,3},\mathcal{C}^{yy}_{1,2} = \mathcal{C}^{xx}_{2,3},  \mathcal{C}^{zz}_{1,2} = \mathcal{C}^{zz}_{2,3}.
\end{eqnarray}
However, this joint symmetry is broken by the emergence of the dimer phase when three-component SIAs are introduced with $A_r > A_{rc}$ for $\phi = \pi/4$ and $\theta_0 < \theta < \pi + \theta_0$. The dimer phase is distinctly marked by alternating patterns in the nearest-neighbor spin-spin correlations. This is characterized by the dimer order parameter defined as $O_D =\vert \langle \boldsymbol{S}_{2} \cdot \boldsymbol{S}_{3} \rangle  -  \langle \boldsymbol{S}_{1} \cdot \boldsymbol{S}_{2} \rangle  \vert$. %However,  our previous research has shown that the dimer phase will emerge upon turning on  three-component SIAs  $A_r>A_{rc}$ with $\phi=\pi/4,\theta_0<\theta<\pi+\theta_0$. The dimer phase  is distinctly marked by a pattern of alternating nearest-neighbor spin-spin correlations and thus breaks the \black{rotational and the} translational symmetry, which is characterized by the dimer order parameter defined by $O_D = \vert \langle \boldsymbol{S}_{1} \cdot \boldsymbol{S}_{2} \rangle  -  \langle \boldsymbol{S}_{2} \cdot \boldsymbol{S}_{3} \rangle  \vert$.
 For a more detailed analysis, it is insightful to examine the dimer order parameter into its component-wise contributions along 
 the $x$, $y$, and $z$  axes, such as
 \begin{eqnarray}
\label{dimerorderparametercomponents}
O_D^x=\mathcal{C}^{yy}_{2,3} \!-\mathcal{C}^{xx}_{1,2},\! 
O_D^y=\mathcal{C}^{xx}_{2,3} \!-\mathcal{C}^{yy}_{1,2},\!  
O_D^z=\mathcal{C}^{zz}_{2,3} \!-\mathcal{C}^{zz}_{1,2}.  
\end{eqnarray}
%\begin{eqnarray}
%\label{dimerorderparametercomponents}
%O_D^x&=&\vert \langle {S}_{1}^x {S}_{2}^x \rangle  -  \langle {S}_{2}^y   {S}_{3}^y \rangle  \vert, \nonumber\\
%O_D^y&=&\vert \langle {S}_{1}^y {S}_{2}^y \rangle  -  \langle {S}_{2}^x   {S}_{3}^x \rangle  \vert, \nonumber\\
%O_D^z&=&\vert \langle {S}_{1}^z {S}_{2}^z \rangle-  \langle {S}_{2}^z  {S}_{3}^z \rangle  \vert .
%\end{eqnarray}
Meanwhile,  we consider the magnetic order in the KSL phase. \black{The KSL is characterized by the quadrupolar order,
 % \sout{ in which the spin-rotational symmetry is broken whereas  both translational} 
where both the screw symmetry  (\ref{jointsymmetry})  and the time-reversal symmetry are preserved.} A symmetric and traceless rank-2 quadrupole tensor operator is given by
\begin{eqnarray}
\label{eq:tensor}
     Q_j^{\alpha\beta}  &=& \frac{1}{2}\left( S_j^\alpha S_j^\beta + S_j^\beta S_j^\alpha \right)-\frac{1}{3} \boldsymbol{S}_j^2 \delta_{\alpha\beta},
\end{eqnarray}
with $\alpha,\beta \in \{ x, y, z\}$ and the Kronecker delta  $\delta_{\alpha\beta}$.
% \sout{at site $j$}. 
\black{Due to the $D_{2h}$ symmetry in Eq. (\ref{eq:HKED}), which involves a 180-degree rotation about three mutually perpendicular axes, the off-diagonal quadrupolar tensor components must vanish}, i.e., $\langle Q_j^{xy} \rangle$ =$\langle Q_j^{yz} \rangle$=$\langle Q_j^{zx} \rangle$=0.
We then focus on the two remaining diagonal components of the quadrupole tensor:
\begin{eqnarray}
\label{eq:Qjxy}
  Q_j^{x^2-y^2} &=& Q_j^{xx}-Q_j^{yy},   \quad
  Q_j^{3z^2-r^2}=\sqrt{3} Q_j^{zz}, 
\end{eqnarray}
where $Q_j^{x^2-y^2}$ corresponds to biaxial quadrupole order, and $Q_j^{3z^2-r^2}$ corresponds to uniaxial quadrupole order. We also examine their respective quadrupole correlation functions:
\begin{eqnarray}
\label{eq:Qijxy}   
 \mathcal{Q}_{i,j}^{\gamma}  &=& \langle Q_i^{\gamma} Q_j^{\gamma} \rangle - \langle Q_i^{\gamma} \rangle \langle Q_j^{\gamma} \rangle, \gamma=x^2-y^2, 3z^2-r^2. \quad \quad
\end{eqnarray}
\black{Nonzero quadrupole moments characterize ground states that reduce the  $D_{2h}$ symmetry to  $C_{2z}$  symmetry
 through anisotropic spin fluctuations.} %, while maintaining both screw and time reversal symmetries.  
 This symmetry reduction is akin to the behavior observed in liquid crystals, where the system exhibits a preferred axis of alignment, indicating that our spin-1 systems are in spin nematic quadrupole phases.

%Nonzero quadrupole moments characterize ground state where spin fluctuations are anisotropic along preferred axis in the absence of  zero net magnetization, while maintaining  $C_2$​ symmetry about the $z$-axis. It can be visualized as precession of the magnetic moment  in the plane perpendicular to this specific axis. For uniaxial quadrupole order, the system retains $C_2$​ symmetry about the $z$-axis, while for biaxial quadrupole order  the symmetry breaks down to $C_2$​ symmetry about the $z$-axis.

%Nonzero quadrupole moments characterize ground states that break  $D_{2h}$  symmetry through anisotropic spin fluctuations, while maintaining  $C_2$​ symmetry about the $z$-axis. It can be visualized as precession of the magnetic moment of every atom in the plane perpendicular to the specific axis, ﬁx the position of the quadrupole-ordering plane hence breaking the axial symmetry.

%Equation (\ref{eq:Qjxy}) describes such a quadrupole ordering that the projection of the magnetic moment of every atom on the $OP$-axis is equal zero, i.e., $\langle S_j^{\epsilon} \rangle = 0$, where the $OP$-axis is the axial symmetry axis, thus   breaking the axial symmetry.  It can be visualized as precession of the magnetic moment of every atom in the plane perpendicular to the $OP$-axis. The presence of quadrupole-ordering plane perpendicular to the symmetry axis of the   in this spin nematic phase disturbs the axial symmetry of the system. Coexistence implies with a noncollinear  plane of quadrupole ordering. 

 In this study, we employ the infinite time-evolving block decimation (iTEBD) technique with the bond dimension set to $\chi=120$~\cite{Vidal2007}, to compute \black{ground-state expectation values of various quantities} for the Hamiltonian Eq. (\ref{eq:HKED}). Specifically, we investigate the quadrupole \black{moments} $\langle Q_j^{\gamma} \rangle$ and  quadrupole correlations $\mathcal{Q}_{j,j+1}^{\gamma}$, particularly at $A_r=1$—as delineated by the dashed line in Fig. \ref{fig:phase_diagram3D}(c)—as a function of the angle $\theta$. These findings are illustrated in Fig. \ref{fig:quadrupole}(a). Notably, the analysis of quadrupole correlations on both odd and even bonds reveals a pronounced discontinuity at $\theta=\pi$, indicating the occurrence of first-order quantum phase transitions. Adjacent to $\theta = \pi/2$, the anisotropy terms  in Eq. (\ref{eq:HKED})  are predominantly characterized by a finite rhombic SIA with a negative coefficient,  while the uniaxial SIA becomes negligible,  which leads to  $\langle Q_j^{x^2-y^2}\rangle$ approaches unity, heralding  the ground state is characterized by $\vert S_j^y=0\rangle$ and  dominant spin fluctuations are within the $xz$ plane. This \black{biaxial} spin nematic state is defined by $y$-ferroquadrupole ordering and persists over the range $\theta \in (\pi/2, \pi)$. For the interval $\theta \in (\pi, 3\pi/2)$, the system's evolution can be understood by a dual mapping $U' = \prod_j \exp \left(i \frac{\pi}{2} S_j^z\right)$, leading to a reorientation of the $x$ and $y$ axes. This transformation coincides with the presence of a finite rhombic SIA bearing a positive sign, under which conditions $\langle Q_j^{x^2-y^2}\rangle$ tends toward -1. Consequently, the system's ground state transitions to a configuration characterized by $\vert S_j^x=0\rangle$, denoting  $x$-ferroquadrupole ordering. Although the quadrupole \black{moment} $\langle Q_j^{x^2-y^2}\rangle$ %is 
 \black{remains} continuous across the $\theta=\pi$, the corresponding correlation functions 
 exhibit a jump. One can find that as $\theta$ varies from $\pi/2$ to $3\pi/2$,  $\mathcal{Q}_{1,2}^{3z^2-r^2}$  changes from  a finite value to zero while 
   $\mathcal{Q}_{1,2}^{3z^2-r^2}$ shows an opposite trend.  $\mathcal{Q}_{i, j}^{x^2-y^2}$ also displays discontinuity at the critical point, yet remains nonzero in both phases. This demonstrates that \black{the two-point correlations $\mathcal{Q}_{i, j}^{x^2-y^2}$ and $\mathcal{Q}_{i, j}^{3z^2-r^2}$} serve as effective order parameter  for identifying \black{biaxial} spin nematic states.  Figure \ref{fig:quadrupole}(b) shows that when $\theta = 0$, the dominant term in the correlation functions undergoes a continuous \black{change}, indicating that the transition from 
   %only the dominant term of the correlation functions undergo a continuous change, indicating that the \black{change} from 
   $x$-ferroquadrupole to $y$-ferroquadrupole is %an intra-phase shift, signifying 
   a crossover rather than a phase transition.

 \section{Effective Hamiltonian for $\phi=3\pi/4$}
\label{sec:Schrieffer-Wolff transformation}

To enhance the understanding of this model, \black{we apply a unitary transformation on the even sites}~\cite{Sen2010}:
\begin{eqnarray}
U &=& \prod_j \exp (i \pi S_{2j}^x) \exp (i \frac{\pi}{2} S_{2j}^z).\label{e_rot}
\end{eqnarray}
\black{This transformation converts the biperiodic Kitaev interactions in Eq. (\ref{eq:HKED}) into a translationally invariant form. The transformed Hamiltonian is given by:}
%\sout{which will rewrite the biperiodic \black{Kitaev} interactions in Eq. (\ref{eq:HKED}) into a translationally invariant form.
%After the unitary transformation, the Hamiltonian Eq. (\ref{eq:HKED}) can be rewritten:}
\begin{eqnarray}
\label{eq:tilde_HKED}
    \tilde{H}&=&K\sum_{j=1}^{N} S_{j}^x S_{j+1}^y+ \frac{A_{r} \sin \theta}{\sqrt{2}}  \sum_{j=1}^N (-1)^{j} \left[(S_j^x)^2 - (S_j^y)^2\right] \nonumber \\
    &+&  A_{r} \cos \theta  \sum_{j=1}^N (S_j^z)^2.
\end{eqnarray}
\black{It is evident that the uniaxial SIA term remains unchanged, while the rhombic SIA term acquires a staggered phase factor due to the transformation. Under the rotation  (\ref{e_rot}),  the local bond parity operators are expressed as}
%\sout{It is easy to see that the uniaxial SIA term remains in its original form, but the rhombic SIA term could get a staggered phase factor. 
%Interestingly, under the rotation (\ref{e_rot}) the local bond parity operators are defined by}
\begin{eqnarray}
\tilde{W}_j = \Sigma_j^y \Sigma_{j+1}^x, \label{equ:tilde W}
\end{eqnarray}
where the site parity operator is defined as 
\begin{eqnarray}
\Sigma_j^a \equiv e^{i \pi S_j^a}. 
 \label{equ:tilde Sigma}
\end{eqnarray}
These operators are diagonal in the bases specified by Eq. (\ref{xyzbases}).
One can readily find that $\tilde{W}_j$ is invariant by inspecting $[\tilde{W}_j,\tilde{H}]=0$. 
Since the eigenvalues of $\Si_j^a$ are $\pm 1$, 
the eigenvalues of $\tilde{W}_j$ in Eq. (\ref{equ:tilde W}) are related to $\mathbb{Z}_2$-valued gauge charges, i.e., $w_j=\pm1$. 
It is straightforward to deduce that for a pair of nearest-neighbor sites $\langle j,j{+}1\rangle$,  total $3{\times}3=9$ allowed states can be distinguished into the  $w_j=1$ sector  spanned by $\vert xy\rangle$, $\vert xz\rangle$, $\vert yx\rangle$, $\vert zy\rangle$, $\vert zz\rangle$ and the $w_j=-1$ sector spanned by $\vert xx\rangle$, $\vert yy\rangle$, $\vert yz\rangle$, $\vert zx\rangle$. 
The Hilbert space $\mathcal{H}$ decomposes into $2^N$ dynamically disconnected Krylov subspaces of varying sizes, $\mathcal{H}=\bigoplus_n \mathcal{K}_n$, each characterized by a configuration vector $\vec{w}=\{w_1, w_n, \ldots, w_N\}$.

\subsection{The north-south direction}
\label{sec:north-south-direction}
We further consider the limiting cases for the Hamiltonian in Eq. (\ref{eq:tilde_HKED}). %$\theta=0$ or $\pi$ (\black{$A_{x}=A_y=0$}).
In the simplest limiting case,  $\theta=0$ and  $A_r \to \infty$ ($A_z \gg K$), all spins are confined to the $\vert 0 \rangle $ state, \black{ resulting in the KSL phase as the ground state. }% and the ground state is obviously the KSL phase. 
 For $\theta=\pi$ and  $A_r \to \infty $ ($A_z \ll -K$), the spins can occupy the $\vert \pm1 \rangle$ states, \black{leading to a phase transition in the ground state}. %Therefore the ground state is expected to undergo a phase transition. 
 The three components (\ref{dimerorderparametercomponents}) of dimer order parameter as a function of $A_r$ at  $\theta=\pi$ are shown in Fig. \ref{fig:quadrupole}(c). \black{As the parameter $A_r$ surpasses the critical value, the $x$ component of the dimer order parameter progressively rises from zero to a non-zero value, signaling a second-order transition at $\theta=\pi, A_{rc}=0.655$.  It is observed that the screw symmetry (\ref{jointsymmetry}) is broken after $A_r > A_{rc}$ due to the imbalance between $\mathcal{C}^{yy}_{2,3}$ and $\mathcal{C}^{xx}_{1,2}$ in the dimer phase. The other components of the two-point spin-spin correlations remain zero and thus  result in the $ y$ and $z$ components of the dimer order parameter staying zero. The two-spin correlation functions shown in Fig. \ref{fig:quadrupole}(c)  demonstrate that the local gauge symmetries of the model strictly determine which expectation values can be zero or non-zero, as explained in Appendix \ref{APPENDIX B}.} 
 %\sout{The $x$ component of the dimer order parameter increases smoothly from zero to a finite value as the parameter $A_r$ increase across the critical value,  indicating a second-order transition occurs at $\theta=\pi, A_r=0.655$. The presence of nonvanishing dimer correlations before the critical point in dimer phase can be attributed to the limitations imposed by the finite bond dimension. 
%The dimer orders are associated with the spontaneous breaking of translational symmetry of $O_D$ in an infinite system. The emergence of the dimer ordering is distinct from the general mechanism for the formation of dimerized phases, which is typically induced by inherent bond alternation and the resulting breaking of translational symmetry. The ground state is two-fold degenerate for dimer phase in the thermodynamic limit, which is in contrast to the gapped ground state for the KSL phase.}
\black{The detection of non-zero dimer correlations prior to reaching the critical point within the dimer phase can be attributed to the limitations imposed by the finite bond dimension. %These dimer orders are linked to the spontaneous breaking of the rotational and the translational symmetry of $O_D$ in an unbounded system. The emergence of this dimer ordering deviates from the conventional pathway leading to dimerized phases, which is usually triggered by inherent bond alternation and the subsequent disruption of the rotational and the translational symmetry. 
In the thermodynamic limit, the ground state for the dimer phase exhibits a two-fold degeneracy,  contrasting with the gapped ground state observed in the KSL phase. }

\subsection{The east-west direction}%$\theta= \pi/2$ and $3\pi/2$}
\label{sec:east-west-direction}
Next,  we consider another limit $A_r \to \infty$ (equivalently $K \to 0$), \black{focusing on  %We analyze 
the scenario  where $\theta=3\pi/2$.  In this case,  %here 
the Hamiltonian (\ref{eq:tilde_HKED}) simplifies to a form  due to the dominance of the staggered rhombic SIA  ($A_{x}=-A_y \gg K$ and $A_z=0$)}:  
\begin{eqnarray}
\label{eq:tilde_HKED2}
    \tilde{H}&=&\sum_{j=1}^{N} S_{j}^x S_{j+1}^y+\frac{A_{r}}{\sqrt{2}}  \sum_{j=1}^N (-1)^{j+1} \left[(S_j^x)^2 - (S_j^y)^2\right]. \quad\quad
\end{eqnarray}
Notably, the condition $ \theta=\pi/2$  can similarly be obtained by a direct spatial translation from $\theta=3\pi/2$. 
%can be obtained through a direct spatial translation related to  $\theta=3\pi/2$.
In the limit %of 
$A_r \to \infty$, the system can be analyzed perturbatively using an effective Hamiltonian approach based on the Schrieffer-Wolff transformation~\cite{Schrieffer1966}. Considering $(S_j^a)^2 =(1-\Sigma_j^a )/2 $ $(a=x,y,z)$,
 we rewrite Eq. (\ref{eq:tilde_HKED2}) as $\tilde{H}=  \tilde{H}_0+\tilde{V}$ ,
where
\begin{eqnarray}
\label{eq:tildeH0}
 \tilde{H}_0 
 &=& \frac{A_{r}}{2\sqrt{2}}  \sum_{j=1}^N (-1)^{j+1} (\Sigma_j^y-\Sigma_j^x),
\end{eqnarray}
and 
\begin{eqnarray}
\label{eq:tildeV}
\tilde{V} = \sum_{j=1}^{N} S_{j}^x S_{j+1}^y.
\end{eqnarray}
To obtain the effective Hamiltonian, \black{ we apply a unitary transformation generated by} %a unitary transformation is applied, which is generated by 
an anti-Hermitian operator $S$:
\begin{eqnarray}
\label{eq:Heff_formula}
   \tilde{H}_{\rm{eff}}=e^S \tilde{H} e^{-S}\equiv \tilde{H}_0 + \sum_{n=1}^{\infty} \tilde{H}_{\rm{eff}}^{(n)},
\end{eqnarray}
where each term  $\tilde{H}_{\rm{eff}}^{(n)}$ is of order $(1/A_{r})^n$. 
The Schrieffer-Wolff generator $S$ is formulated to ensure  
$[\tilde{H}_{\rm{eff}}^{(n)},\tilde{H}_0]=0$ for any $n$.
This is achieved by expanding  $S=\sum_{m=1}^{{\infty}}S^{(m)}$ where each $S^{(m)}$ takes the order $(1/A_{r})^m$.  
Typically, selecting $S \equiv  S^{(1)}$ simplifies the first order term 
$\tilde{H}_{\rm{eff}}^{(1)}=0$,  and allows the computation of higher-order terms based on $S^{(1)}$ and $\tilde{H}_0$. 
In this case, we can obtain  
\begin{eqnarray}
\label{eq:Heff_expansion}
\tilde{H}_{\rm{eff}}\!&=&\tilde{H}_0+  \mathcal{P}([S^{(1)},\tilde{H}_0]+\tilde{V})\mathcal{P} \nonumber \\
    &&+\! \mathcal{P}([S^{(1)},\! \tilde{V}]\!+\!\frac{1}{2!}[S^{(1)},\! [S^{(1)},\! \tilde{H}_0]])\mathcal{P}\!+ \!\cdots \!,
\end{eqnarray}
where $\mathcal{P}$  is the projection operator that maps each term in the expression above onto 
the eigenbasis of $\tilde{H}_0$,  
 \black{specifically projecting onto the subspace where all $\mathbb{Z}_2$ gauge charges are positive, corresponding to the diagonal entries in the site parity matrices defined in Eq. (\ref{equ:tilde Sigma}).}
The requirement for $\tilde{H}_{\rm{eff}}^{(1)}=0$ implies that $S^{(1)}$ must satisfy 
\begin{eqnarray}
\label{eq:S_H_Commutation_relation}
    [S^{(1)},\tilde{H}_0]+\tilde{V}=0.
\end{eqnarray}
%Direct verification shows that Eq. (\ref{eq:S_H_Commutation_relation}) is satisfied  by choosing:
\black{This is satisfied by choosing:}
\begin{eqnarray}
\label{eq:S(1)}
    S^{(1)}=\frac{1}{\sqrt{2}A_{r}} \sum_{j=1}^N (-1)^{j+1} S_j^x \Sigma_j^z S_{j+1}^y.
\end{eqnarray}
%It is readily find that 
It is straightforward to verify that $[S^{(1)},\tilde{W}_j]=0$.
%\sout{The leading-order effective Hamiltonian is then determined as up to the second order:}
\black{The leading-order effective Hamiltonian, up to second order, then becomes:}
\begin{eqnarray}
\label{eq:Heff2}
\tilde{H}_{\rm{eff}}^{(2)} 
&=&  \frac{1}{4 \sqrt{2}  A_{r}} \sum_{j=1}^N (-1)^j (\Sigma_j^z-\Sigma_j^y)(1-\Sigma_{j+1}^y).
\end{eqnarray}
 See Appendix \ref{APPENDIX A} for details. When expanded up to the second order, the effective Hamiltonian (\ref{eq:Heff2}) exhibits a diagonal structure in the specified bases (\ref{xyzbases}), representing an emergent conserved quantity enforced by the Schrieffer-Wolff transformation. The underlying reasoning for this will be further elucidated in the spin-1/2 representation discussed in Sec.\ref{sec:Effect—staggered}.

\begin{figure}[tb]
 \centering
 \includegraphics[width=\columnwidth]{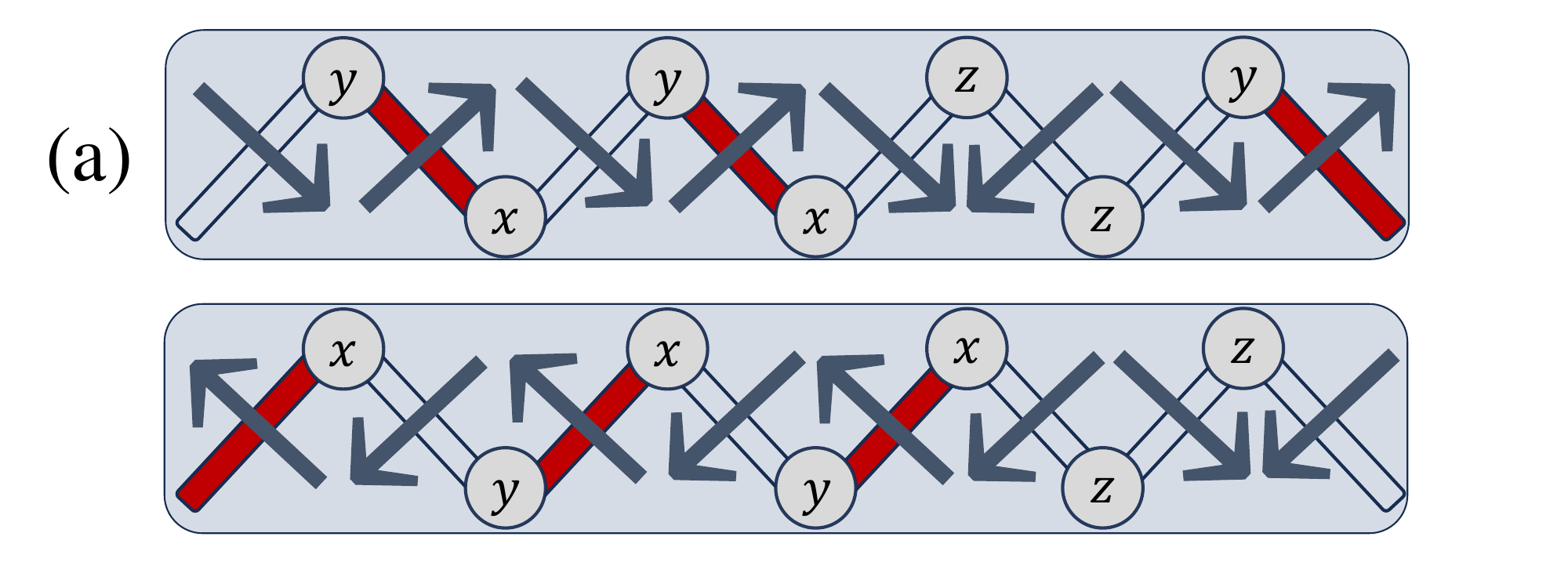}
 \includegraphics[width=\columnwidth]{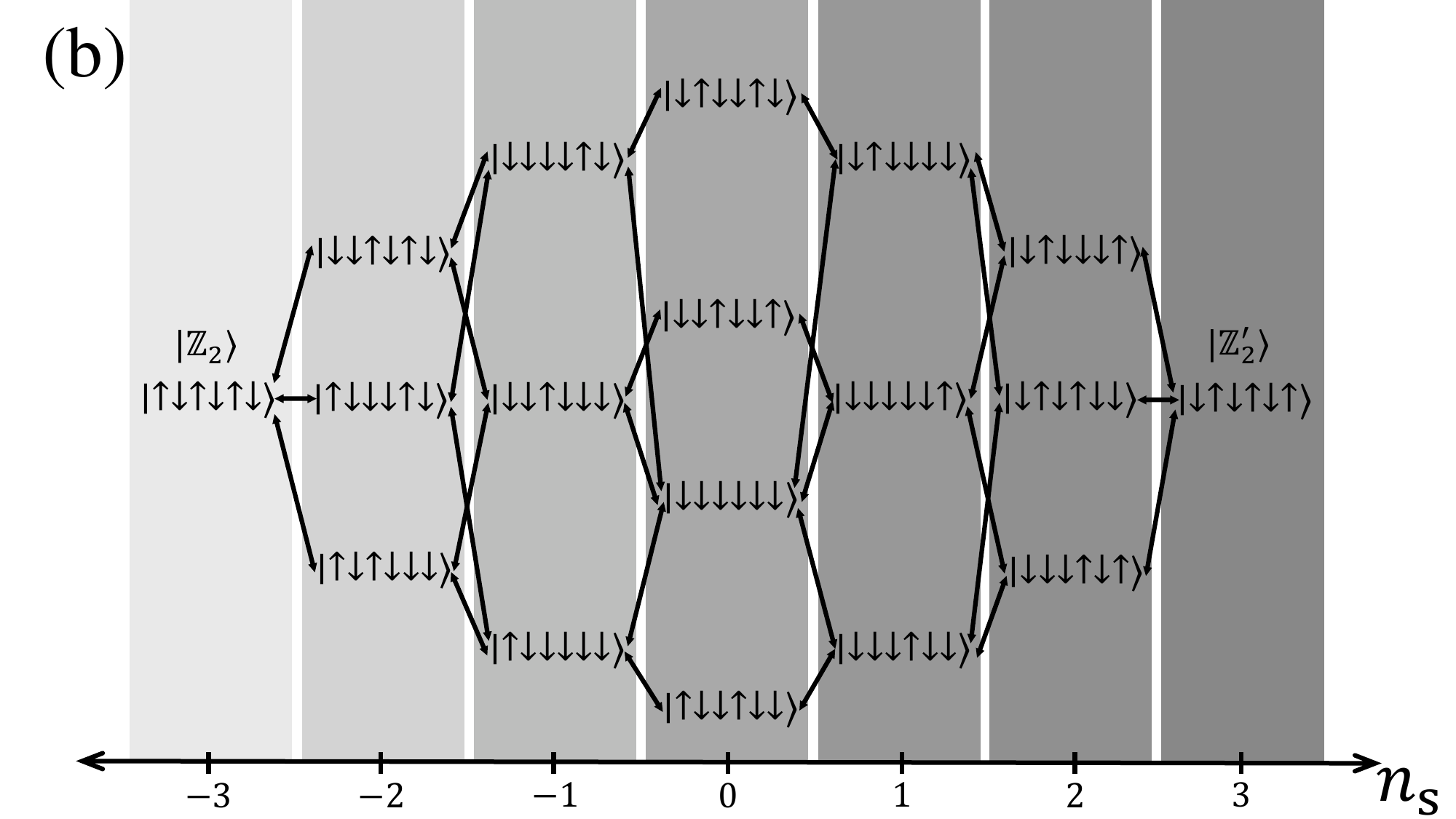}
  \caption{ (a) Illustration of the topological mapping between the spin-1 Kitaev chain and the spin-1/2 PXP model. The primary lattice defined by the spin-1 sites $\{j\}$ is mapped to the dual lattice with spin-1/2 degrees of freedom located at the bond centers $\{j+1/2\}$. The upper figure shows how the spin-1 state $\vert \cdots yxyxzzy \cdots \rangle$ is transformed into the spin-1/2 state $\vert \cdots \downarrow \uparrow \downarrow \uparrow \downarrow \downarrow \downarrow \uparrow\cdots \rangle$, and the lower figure shows the mapping of the spin-1 state $\vert \cdots xyxyxzz \cdots \rangle$ to the spin-1/2 state $\vert \cdots \uparrow \downarrow \uparrow \downarrow \uparrow \downarrow \downarrow \downarrow\cdots \rangle$. (b) The Hilbert space graph of $\hat{H}_{\rm{sdPXP}}$ in Eq. (\ref{eq:sdPXP}) with a sufficiently large value of $A_r$ for $N = 6$ sites with periodic boundary conditions. Each cluster is the system's eigenfunction, and the number of wave functions represents the degeneracy of the system at this energy level. Nodes of the graph are grouped according to the staggered number $n_{\rm s}$, from the $\vert \mathbb{Z}_2 \rangle$ or $\vert \mathbb{Z}^\prime_2 \rangle$ states.  
  }
  \label{fig:mapping} 
 \end{figure}

\section{Mapping to spin-1/2 extended PXP model}
\label{sec:spin-1/2}
In the context of a spin-1 Kitaev chain, \black{it was established that the ground state resides in the sector where all $\mathbb{Z}_2$ gauge charges are positive, represented by $\vec{w}=\{1,1,\cdots,1\}$. Interestingly, this fractal dimension-based subspace is the largest among all possible sectors. In this constrained Hilbert space, there is a one-to-one mapping between the permissible two-site configurations $\langle j, j+1 \rangle$ on the \black{primary} lattice and the corresponding dual lattice~\cite{Zhangwy2023}, where a spin-$1/2$ degree of freedom is located at the bond center, denoted as $i \equiv j+1/2$. This site-to-bond mapping rule schematically illustrated in Fig. \ref{fig:mapping}(a). Within the sector with all positive gauge charges, the Hamiltonian in Eq. (\ref{eq:tilde_HKED}) can be exactly mapped to a compact form in the spin-1/2 representation:}
\begin{eqnarray}\label{eq:H_PXP}
\hat{H}_{\rm ePXP} &=& \sum_{i=1}^N P_{i-1} X_i P_{i+1}+ 2A_r \cos \theta \sum_{i=1}^N  P_{i-1}n_iP_{i+1}\nonumber \\
&-&\sqrt{2}A_{r}\sin \theta \sum_{i=1}^N (-1)^{i}  P_{i-1}n_iP_{i+1},
\end{eqnarray}
where $X=\vert \downarrow \rangle \langle \uparrow \vert +\vert \uparrow \rangle \langle \downarrow \vert $ and $P  = \vert \downarrow \rangle \langle \downarrow \vert $ acts as a projector that ensures neighboring sites are not both in the excited state, and $n =1-P= \vert \uparrow\rangle \langle \uparrow \vert$. \black{Here, the uniform detuning in Eq. (\ref{eq:H_PXP}) is equivalent to the uniaxial SIA in the spin-1 Kitaev model, while the staggered detuning corresponds to the rhombic SIA. }

%it inherits a richness of physical phenomena from the paradigmatic Schwinger model
 \subsection{Effect of uniform detuning}
 \label{sec:Effect—uniform}
Parallel to the discussion in  Sec.\ref{sec:north-south-direction},  tuning $A_r$ along the axis where $\theta=\pi$ in Hamiltonian  Eq. (\ref{eq:H_PXP}) reveals two distinct phases, which become apparent when considering two limiting cases. In the limit of large uniform detuning $A_r$, the system energetically favors atoms in the excited state.
However, the Rydberg blockade prevents neighboring atoms from simultaneously occupying the \black{excited} state. As a result, the most energetically favorable configuration is for half of the atoms to be in the excited state, with the Rydberg atoms occupying every other site in an alternating sequence. This configuration leads to an antiferromagnetic spin configuration in the ground state  $\vert \mathbb{Z}_2  \rangle = \vert  \uparrow \downarrow \uparrow \downarrow  \cdots \rangle$ or $\vert \mathbb{Z}_2^{\prime} \rangle=\vert \downarrow \uparrow \downarrow \uparrow   \cdots \rangle$, which spontaneously breaks the global $\mathbb{Z}_2 $ symmetry, resulting in double degeneracy. 

In the opposite limit, where $A_r=0$,  the ground state of the pure PXP model is gapped and exhibits exponentially decaying correlations in local observables. The Hamiltonian for the pure PXP model is given by
 \begin{eqnarray}
 \label{eq:PXP}
  \hat{H}_{\rm{PXP}}=\sum_{i=1}^N P_{i-1} X_i P_{i+1}.
  \end{eqnarray}
In this limit,  the N\'{e}el state  shows significant overlap with the high-energy scarred eigenstates. The Rydberg blockade imposes kinetic constraints crucial for the atypical dynamics observed in QMBS states.  At  $t = 0$, the system is initialized in product states $\vert \psi(0) \rangle$ and evolves as   $\vert \psi(t) \rangle = \exp(-i\hat{H}_{\rm PXP} t) \vert \psi(0)\rangle$. Quantum quenches from $\vert \mathbb{Z}_2 \rangle$ or $\vert \mathbb{Z}_2^{\prime}\rangle$  reveal periodic revivals in quantum fidelity:
\begin{eqnarray}
\label{eq:fidelity}
F(t) = \vert \langle \psi(0) \vert   \psi(t) \rangle \vert^2.
\end{eqnarray}
These oscillations and non-ergodic behavior stem from the equal spacing of QMBS eigenstates in Eq. (\ref{eq:PXP}). A uniform detuning field transitions the system from a QMBS regime to a quantum critical regime as its strength increases. The initially non-thermal scar states can be tracked into the quantum critical regime by tuning the field. As the system approaches the critical point, it shows increased thermalization tendencies, with scar states evolving into low-energy critical states with low entanglement entropy~\cite{cui2022,zhaih2022}. The von Neumann entanglement entropy for a subsystem consisting of $\ell$ sites is defined as 
\begin{eqnarray}
\label{eq:SL}
S_{\ell} = -{\rm tr} (\rho_\ell^{(n)} \log_2 \rho_\ell^{(n)}),
\end{eqnarray}
 where $\rho_\ell^{(n)}$ is the reduced density matrix obtained by tracing out the $N-\ell$ sites of the complement system from the density matrix $\rho^{(n)} = |\chi_n \rangle \langle \chi_n |$  of each $n$th eigenstate $|\chi_n \rangle$.

% the CP-invariant point of the Schwinger model), the system undergoes a quantum phase transition towards regions of the parameter space in which the parity and charge conjugation symmetries are spontaneously broken

%The two-fold degeneracy is lifted 
When $\theta$ in Eq. (\ref{eq:H_PXP}) deviates from  $\pi$, the staggered potential, which differentiates on-site energies between even and odd sites, selectively stabilizes one of the degenerate states. This asymmetry inhibits the occurrence of a quantum phase transition. \black{The phase structure of the spin-1/2 extended PXP model is in perfect correspondence with  the phase diagram of the spin-1 Kitaev model, as depicted in Fig. \ref{fig:phase_diagram3D}(c).} 
\black{Thus, the Kitaev model is an ideal candidate for simulating the lattice Schwinger model, where the parameter $\theta$ acts like the topological angle. The exact site-to-bond mapping between the Kitaev and PXP models within the ground-state manifold, where the $\mathbb{Z}_2$ gauge charge $w_j=1$ for all bonds, ensures that the $\mathbb{Z}_2$ gauge condition in LGT is automatically met, thus establishing equivalence between the lattice Schwinger model and the Kitaev model. 
According to Eq. (\ref{eq:H_PXP}), both the mass of the matter field and the topological angle can be experimentally controlled in the Kitaev model, enabling exploration of the phase diagram from multiple perspectives. }

\begin{figure}[tb]
\centering
\includegraphics[width=\columnwidth]{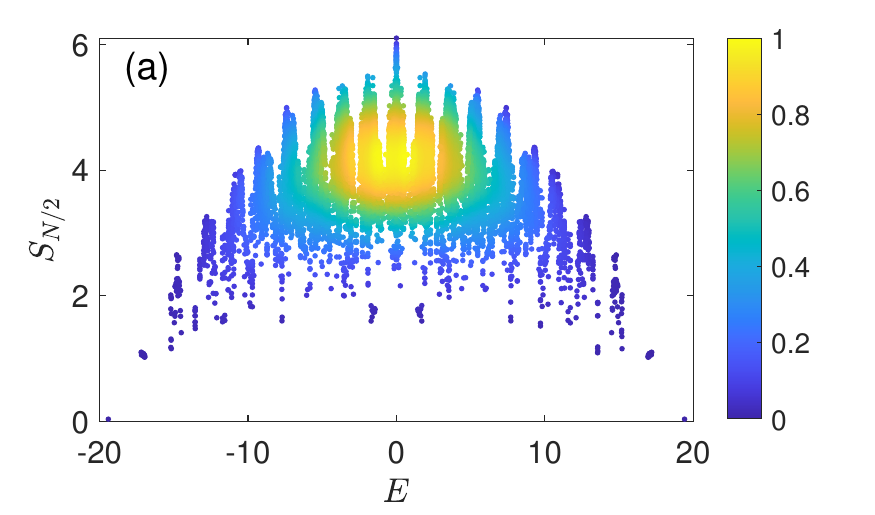}
\includegraphics[width=\columnwidth]{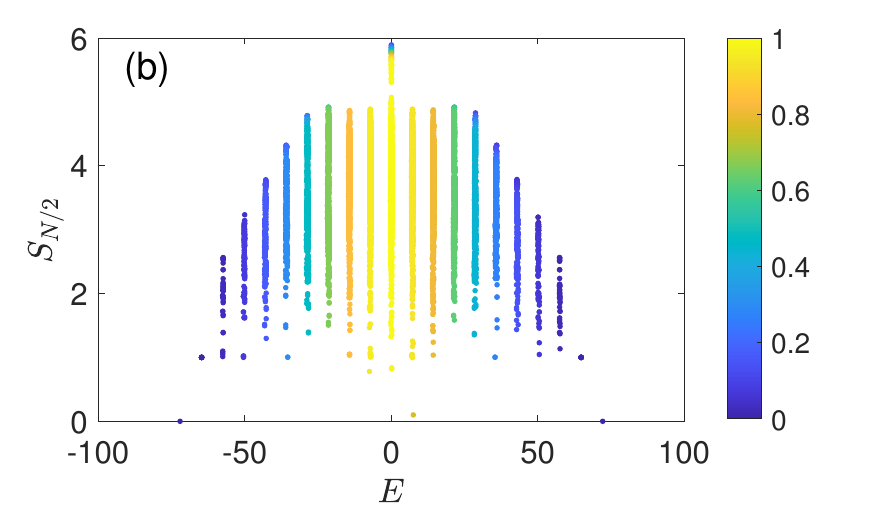}
\includegraphics[width=\columnwidth]{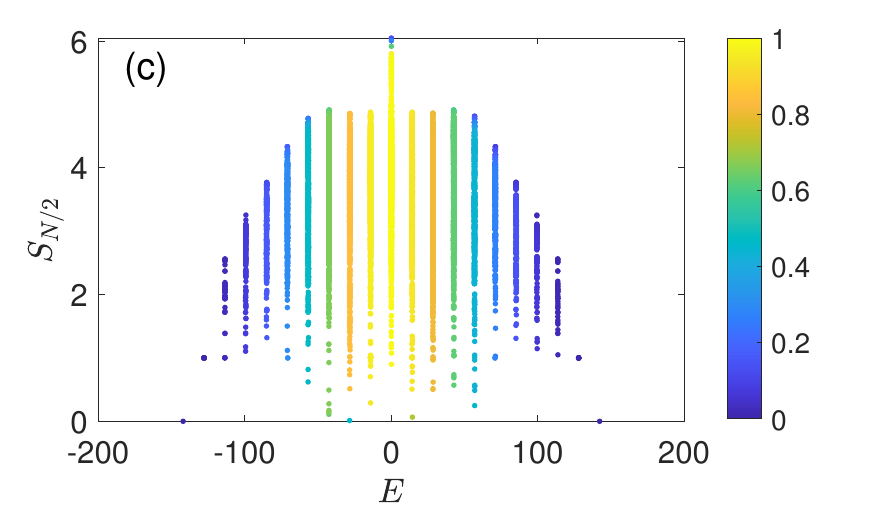}
\caption{Half-chain entanglement entropies $S_{N/2}$ as a function of energy $E$ for the staggered detuned PXP model in Eq. (\ref{eq:sdPXP}). (a) $A_{r}=1$, (b) $A_{r}=5$, and (c) $A_{r}=10$. All data presented are for systems with length $N = 20$.
}
\label{fig:SA_E}
\end{figure}

 \begin{figure}[tb]
\centering
\includegraphics[width=\columnwidth]{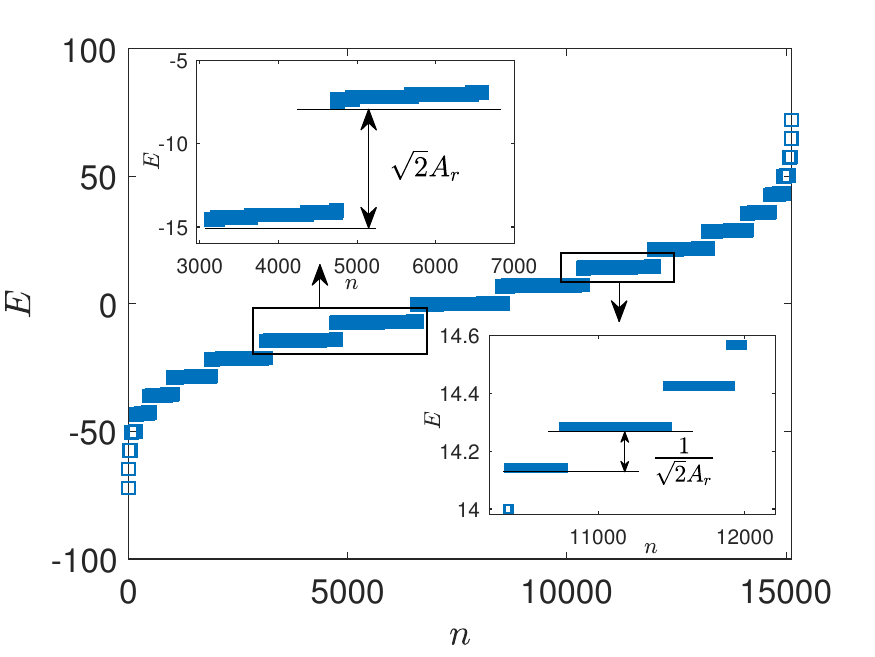}
 \caption{Energy level distribution for $\hat{H}_{\rm{sdPXP}}$ in Eq. (\ref{eq:sdPXP}) with $N=20$ sites. The upper inset shows two sets of energy levels spaced by $\sqrt{2}A_r$, corresponding to the primary clusters. The lower inset shows the energy level spacing $1/\sqrt{2}A_r$ of nearly degenerate states within each cluster, corresponding to the secondary clusters.}
 \label{fig:energy level}
\end{figure}

%\sout{As discussed in Eq. (\ref{eq:H_PXP}), the spin-1 Kitaev model with rhombic and uniaxial single-ion anisotropies can be perfectly mapped to the spin-1/2 PXP model with staggered detuning and uniform detuning within the flux-free subspace~\cite{Zhangwy2023}. The detuning term is ubiquitous in actual experiments. For example, the uniform detuning of the driving laser from the excited state can be finely tuned in cold-atom platforms. The static uniform detuning has been discussed in detail in our previous work, and some interesting results have been obtained. For example, the quench dynamics starting from the product states is symmetric between positive and negative values of the static uniform detuning.  We find that the coherent oscillations of quantum fidelity and certain local observables are sustained against small static uniform detuning perturbations in a quantum quench from special initial states. While the oscillation amplitudes of these observables decay with time as the static uniform detuning strength is increased, the system completely thermalizes upon approaching the critical point.} 

%\sout{\black{In our investigation, we observe that the coherent oscillations of quantum fidelity and specific local observables remain robust against minor static uniform detuning disturbances during a quantum quench originating from particular initial states. Although the amplitude of these oscillations decreases over time as the strength of the static uniform detuning grows, the system attains full thermalization as it nears the critical point.} }

 \subsection{Effect of staggered detuning}
  \label{sec:Effect—staggered}
Next, we \black{focus} on the effect of the staggered detuning, specifically at $\theta=3\pi/2$, \black{as discussed in  Sec. \ref{sec:east-west-direction}.} In this case, the Hamiltonian for the staggered detuned PXP model is given by
\begin{eqnarray}
\label{eq:sdPXP}
\hat{H}_{\rm{sdPXP}}&=&\sum_{i=1}^N \! P_{i-1} X_i P_{i+1} \!+ \! \sqrt{2}A_{r} \! \sum_{i=1}^N \! (-1)^{i} \!  P_{i-1}n_iP_{i+1}. \quad \quad
\end{eqnarray}
Figure \ref{fig:SA_E} shows the half-chain entanglement entropy $S_{N/2}$ as a function of eigenenergy $E$ for different values of the staggered detuning $A_r$. At low $A_r$ [Fig. \ref{fig:SA_E}(a)],  the spectrum is populated by a large number of non-degenerate states. As $A_r$ increases, the energy levels become more uniformly spaced~\cite{Mukherjee2022}. Figure \ref{fig:SA_E}(b) shows a mix of states, including anomalous states with low $S_{N/2}$ and typical thermal state with high $S_{N/2}$, within each primary cluster. 
At sufficiently large $A_r$, the eigenstates of the effective Hamiltonian in Eq. (\ref{eq:sdPXP}) organize into nearly equidistant clusters, as depicted in Fig. \ref{fig:SA_E}(c).  
Figure \ref{fig:energy level} shows the energy level distribution for $\hat{H}_{\rm{sdPXP}}$  at a moderate $A_r$, which clearly identifies several primary sets of energy levels, each separated by $\sqrt{2}A_r$. 
%For systems with an even number of sites $N$, 
The energy levels are organized into $N+1$ distinct sectors. This classification hinges on the staggered occupancy number defined by:
\begin{eqnarray}
\label{eq:ns}
n_{\rm s}=\sum_{j=1}^N (-1)^j n_{j}.
\end{eqnarray}
The staggered occupancy number acts as a metric analogous to the Hamming distance, differentiating states across these sectors, as illustrated in Fig. \ref{fig:mapping}(b).  %where $n_{\rm s}=$ effectively acts as a metric for spatial coordinates, 
It effectively defines physical boundaries between  $\vert \mathbb{Z}_2 \rangle$ and $\vert \mathbb{Z}^\prime_2 \rangle$ states.  
%where $\vert \mathbb{Z}_2^{\prime} \rangle$ is obtained by translating one lattice spacing on $\vert \mathbb{Z}_2 \rangle$. 
\black{Recently the non-Hermitian scar has been proposed~\cite{shen2024},} where the Hamming distance serves as a proxy for spatial coordinates, and these states function as physical boundaries. 
Within primary clusters, secondary clustering occurs with states spaced at intervals of $1/\sqrt{2}A_r$. As $A_r$ increases towards infinity, the degeneracy of the system markedly increases. 
%Despite partial resolution of these degeneracies by the Schrieffer-Wolff Hamiltonian, the spectrum continues to exhibit a considerable degree of degeneracy at finite $A_r$.

\black{Eigenstate clustering in the nonintegrable model of Eq. (\ref{eq:sdPXP}) under large} staggered detuning $A_r$, which leads to violation of ETH at intermediate frequencies, \black{can be understood through the Schrieffer-Wolff Hamiltonian.}
%can be comprehended via the effective spin-1/2 Hamiltonian.  
According to the site-to-bond mapping rule, Eq. (\ref{eq:tildeH0}) can be rewritten as
\begin{eqnarray}
\label{eq:staggered_PnP}
    \hat{H}_{\rm{eff}}^{(0)}\! &=& \! \sqrt{2}A_r  \sum_{i=1}^N (-1)^{i}  P_{i-1}n_iP_{i+1} \equiv \sqrt{2}A_r \hat{H}_{\rm{sPnP}},\quad
\end{eqnarray}
and  Eq. (\ref{eq:Heff2}) is recast into
\begin{eqnarray}
\label{eq:Heff2_spin-1/2}
  \hat{H}_{\rm{eff}}^{(2)}\!=\! \frac{1 }{\sqrt{2} A_{r}} \! \sum_{i=1}^N (-1)^i \!P_{i-1}\!Z_iP_{i+1} \! \equiv \! \frac{1}{2\sqrt{2}A_r} \!\hat{H}_{\rm{sPZP}}.
\end{eqnarray}
Meanwhile, the Schrieffer-Wolff generator $S^{(1)}$  to the spin-1/2 system can also be mapped as
\begin{eqnarray}
\label{eq:S(1)_spin-1/2}
    S^{(1)}\!=\! \frac{i }{ \sqrt{2}A_{r}} \!\sum_{i=1}^{N} (-1)^iP_{i-1}Y_iP_{i+1} \!\equiv\! \frac{i }{ \sqrt{2}A_{r}} \hat{H}_{\rm{sPYP}}.
\end{eqnarray}
\black{The formation of primary clusters is attributed to the nearly integrable nature of the zero-order effective Hamiltonian $\hat{H}_{\rm{eff}}^{(0)}$ in Eq. (\ref{eq:staggered_PnP}) under large staggered detuning.  Secondary clusters, appearing at a much smaller quasi-energy scale, are generated by the second-order effective Hamiltonian $\hat{H}_{\rm{eff}}^{(2)}$ in Eq. (\ref{eq:Heff2_spin-1/2}) at a large $A_r$.   In fact, the diagonal structure can be interpreted as resulting from the roles of Eqs. (\ref{eq:PXP}), (\ref{eq:staggered_PnP}), and (\ref{eq:S(1)_spin-1/2}) as the generators "X", "Z", and "Y"} of a nearly SU(2) algebra.  Specifically, the commutation relations are given by:
\begin{subequations}
\begin{eqnarray}
\label{eq:nearly-SU2}
\left[\hat{H}_{\rm{sPnP}}, \hat{H}_{\rm{PXP}}\right] &=& i \hat{H}_{\rm{sPYP}},   \\
\left[\hat{H}_{\rm{sPnP}}, \hat{H}_{\rm{sPYP}}\right]&=& -i \hat{H}_{\rm{PXP}},     \\
\left[\hat{H}_{\rm{PXP}}, \hat{H}_{\rm{sPYP}}\right] &=& i \hat{H}_{\rm{sPZP}}.  \label{eq:nearly-SU2-b}
\end{eqnarray}
\end{subequations}
It is interesting to note that the right-hand side of Eq. (\ref{eq:nearly-SU2-b}) does not adhere strictly to the SU(2) algebra. This deviation is encapsulated by the expression:
\begin{eqnarray}
\label{eq:zzz}
\hat{H}_{\rm{sPZP}} = (\hat{H}_{\rm{sPnP}}+O_{zzz}),
\end{eqnarray}
 where  $O_{zzz}=\frac{1}{2}\sum_{i=1}^N (-1)^i Z_{i-1} Z_i Z_{i+1}$ represents a residual contribution. Such
nearly SU(2) algebra was studied in the context of QMBS in the PXP model
in Refs.~\cite{Dmitry2019,Iadecola2019,Bull2020}. \black{The term "nearly" SU(2) refers to the fact that while certain aspects of the system exhibit similarities to an SU(2) algebra, deviations occur due to the presence of additional generators.}  The algebraic counterpart for Eq. (\ref{eq:tilde_HKED2}) is detailed in Appendix \ref{APPENDIX A}.

\begin{figure}[tb]
\centering
\includegraphics[width=\columnwidth]{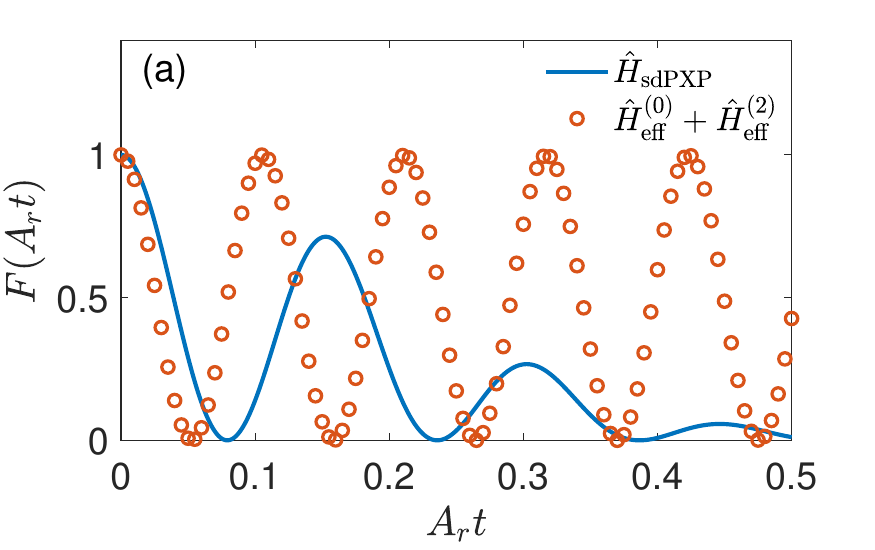}
\includegraphics[width=\columnwidth]{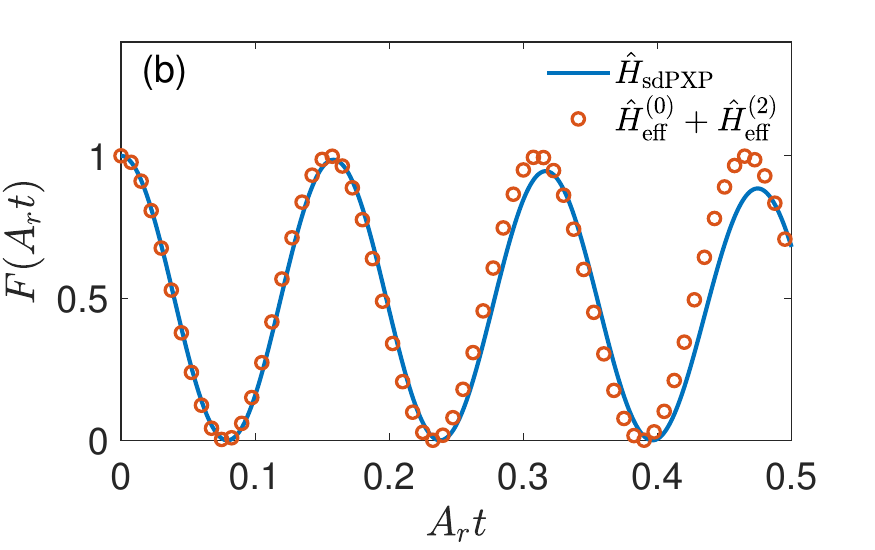}
\includegraphics[width=\columnwidth]{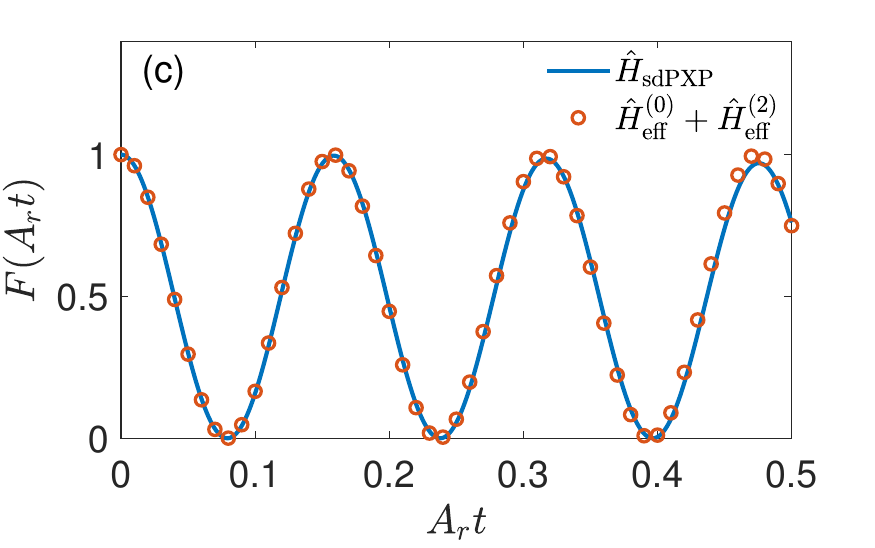}
 \caption{Dynamics of quantum fidelity \black{(\ref{eq:fidelity})}, for the staggered detuned PXP model in Eq. (\ref{eq:sdPXP}) and the effective Hamiltonian $\hat{H}_{\rm{eff}}^{(0)} +\hat{H}_{\rm{eff}}^{(2)}$ [cf. Eqs. (\ref{eq:staggered_PnP}-\ref{eq:Heff2_spin-1/2})] with $N=28$ sites starting from initial state $(\vert \mathbb{Z}_2 \rangle + \vert \mathbb{Z}_2^{\prime} \rangle)/\sqrt{2}$. (a) $A_{r}=1$, (b) $A_{r}=5$, and (c) $A_{r}=10$. Here, the first three periods are chosen for comparison.}
 \label{fig:Ft_H_Heff_comparison}
\end{figure}

\black{The  clustering behaviors  observed  in the staggered detuned PXP chain, which lead to violations of ETH, cannot be attributed to} conventional Hilbert space fragmentation, as Eq. (\ref{eq:sdPXP}) does not exhibit dynamically disconnected sectors for any finite $A_r$. We explore the dynamics at short timescales, focusing on an effective Hamiltonian up to second order, induced by deformations  in Eq. (\ref{eq:sdPXP}). Figure \ref{fig:Ft_H_Heff_comparison} \black{compares} the quantum fidelity between the original Hamiltonian $\hat{H}_{\rm{sdPXP}}$ in Eq. (\ref{eq:sdPXP}) and the effective Hamiltonian $\hat{H}_{\rm{eff}}^{(0)} + \hat{H}_{\rm{eff}}^{(2)}$ [cf. Eqs. (\ref{eq:staggered_PnP}-\ref{eq:Heff2_spin-1/2})]  for various values of $A_r$ at $N=28$. The initial state used for these measurements is the cat state $(\vert \mathbb{Z}_2 \rangle + \vert \mathbb{Z}_2^{\prime} \rangle)/\sqrt{2}$.  Figure \ref{fig:Ft_H_Heff_comparison}(a) shows the effective Hamiltonian exhibits no decay in fidelity during short-time evolution, indicating slow dynamics. At very small values of $A_r$, the fidelity evolution of the effective Hamiltonian diverges clearly from that of the original model in Eq. (\ref{eq:sdPXP}) due to the absence of off-diagonal terms. As $A_r$ increases, the trajectories of the two models begin to converge, but a noticeable deviation appears after the second oscillation period, as illustrated in Fig. \ref{fig:Ft_H_Heff_comparison}(b). Notably, Fig. \ref{fig:Ft_H_Heff_comparison}(c) shows that for $A_r=10$, the periodic revivals in quantum fidelity for the effective Hamiltonian, starting from the superposed initial state $(\vert \mathbb{Z}_2 \rangle + \vert \mathbb{Z}_2^{\prime} \rangle)/\sqrt{2}$, nearly perfectly align with those of the original Hamiltonian $\hat{H}_{\rm{sdPXP}}$ in Eq. (\ref{eq:sdPXP}). This suggests that when $A_r$ reaches around 10, the effective Hamiltonian provides an accurate description of the dynamics of the original model. \black{The oscillation period $T$ is inversely proportional to the energy spacing $\Delta E$ between the $N+1$ primary clusters,  analogous to the periodic revivals of the $\mathbb{Z}_2$ state with period $T=2\pi/\Delta E$ in the PXP Hamiltonian [Eq. (\ref{eq:PXP})], where $\Delta E$ corresponds to the nearly uniform energy spacing between the $N+1$ scar eigenstates~\cite{TurnerPRB2018}.}

\black{In fact, the glassy dynamics of Eq. (\ref{eq:sdPXP}) is directly attributed to the underlying Hilbert space fragmentation (HSF),  which arises from the interplay between gauge symmetry and the emergent “nearly SU(2)” algebra~\cite{Chen2021}.  The fragmentation leads to an extensive division of the many-body Hilbert space into numerous smaller invariant subspaces, with the total number of these subspaces growing exponentially with system size. These disconnected sectors vary in size, ranging from one-dimensional frozen states to finite-dimensional subspaces.  
We have previously identified a hierarchical fragmentation within the spin-1 Kitaev model’s Hilbert space. At the primary level, this fragmentation divides the entire Hilbert space into $2^N$ dynamically isolated Krylov subspaces, characterized by the symmetry sector $\vec{w}=\{w_1,w_n,\cdots,w_N\}$  of gauge charges. The largest of these subspaces, where all $\mathbb{Z}_2$  gauge charges are positive, is referred to as the Fibonacci Hilbert space~\cite{You2022}. Within this subspace, connectivity graphs of the spin-1/2 representation further elucidate the hierarchical fragmentation structure, as illustrated in Fig. \ref{fig:mapping}(b). Thus, subsectors are further distinguished  by their degeneracy levels and clustering, as shown in Table \ref{Tab:Degeneracy}.}

%\sout{The glassy dynamics with memory effects, which imply that the system retains memory of its initial state for longer times due to the restricted dynamics within isolated subspaces, %observed in Fig. \ref{fig:Ft_H_Heff_comparison} highlights the role of    in shaping the system's behavior.  
%subspaces in the local bases (\ref{xyzbases}), arising from the interplay of gauge symmetry and emergent nearly-SU(2) algebra.
% We have previously identified a hierarchical fragmentation within the spin-1 Kitaev model’s Hilbert space.  The primary level of fragmentation divides the entire Hilbert space into $2^N$ dynamically isolated Krylov subspaces. These subspaces vary in size and are characterized by symmetry sector $\vec{w}=\{w_1,w_n,\cdots,w_N\}$ of  gauge charge. The largest of these subspace, imposed by all positive $\mathbb{Z}_2$ gauge charges, is termed the Fibonacci Hilbert space~\cite{You2022}. Connectivity graphs of the spin-1/2 representation  further delineate the hierarchical fragmentation structure  in Fig. \ref{fig:mapping}(b),  where the subsectors are further differentiated by their degeneracy levels and clustering, as shown in Table \ref{Tab:Degeneracy}.  
% } 

\begin{figure}[tb]
\centering
\includegraphics[width=\columnwidth]{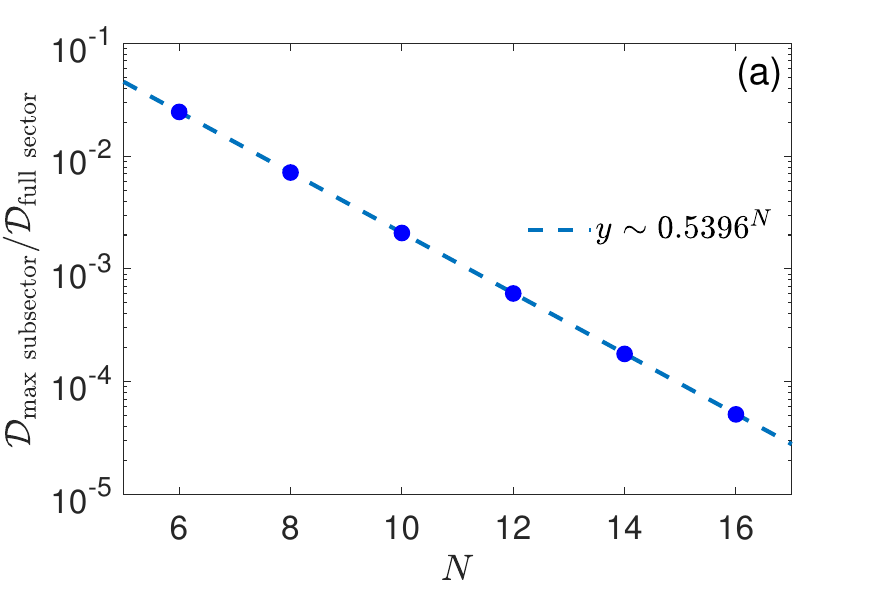}
\includegraphics[width=\columnwidth]{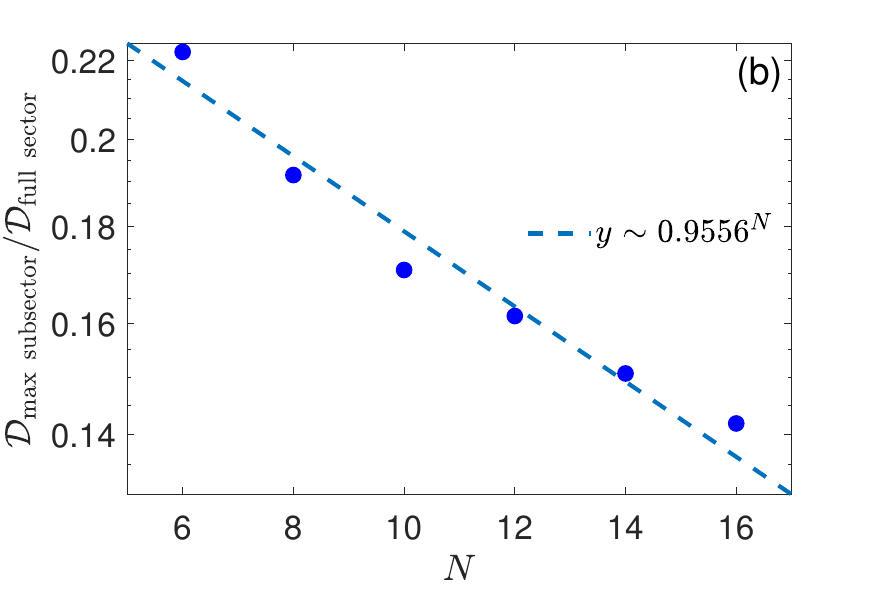}
 \caption{$\mathcal{D}_{\rm max \ subsector}/\mathcal{D}_{\rm full \ sector}$ as the function of $N$ for different spin. Strong fragmentation for (a) the spin-1 Kitaev chain in Eq. (\ref{eq:tilde_HKED2}) and (b) the spin-1/2 Hamiltonian $\hat{H}_{\rm{sdPXP}}$ in Eq. (\ref{eq:sdPXP}) with a sufficiently large value of $A_r$.}
 \label{fig:fragment}
\end{figure}

One can find  exponential growth in %both the number of clusters and 
the dimensionality of the Hilbert space with increasing system size $N$ in each cluster. 
In  particular,  the largest symmetry sector within the Fibonacci space is identified by $n_s= 0$.  This hierarchy is distinguished by how the number of disconnected subspaces within a symmetry sector scales with the system size $N$, offering insights into the varieties of fragmentation. In systems exhibiting strong fragmentation, the largest connected subspace within any given symmetry sector is exponentially smaller than the entire sector. The ratio of the dimension of the largest connected subspace $\mathcal{D}_{\rm max \ subsector }$  to the dimension of the full symmetry sector   ${\mathcal{D}_{\rm full \ sector}}$ decreases exponentially with the system size $N$~\cite{Sala2020,Yang2020}
\begin{eqnarray}
\label{eq:HSF}
    \frac{\mathcal{D}_{\rm max \ subsector}}{\mathcal{D}_{\rm full \ sector}} \sim {\rm exp}(-\kappa N),
\end{eqnarray}
where $\kappa > 0$. Otherwise, the system is said
to be weakly fragmented. Figure \ref{fig:fragment} shows both the spin-1 Kitaev chain in Eq. (\ref{eq:tilde_HKED2}) and the staggered detuned PXP model in  Eq. (\ref{eq:sdPXP}) with a sufficiently large value of $A_r$ display strong HSF. Remarkably, the decay rate in the former case is more pronounced, suggesting that the Hilbert space of the spin-1 Kitaev chain is indeed  more fragmented  than the staggered detuned PXP model.

  \begin{table}
%\footnotesize
\centering
\begin{tabular}{|c|ccccccccccccc|c|c|}
  \hline
  $N$&\multicolumn{13}{c|}{Degeneracy} & $n_{\rm clustering}$ & Dimension \\\hline
  $4$  & & & & &$1$&$2$&$1$&$2$&$1$& & & &     &$5$&$7$\\ 
  $6$  & & & &$1$&$3$&$3$&$4$&$3$&$3$&$1$& & &      &$7$&$18$\\ 
  $8$  & & &$1$&$4$&$6$&$8$&$9$&$8$&$6$&$4$&$1$& &    &$9$&$47$\\ 
  $10$ & &$1$&$5$&$10$&$15$&$20$&$21$&$20$&$15$&$10$&$5$&$1$&    &$11$&$123$\\ 
  $12$ &$1$&$6$&$15$&$26$&$39$&$48$&$52$&$48$&$39$&$26$&$15$&$6$&$1$     &$13$&$322$\\\hline  
\end{tabular}
\caption{The degenerate manifold encoded in Eq. (\ref{eq:sdPXP}) for a sufficiently large value of $A_r$, clustering number and the dimension of Hilbert space for different even system sizes $N$.}
\label{Tab:Degeneracy}
\end{table}

\begin{figure}[h!]
\centering
\includegraphics[width= \columnwidth]{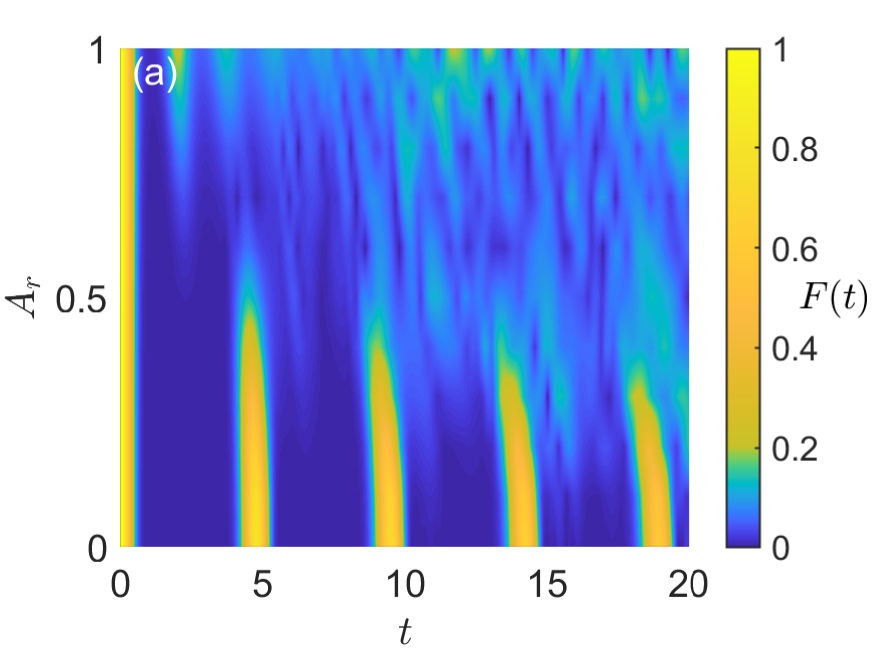}  
\includegraphics[width= \columnwidth]{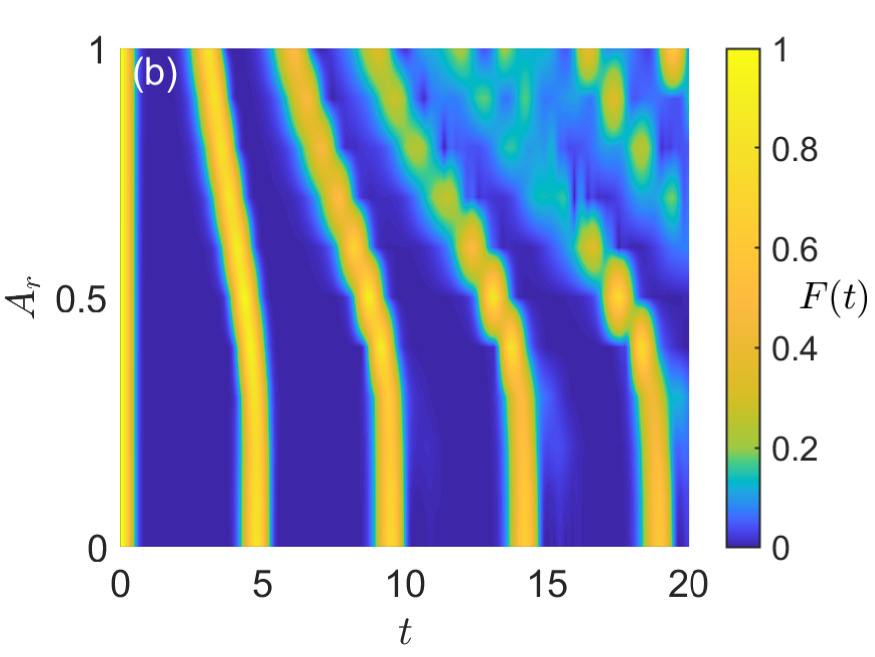}  
 \caption{
 Contour maps showing the time evolution of fidelity $F(t)$ (\ref{eq:fidelity})  for the extended PXP model given by Eq. (\ref{eq:H_PXP}). Panel (a) shows results for a uniform detuning with $\theta=\pi$, while panel (b) presents results for a staggered detuning with $\theta=3 \pi/2$. The data were obtained by ED for a system size of $N=28$, with the quench initiated from the $\vert \mathbb{Z}_2 \rangle$ state. 
 \label{fig:Ftcontour}
 }
\end{figure}

\black{Finally,  we compare the effects of the uniform and staggered detunings on the non-ergodicity behavior of the PXP model in Fig. \ref{fig:Ftcontour}. Since uniaxial SIA corresponds to uniform detuning and rhombic SIA to staggered detuning, these effects can be seen as the influence of two types of SIAs on the non-ergodic behavior of the spin-1 Kitaev chain. Figure \ref{fig:Ftcontour}(a) shows that as $A_r$ approaches $A_{rc}$, periodic fidelity revivals weaken and vanish, indicating complete thermalization. Conversely, Fig.~\ref{fig:Ftcontour}(b) shows that staggered detuning sustains revivals, preserving non-thermal behavior. This contrast illustrates how quantum criticality transitions the system from non-ergodic to thermalizing dynamics as detuning increases.
}

\section{Summary}
\label{Summary and conclusions}
In summary, we study a one-dimensional spin-1 Kitaev model with three-component single-ion anisotropies (SIAs). To effectively describe the effects of SIAs on ground-state properties and far-from-equilibrium dynamics, we use a three-dimensional coordinate, i.e., $A_x\!=\!A_r \sin \theta \cos \phi$, $A_y\!=\!A_r \sin \theta \sin \phi$ and $A_z\!=\!A_r \cos \theta$. At zero SIA field ($A_r=0$), the local bond operators defined in (\ref{equ:tilde W}) of the pure Kitaev chain are conserved with eigenvalues $w_j=\pm1$. Even in the presence of SIAs, these local bond operators retain their conservation, playing a role of the gauge charge. In terms of the infinite time evolving block decimation (iTEBD) method, the comprehensive phase diagram is given. In the zero SIA field limit, the Kitaev chain hosts the Kitaev spin liquid (KSL) ground state, characterized by extremely short-ranged spin-spin correlations. At $\theta=\pi$, where the $\phi$ is irrelevant, a phase transition is triggered by increasing $A_r$ from the KSL phase to the dimer phase. Due to the spontaneous breaking of \black{global $\mathbb{Z}_2 $} symmetry, the two-fold degenerate dimer phase is analogous to the deconfined phase in the lattice Schwinger model, which incorporates a celebrated topological $\theta$ angle of $\pi$. To this end, the KSL-dimer transition is equivalent to the confinement-deconfinement transition. When a translation-symmetry-breaking term is added, the topological angle $\theta$ starts to deviate from $\pi$. In this case, such Coleman's transition no longer exists and the deconfined line disappears as a consequence of charge-parity symmetry breaking. We here find that the polar angle $\theta$ can act as the topological angle in the U(1) lattice gauge theory. When $\phi = \pi/4$, the deconfined line expands into a semi-infinite region, while for other $\phi$ values, it remains a simple deconfined line.

We focus on the scenario where $\phi = 3\pi/4$, where the SIAs can be recast into the mixture of uniaxial and rhombic SIAs. Under these circumstances, such two types of SIAs can be tuned independently. Notably, the varying uniaxial SIA induces the KSL-dimer transition at $\theta = \pi$ and a finite rhombic SIA can modify $\theta$ away from $\pi$. Considering the local quadrupole \black{operator} (\ref{eq:tensor}) coupled with the rhombic SIA in Hamiltonian (\ref{eq:HKED}), the spin nematic order, as a consequence of lowering $D_{2h}$ symmetry down to  $C_{2z}$ , will be induced when the rhombic SIA becomes predominant. Consequently, a positive rhombic SIA leads to the emergence of the $x$-ferroquadrupole phase, whereas a negative sign of the rhombic SIA stabilizes the $y$-ferroquadrupole phase. Interestingly, as the rhombic SIAs are tuned from negative to positive values, the system from the $y$-ferroquadrupole to the $x$-ferroquadrupole phases can either pass through the deconfined line via a first-order phase transition or undergo a crossover through the KSL phase, which gives us an example of unnecessary criticality.

It has been recognized the spin-1/2 PXP model is embedded in the ground-state manifold of the spin-1 Kitaev model, where the topological charges are $+1$.
The quantum quench from either $\vert \mathbb{Z}_2 \rangle$ or $\vert \mathbb{Z}_2^{\prime} \rangle$ states exhibits periodic revivals in the quantum fidelity at the PXP model, referred to as quantum many-body scars (QMBSs)\black{~\cite{Zhangprl2023,Windt2022}}. To investigate the dynamical properties far from the pure Kitaev limit, we utilize a Schrieffer-Wolff transformation to develop a perturbative description of the spin-1 Kitaev chain in the large limit of rhombic SIA. We find that for a sufficiently large value of $A_r$, the effective Hamiltonian, expanded up to the second order, remains diagonal. In order to provide a clearer explication, we map the spin-1 Hamiltonian (\ref{eq:tilde_HKED}) to the spin-1/2 extended PXP Hamiltonian (\ref{eq:sdPXP}) exactly in the ground-state manifold characterized by all positive $\mathbb{Z}_2$ gauge charges, where the uniaxial SIA is equivalent to the uniform detuning while the rhombic SIA plays a role of staggered detuning. For the PXP model with spatially uniform detuning, the mapping to U(1) lattice gauge theory is that the system hosts a regime exhibiting quark-antiquark confinement, which provides a route to investigate confinement in continuous gauge theories with quantum simulators. Intriguingly, for the case of the PXP model with staggered detuning, a Schrieffer-Wolff transformation adopted in the large staggered detuning enforces an additional conservation law for the staggered occupation number $n_s$ (\ref{eq:ns}). We show that when the staggered detuning is sufficiently large, the eigenspectrum is strongly clustered. In the large staggered detuning limit, an additional conservation law for the staggered occupation number $n_s$ emerges and the energy level exhibit a equally spaced distributions.

Returning to the spin-1 Kitaev chain, due to the existence of the conserved gauge charges $\hat{W}_j$ (\ref{equ:tilde W}), a strong fragmentation of the Hilbert space into exponentially many disconnected sectors. The largest subspace with all positive gauge charges corresponds to the Fibonacci Hilbert space, reminiscent of a spin-1/2 system with Rydberg blockade. We trace the origin of the emergent conserved quantity (\ref{eq:Heff2_spin-1/2}) to the nearly SU(2) algebra, which is also believed to underlie QMBS in the PXP model at zero staggered field. Furthermore, the emergent symmetry (\ref{eq:Heff2_spin-1/2}) leads, at the second order in the effective Hamiltonian description, to a substantially more fragmented Hilbert space. The interplay of these two conservation laws results in hierarchical Hilbert space fragmentation (HSF), leading to glassy dynamics in specific limiting cases, which reveals no decay in fidelity over the short-time evolution. Despite the surprising algebraic connection, such slow dynamics belongs to a scenario distinct from the QMBS arising from the Fibonacci constraint. Our work not only introduces a theoretical perspective in simulating the topological $\theta$ angle and ergodicity-breaking dynamics, but also paves the way for exploring higher-spin generalizations of the scarred model. Such solid-state-based quantum systems could be experimentally realized by state-of-the-art cold-atom quantum simulators.

%\black{In general, QMBS states are special non-thermal eigenstates found in certain many-body systems. These states are embedded within a spectrum of otherwise thermalizing states and are characterized by atypically low entanglement entropy. The existence of QMBS states leads to violations of the eigenstate thermalization hypothesis (ETH), which typically governs the thermalization process in isolated quantum systems. Ergodicity breaking in systems with QMBS states occurs because these special states do not follow the typical thermalization dynamics. Instead of the system exploring the entire available phase space (ergodic behavior), the presence of QMBS states causes the system to exhibit long-lived coherent oscillations and periodic revivals, thus preventing complete thermalization. This behavior is in stark contrast to what is expected in ergodic systems where all eigenstates contribute equally to thermal equilibrium.}

%\black{Both types of SIA lead to non-ergodic behavior in the system. Uniaxial SIA corresponds to uniform detuning, which causes the system to fully thermalize near the phase transition point. Rhombic SIA corresponds to staggered detuning, which induces slow dynamics different from those in the PXP model. It should be noted that the large correlation length near the quantum critical point can enhance quantum ergodicity, leading to a system that is more inclined towards thermalization~\cite{cui2022,zhaih2022}. This enhanced quantum ergodicity helps achieve a transition from non-thermal to thermal states.} 

\begin{acknowledgments}
We thank Gaoyong Sun, Chengjie Zhang, Xuejia Yu, \black{Yi Lu and Zhiling Wei} for valuable discussions on related work. This work is supported by the National Natural Science Foundation of
China (NSFC) under Grant No. 12174194, Postgraduate Research \&
Practice Innovation Program of Jiangsu Province, under Grant No. KYCX23\_0347, Opening Fund of the Key Laboratory of Aerospace Information Materials and Physics (Nanjing University of Aeronautics and Astronautics), MIIT, Top-notch Academic Programs Project of Jiangsu Higher Education Institutions (TAPP), and stable supports for basic institute research under Grant No. 190101. 
\end{acknowledgments}

\appendix

\section{ Symmetry Considerations on Dimer Order Parameters }
\label{APPENDIX B}

We here show the ground-state expectation values of certain operators can be analysed through symmetry considerations. Under the unitary transformation (\ref{e_rot}), the three components of the dimer order parameter in Eq.(\ref{dimerorderparametercomponents}) can be expressed by 
 \begin{eqnarray}
 \tilde{O}_D^x&=&  \langle {S}_{2}^x {S}_{3}^y \rangle -  \langle {S}_{1}^x   {S}_{2}^y \rangle, \nonumber \\
 \tilde{O}_D^y&=&   \langle{S}_{2}^y {S}_{3}^x \rangle -  \langle {S}_{1}^y   {S}_{2}^x \rangle, \nonumber \\
 \tilde{O}_D^z&=&  \langle {S}_{2}^z {S}_{3}^z \rangle -  \langle {S}_{1}^z  {S}_{2}^z \rangle.
 \end{eqnarray}
Recalling that $\tilde{W}_j = e^{i \pi S_{j}^y} e^{i \pi S_{j+1}^x}$  commutes components in the dimer order parameter, such as $S_j^x S_{j+1}^y$, $S_j^y S_{j+1}^x$ and  $S_j^z S_{j+1}^z$. To this end, we can apply the action of  
\begin{eqnarray}
\tilde{W}_{j+1}=e^{i \pi S_{j+1}^y} e^{i \pi S_{j+2}^x} \nonumber
\end{eqnarray}
 on the components in the dimer order parameter,  which arrives
  \begin{eqnarray}
   \tilde{W}_{j+1}S_j^x S_{j+1}^y \tilde{W}_{j+1}^{-1} &=& S_j^x S_{j+1}^y, \label{Wxy}   \\
  \tilde{W}_{j+1}S_j^y S_{j+1}^x \tilde{W}_{j+1}^{-1} &=& -S_j^y S_{j+1}^x, \label{Wyx} \\
   \tilde{W}_{j+1}S_j^z S_{j+1}^z \tilde{W}_{j+1}^{-1}&=& -S_j^z S_{j+1}^z. \label{Wzz}
  \end{eqnarray}
Since $\tilde{W}_{j+1}$ commutes with the Hamiltonian Eq. (\ref{eq:tilde_HKED}),  the ground state $\vert \psi_0 \rangle$ must be invariant under the action of $\tilde{W}_{j+1}$. In other words, $\vert \psi_0 \rangle$ is an eigenstate of $\tilde{W}_{j+1}$, i.e., $\tilde{W}_{j+1} \vert \psi_0 \rangle = \vert \psi_0 \rangle$. Therefore, according to Eq. (\ref{Wyx}), it implies
\begin{eqnarray}
\langle \psi_0 \vert W_{j+1}S_j^y S_{j+1}^x W_{j+1}^{-1} \vert \psi_0 \rangle = - \langle \psi_0 \vert S_j^y S_{j+1}^x \vert \psi_0 \rangle.
\end{eqnarray}
It is easy to find $\langle \psi_0 \vert S_j^y S_{j+1}^x \vert \psi_0 \rangle =0$.
 Similarly, according to Eq. (\ref{Wzz}), we have
 \begin{eqnarray}
 \langle \psi_0  \vert  S_j^z S_{j+1}^z \vert \psi_0  \rangle=0.
 \end{eqnarray}
  However, since $\tilde{W}_{j+1}$ commutes with  $ S_j^x S_{j+1}^y $ according to Eq. (\ref{Wxy}),  the ground-state expectation value $\langle S_j^x S_{j+1}^y \rangle$ is not forced to be zero by the gauge symmetry. In a word, the ground-state expectation values of $S_j^y S_{j+1}^x$ and $S_j^z S_{j+1}^z$ must be zero, while those of $S_j^x S_{j+1}^y$ can be nonzero.

\section{Schrieffer-Wolff transformation and nearly SU(2) algebra in spin-1 systems}
\label{APPENDIX A}

The derivation process is as follows. First, we define the projection operator that projects the Hamiltonian to the flux-free sector $\vec{w}{=}\{1,1,\cdots,1\}$
\begin{eqnarray}
    \mathcal{P} = \prod_j (\frac{\Sigma_j^y \Sigma_{j+1}^x +\mathbbm{1}}{2}).
\end{eqnarray}
The requirement for $\tilde{H}_{\rm{eff}}^{(1)}=0$ implies that $S^{(1)}$ must satisfy $[S^{(1)},\tilde{H}_0]+\tilde{V}=0.$ Next, we calculate the commutation relations in the constrained Hilbert space
\begin{eqnarray}
    &&\mathcal{P} [S^{(1)},H_0] \mathcal{P} \nonumber \\
    =&& \frac{1}{2}\sum_{j=1}^N\left(\mathcal{P} \left[(-1)^{j+1} S_j^x \Sigma_j^z S_{j+1}^y,   (-1)^{j+1} \left( (S_j^x)^2 - (S_j^y)^2 \right) \right] \right. \mathcal{P} \nonumber \\
     &&+\left.  \mathcal{P} \left[(-1)^{j+1} S_j^x \Sigma_j^z S_{j+1}^y,   (-1)^{j+2} \left( (S_{j+1}^x)^2 - (S_{j+1}^y)^2 \right) \right]\mathcal{P}\right) \nonumber \\
    =&& \frac{1}{2} \left(-S_j^xS_{j+1}^y -S_j^xS_{j+1}^y \right)\nonumber \\
    =&& -\tilde{V}.
\end{eqnarray}
Therefore, the selected $S^{(1)}$ satisfies $[S^{(1)},H_0]+V=0$. The second-order effective Hamiltonian can be calculated as
\begin{eqnarray}
    \tilde{H}_{\rm{eff}}^{(2)}&=& \frac{1}{2}\mathcal{P} [S^{(1)},V]\mathcal{P} \nonumber \\
    &=&\frac{1}{2} \mathcal{P}\left[\frac{1}{\sqrt{2}A_{r}} \sum_{j=1}^N (-1)^{j+1} S_j^x \Sigma_j^z S_{j+1}^y,\sum_{j=1}^{N} S_{j}^x S_{j+1}^y \right] \mathcal{P} \nonumber \\
    &=&\frac{1}{\sqrt{2}A_{r}}\sum_{j=1}^N (-1)^{j}(S_j^x)^2\Sigma_j^z(S_{j+1}^y)^2 \nonumber \\
    &=&  \frac{1}{4 \sqrt{2}  A_{r}} \sum_{j=1}^N (-1)^j (\Sigma_j^z-\Sigma_j^y)(1-\Sigma_{j+1}^y).
\end{eqnarray}

The diagonality of $\tilde{H}_{\rm{eff}}^{(2)}$ in Eq. (\ref{eq:Heff2}) is a consequence of the fact that $\tilde{H}_0$ in Eq. (\ref{eq:tildeH0}) and $\tilde{V}$ in Eq. (\ref{eq:tildeV}) can be viewed as generators of a “nearly SU(2)” algebra. The nearly SU(2) algebra is generated by the operators $\tilde{H}_{\rm n}$, $\tilde{H}_{\rm X}$, and $\tilde{H}_{\rm Y}$, defined by
\begin{eqnarray}
\tilde{H}_{\rm n} &\equiv& \frac{\tilde{H}_0}{\sqrt{2}A_r} = \frac{1}{4} \sum_{j=1}^N (-1)^{j+1} (\Sigma_j^y-\Sigma_j^x), \\
\tilde{H}_{\rm X} &\equiv& \tilde{V} = \sum_{j=1}^{N} S_{j}^x S_{j+1}^y, \\
\tilde{H}_{\rm Y} &\equiv& -i\sqrt{2}A_r S^{(1)} = i \sum_{j=1}^N (-1)^{j} S_j^x \Sigma_j^z S_{j+1}^y,
\end{eqnarray}
and  
\begin{eqnarray}
\tilde{H}_{\rm Z}\! &\equiv&\! 2\!\sqrt{2}\!A_r\! \tilde{H}_{\rm{eff}}^{(2)} \!=\! \frac{1}{2}\!\sum_{j=1}^N \!(-1)^j \!(\!\Sigma_j^z\!-\!\Sigma_j^y\!)\!(\!1\!-\!\Sigma_{j+1}^y\!). 
\end{eqnarray}
For periodic boundary condition, the commutation relations among these generators are given by 
\begin{eqnarray}
\left[\tilde{H}_{\rm n}, \tilde{H}_{\rm X}\right] &=& i \tilde{H}_{\rm Y},   \\
\left[\tilde{H}_{\rm n}, \tilde{H}_{\rm Y}\right]&=& -i \tilde{H}_{\rm X},     \\
\left[\tilde{H}_{\rm X}, \tilde{H}_{\rm Y}\right] &=& i \tilde{H}_{\rm Z}.
\end{eqnarray}
From the above, we see that $\tilde{H}_0$ and $\tilde{V}$ are the "$Z$" and "$X$" generators of the nearly SU(2) algebra. The Schrieffer-Wolff generator $S^{(1)}$ in Eq. (\ref{eq:S(1)}) that eliminates off-diagonal matrix elements of Eq. (\ref{eq:tilde_HKED2}) to leading order is related to the "$Y$" generator of this algebra.

\bibliography{KAref}

\begin{thebibliography}{77}%
\makeatletter
\providecommand \@ifxundefined [1]{%
 \@ifx{#1\undefined}
}%
\providecommand \@ifnum [1]{%
 \ifnum #1\expandafter \@firstoftwo
 \else \expandafter \@secondoftwo
 \fi
}%
\providecommand \@ifx [1]{%
 \ifx #1\expandafter \@firstoftwo
 \else \expandafter \@secondoftwo
 \fi
}%
\providecommand \natexlab [1]{#1}%
\providecommand \enquote  [1]{``#1''}%
\providecommand \bibnamefont  [1]{#1}%
\providecommand \bibfnamefont [1]{#1}%
\providecommand \citenamefont [1]{#1}%
\providecommand \href@noop [0]{\@secondoftwo}%
\providecommand \href [0]{\begingroup \@sanitize@url \@href}%
\providecommand \@href[1]{\@@startlink{#1}\@@href}%
\providecommand \@@href[1]{\endgroup#1\@@endlink}%
\providecommand \@sanitize@url [0]{\catcode `\\12\catcode `\$12\catcode
  `\&12\catcode `\#12\catcode `\^12\catcode `\_12\catcode `\%12\relax}%
\providecommand \@@startlink[1]{}%
\providecommand \@@endlink[0]{}%
\providecommand \url  [0]{\begingroup\@sanitize@url \@url }%
\providecommand \@url [1]{\endgroup\@href {#1}{\urlprefix }}%
\providecommand \urlprefix  [0]{URL }%
\providecommand \Eprint [0]{\href }%
\providecommand \doibase [0]{https://doi.org/}%
\providecommand \selectlanguage [0]{\@gobble}%
\providecommand \bibinfo  [0]{\@secondoftwo}%
\providecommand \bibfield  [0]{\@secondoftwo}%
\providecommand \translation [1]{[#1]}%
\providecommand \BibitemOpen [0]{}%
\providecommand \bibitemStop [0]{}%
\providecommand \bibitemNoStop [0]{.\EOS\space}%
\providecommand \EOS [0]{\spacefactor3000\relax}%
\providecommand \BibitemShut  [1]{\csname bibitem#1\endcsname}%
\let\auto@bib@innerbib\@empty
%</preamble>
\bibitem [{\citenamefont {Srednicki}(1999)}]{Srednicki1999}%
  \BibitemOpen
  \bibfield  {author} {\bibinfo {author} {\bibfnamefont {M.}~\bibnamefont
  {Srednicki}},\ }\bibfield  {title} {\bibinfo {title} {The approach to thermal
  equilibrium in quantized chaotic systems},\ }\href
  {https://doi.org/10.1088/0305-4470/32/7/007} {\bibfield  {journal} {\bibinfo
  {journal} {Journal of Physics A: Mathematical and General}\ }\textbf
  {\bibinfo {volume} {32}},\ \bibinfo {pages} {1163} (\bibinfo {year}
  {1999})}\BibitemShut {NoStop}%
\bibitem [{\citenamefont {D'Alessio}\ and\ \citenamefont
  {Rigol}(2014)}]{Rigol2014}%
  \BibitemOpen
  \bibfield  {author} {\bibinfo {author} {\bibfnamefont {L.}~\bibnamefont
  {D'Alessio}}\ and\ \bibinfo {author} {\bibfnamefont {M.}~\bibnamefont
  {Rigol}},\ }\bibfield  {title} {\bibinfo {title} {Long-time behavior of
  isolated periodically driven interacting lattice systems},\ }\href
  {https://doi.org/10.1103/PhysRevX.4.041048} {\bibfield  {journal} {\bibinfo
  {journal} {Phys. Rev. X}\ }\textbf {\bibinfo {volume} {4}},\ \bibinfo {pages}
  {041048} (\bibinfo {year} {2014})}\BibitemShut {NoStop}%
\bibitem [{\citenamefont {Luca~D'Alessio}\ and\ \citenamefont
  {Rigol}(2016)}]{Rigol2016}%
  \BibitemOpen
  \bibfield  {author} {\bibinfo {author} {\bibfnamefont {A.~P.}\ \bibnamefont
  {Luca~D'Alessio}, \bibfnamefont {Yariv~Kafri}}\ and\ \bibinfo {author}
  {\bibfnamefont {M.}~\bibnamefont {Rigol}},\ }\bibfield  {title} {\bibinfo
  {title} {From quantum chaos and eigenstate thermalization to statistical
  mechanics and thermodynamics},\ }\href
  {https://doi.org/10.1080/00018732.2016.1198134} {\bibfield  {journal}
  {\bibinfo  {journal} {Advances in Physics}\ }\textbf {\bibinfo {volume}
  {65}},\ \bibinfo {pages} {239} (\bibinfo {year} {2016})}\BibitemShut
  {NoStop}%
\bibitem [{\citenamefont {Deutsch}(2018)}]{Deutsch2018}%
  \BibitemOpen
  \bibfield  {author} {\bibinfo {author} {\bibfnamefont {J.~M.}\ \bibnamefont
  {Deutsch}},\ }\bibfield  {title} {\bibinfo {title} {Eigenstate thermalization
  hypothesis},\ }\href {https://doi.org/10.1088/1361-6633/aac9f1} {\bibfield
  {journal} {\bibinfo  {journal} {Reports on Progress in Physics}\ }\textbf
  {\bibinfo {volume} {81}},\ \bibinfo {pages} {082001} (\bibinfo {year}
  {2018})}\BibitemShut {NoStop}%
\bibitem [{\citenamefont {Browaeys}\ and\ \citenamefont
  {Lahaye}(2020)}]{Browaeys2020}%
  \BibitemOpen
  \bibfield  {author} {\bibinfo {author} {\bibfnamefont {A.}~\bibnamefont
  {Browaeys}}\ and\ \bibinfo {author} {\bibfnamefont {T.}~\bibnamefont
  {Lahaye}},\ }\bibfield  {title} {\bibinfo {title} {Many-body physics with
  individually controlled {R}ydberg atoms},\ }\href
  {https://doi.org/10.1038/s41567-019-0733-z} {\bibfield  {journal} {\bibinfo
  {journal} {Nature Physics}\ }\textbf {\bibinfo {volume} {16}},\ \bibinfo
  {pages} {132} (\bibinfo {year} {2020})}\BibitemShut {NoStop}%
\bibitem [{\citenamefont {Ho}\ \emph {et~al.}(2019)\citenamefont {Ho},
  \citenamefont {Choi}, \citenamefont {Pichler},\ and\ \citenamefont
  {Lukin}}]{Ho2019}%
  \BibitemOpen
  \bibfield  {author} {\bibinfo {author} {\bibfnamefont {W.~W.}\ \bibnamefont
  {Ho}}, \bibinfo {author} {\bibfnamefont {S.}~\bibnamefont {Choi}}, \bibinfo
  {author} {\bibfnamefont {H.}~\bibnamefont {Pichler}},\ and\ \bibinfo {author}
  {\bibfnamefont {M.~D.}\ \bibnamefont {Lukin}},\ }\bibfield  {title} {\bibinfo
  {title} {Periodic orbits, entanglement, and quantum many-body scars in
  constrained models: Matrix product state approach},\ }\href
  {https://doi.org/10.1103/PhysRevLett.122.040603} {\bibfield  {journal}
  {\bibinfo  {journal} {Phys. Rev. Lett.}\ }\textbf {\bibinfo {volume} {122}},\
  \bibinfo {pages} {040603} (\bibinfo {year} {2019})}\BibitemShut {NoStop}%
\bibitem [{\citenamefont {Giudici}\ \emph {et~al.}(2022)\citenamefont
  {Giudici}, \citenamefont {Lukin},\ and\ \citenamefont
  {Pichler}}]{Giudici2022}%
  \BibitemOpen
  \bibfield  {author} {\bibinfo {author} {\bibfnamefont {G.}~\bibnamefont
  {Giudici}}, \bibinfo {author} {\bibfnamefont {M.~D.}\ \bibnamefont {Lukin}},\
  and\ \bibinfo {author} {\bibfnamefont {H.}~\bibnamefont {Pichler}},\
  }\bibfield  {title} {\bibinfo {title} {Dynamical preparation of quantum spin
  liquids in {R}ydberg atom arrays},\ }\href
  {https://doi.org/10.1103/PhysRevLett.129.090401} {\bibfield  {journal}
  {\bibinfo  {journal} {Phys. Rev. Lett.}\ }\textbf {\bibinfo {volume} {129}},\
  \bibinfo {pages} {090401} (\bibinfo {year} {2022})}\BibitemShut {NoStop}%
\bibitem [{\citenamefont {Yao}\ \emph {et~al.}(2022{\natexlab{a}})\citenamefont
  {Yao}, \citenamefont {Pan}, \citenamefont {Liu},\ and\ \citenamefont
  {Zhang}}]{Yao2022}%
  \BibitemOpen
  \bibfield  {author} {\bibinfo {author} {\bibfnamefont {Z.}~\bibnamefont
  {Yao}}, \bibinfo {author} {\bibfnamefont {L.}~\bibnamefont {Pan}}, \bibinfo
  {author} {\bibfnamefont {S.}~\bibnamefont {Liu}},\ and\ \bibinfo {author}
  {\bibfnamefont {P.}~\bibnamefont {Zhang}},\ }\bibfield  {title} {\bibinfo
  {title} {Bounding entanglement entropy using zeros of local correlation
  matrices},\ }\href {https://doi.org/10.1103/PhysRevResearch.4.L042037}
  {\bibfield  {journal} {\bibinfo  {journal} {Phys. Rev. Res.}\ }\textbf
  {\bibinfo {volume} {4}},\ \bibinfo {pages} {L042037} (\bibinfo {year}
  {2022}{\natexlab{a}})}\BibitemShut {NoStop}%
\bibitem [{\citenamefont {Bernien}\ \emph {et~al.}(2017)\citenamefont
  {Bernien}, \citenamefont {Schwartz}, \citenamefont {Keesling}, \citenamefont
  {Levine}, \citenamefont {Omran}, \citenamefont {Pichler}, \citenamefont
  {Choi}, \citenamefont {Zibrov}, \citenamefont {Endres}, \citenamefont
  {Greiner}, \citenamefont {Vuleti{\'{c}}},\ and\ \citenamefont
  {Lukin}}]{Bernien2017}%
  \BibitemOpen
  \bibfield  {author} {\bibinfo {author} {\bibfnamefont {H.}~\bibnamefont
  {Bernien}}, \bibinfo {author} {\bibfnamefont {S.}~\bibnamefont {Schwartz}},
  \bibinfo {author} {\bibfnamefont {A.}~\bibnamefont {Keesling}}, \bibinfo
  {author} {\bibfnamefont {H.}~\bibnamefont {Levine}}, \bibinfo {author}
  {\bibfnamefont {A.}~\bibnamefont {Omran}}, \bibinfo {author} {\bibfnamefont
  {H.}~\bibnamefont {Pichler}}, \bibinfo {author} {\bibfnamefont
  {S.}~\bibnamefont {Choi}}, \bibinfo {author} {\bibfnamefont {A.~S.}\
  \bibnamefont {Zibrov}}, \bibinfo {author} {\bibfnamefont {M.}~\bibnamefont
  {Endres}}, \bibinfo {author} {\bibfnamefont {M.}~\bibnamefont {Greiner}},
  \bibinfo {author} {\bibfnamefont {V.}~\bibnamefont {Vuleti{\'{c}}}},\ and\
  \bibinfo {author} {\bibfnamefont {M.~D.}\ \bibnamefont {Lukin}},\ }\bibfield
  {title} {\bibinfo {title} {Probing many-body dynamics on a 51-atom quantum
  simulator},\ }\href {https://doi.org/10.1038/nature24622} {\bibfield
  {journal} {\bibinfo  {journal} {Nature}\ }\textbf {\bibinfo {volume} {551}},\
  \bibinfo {pages} {579} (\bibinfo {year} {2017})}\BibitemShut {NoStop}%
\bibitem [{\citenamefont {Turner}\ \emph
  {et~al.}(2018{\natexlab{a}})\citenamefont {Turner}, \citenamefont
  {Michailidis}, \citenamefont {Abanin}, \citenamefont {Serbyn},\ and\
  \citenamefont {Papi{\'{c}}}}]{Turner2018Nature}%
  \BibitemOpen
  \bibfield  {author} {\bibinfo {author} {\bibfnamefont {C.~J.}\ \bibnamefont
  {Turner}}, \bibinfo {author} {\bibfnamefont {A.~A.}\ \bibnamefont
  {Michailidis}}, \bibinfo {author} {\bibfnamefont {D.~A.}\ \bibnamefont
  {Abanin}}, \bibinfo {author} {\bibfnamefont {M.}~\bibnamefont {Serbyn}},\
  and\ \bibinfo {author} {\bibfnamefont {Z.}~\bibnamefont {Papi{\'{c}}}},\
  }\bibfield  {title} {\bibinfo {title} {Weak ergodicity breaking from quantum
  many-body scars},\ }\href {https://doi.org/10.1038/s41567-018-0137-5}
  {\bibfield  {journal} {\bibinfo  {journal} {Nature Physics}\ }\textbf
  {\bibinfo {volume} {14}},\ \bibinfo {pages} {745} (\bibinfo {year}
  {2018}{\natexlab{a}})}\BibitemShut {NoStop}%
\bibitem [{\citenamefont {Su}\ \emph {et~al.}(2023)\citenamefont {Su},
  \citenamefont {Sun}, \citenamefont {Hudomal}, \citenamefont {Desaules},
  \citenamefont {Zhou}, \citenamefont {Yang}, \citenamefont {Halimeh},
  \citenamefont {Yuan}, \citenamefont {Papi\ifmmode~\acute{c}\else
  \'{c}\fi{}},\ and\ \citenamefont {Pan}}]{Pan2023}%
  \BibitemOpen
  \bibfield  {author} {\bibinfo {author} {\bibfnamefont {G.-X.}\ \bibnamefont
  {Su}}, \bibinfo {author} {\bibfnamefont {H.}~\bibnamefont {Sun}}, \bibinfo
  {author} {\bibfnamefont {A.}~\bibnamefont {Hudomal}}, \bibinfo {author}
  {\bibfnamefont {J.-Y.}\ \bibnamefont {Desaules}}, \bibinfo {author}
  {\bibfnamefont {Z.-Y.}\ \bibnamefont {Zhou}}, \bibinfo {author}
  {\bibfnamefont {B.}~\bibnamefont {Yang}}, \bibinfo {author} {\bibfnamefont
  {J.~C.}\ \bibnamefont {Halimeh}}, \bibinfo {author} {\bibfnamefont {Z.-S.}\
  \bibnamefont {Yuan}}, \bibinfo {author} {\bibfnamefont {Z.}~\bibnamefont
  {Papi\ifmmode~\acute{c}\else \'{c}\fi{}}},\ and\ \bibinfo {author}
  {\bibfnamefont {J.-W.}\ \bibnamefont {Pan}},\ }\bibfield  {title} {\bibinfo
  {title} {Observation of many-body scarring in a {B}ose-{H}ubbard quantum
  simulator},\ }\href {https://doi.org/10.1103/PhysRevResearch.5.023010}
  {\bibfield  {journal} {\bibinfo  {journal} {Phys. Rev. Res.}\ }\textbf
  {\bibinfo {volume} {5}},\ \bibinfo {pages} {023010} (\bibinfo {year}
  {2023})}\BibitemShut {NoStop}%
\bibitem [{\citenamefont {Zhang}\ \emph
  {et~al.}(2023{\natexlab{a}})\citenamefont {Zhang}, \citenamefont {Dong},
  \citenamefont {Gao}, \citenamefont {Zhao}, \citenamefont {Hao}, \citenamefont
  {Desaules}, \citenamefont {Guo}, \citenamefont {Chen}, \citenamefont {Deng},
  \citenamefont {Liu}, \citenamefont {Ren}, \citenamefont {Yao}, \citenamefont
  {Zhang}, \citenamefont {Xu}, \citenamefont {Wang}, \citenamefont {Jin},
  \citenamefont {Zhu}, \citenamefont {Zhang}, \citenamefont {Li}, \citenamefont
  {Song}, \citenamefont {Wang}, \citenamefont {Liu}, \citenamefont
  {Papi{\'{c}}}, \citenamefont {Ying}, \citenamefont {Wang},\ and\
  \citenamefont {Lai}}]{Zhang2023}%
  \BibitemOpen
  \bibfield  {author} {\bibinfo {author} {\bibfnamefont {P.}~\bibnamefont
  {Zhang}}, \bibinfo {author} {\bibfnamefont {H.}~\bibnamefont {Dong}},
  \bibinfo {author} {\bibfnamefont {Y.}~\bibnamefont {Gao}}, \bibinfo {author}
  {\bibfnamefont {L.}~\bibnamefont {Zhao}}, \bibinfo {author} {\bibfnamefont
  {J.}~\bibnamefont {Hao}}, \bibinfo {author} {\bibfnamefont {J.-Y.}\
  \bibnamefont {Desaules}}, \bibinfo {author} {\bibfnamefont {Q.}~\bibnamefont
  {Guo}}, \bibinfo {author} {\bibfnamefont {J.}~\bibnamefont {Chen}}, \bibinfo
  {author} {\bibfnamefont {J.}~\bibnamefont {Deng}}, \bibinfo {author}
  {\bibfnamefont {B.}~\bibnamefont {Liu}}, \bibinfo {author} {\bibfnamefont
  {W.}~\bibnamefont {Ren}}, \bibinfo {author} {\bibfnamefont {Y.}~\bibnamefont
  {Yao}}, \bibinfo {author} {\bibfnamefont {X.}~\bibnamefont {Zhang}}, \bibinfo
  {author} {\bibfnamefont {S.}~\bibnamefont {Xu}}, \bibinfo {author}
  {\bibfnamefont {K.}~\bibnamefont {Wang}}, \bibinfo {author} {\bibfnamefont
  {F.}~\bibnamefont {Jin}}, \bibinfo {author} {\bibfnamefont {X.}~\bibnamefont
  {Zhu}}, \bibinfo {author} {\bibfnamefont {B.}~\bibnamefont {Zhang}}, \bibinfo
  {author} {\bibfnamefont {H.}~\bibnamefont {Li}}, \bibinfo {author}
  {\bibfnamefont {C.}~\bibnamefont {Song}}, \bibinfo {author} {\bibfnamefont
  {Z.}~\bibnamefont {Wang}}, \bibinfo {author} {\bibfnamefont {F.}~\bibnamefont
  {Liu}}, \bibinfo {author} {\bibfnamefont {Z.}~\bibnamefont {Papi{\'{c}}}},
  \bibinfo {author} {\bibfnamefont {L.}~\bibnamefont {Ying}}, \bibinfo {author}
  {\bibfnamefont {H.}~\bibnamefont {Wang}},\ and\ \bibinfo {author}
  {\bibfnamefont {Y.-C.}\ \bibnamefont {Lai}},\ }\bibfield  {title} {\bibinfo
  {title} {Many-body {H}ilbert space scarring on a superconducting processor},\
  }\href {https://doi.org/10.1038/s41567-022-01784-9} {\bibfield  {journal}
  {\bibinfo  {journal} {Nature Physics}\ }\textbf {\bibinfo {volume} {19}},\
  \bibinfo {pages} {120} (\bibinfo {year} {2023}{\natexlab{a}})}\BibitemShut
  {NoStop}%
\bibitem [{\citenamefont {Moudgalya}\ \emph {et~al.}(2018)\citenamefont
  {Moudgalya}, \citenamefont {Rachel}, \citenamefont {Bernevig},\ and\
  \citenamefont {Regnault}}]{Moudgalya2018}%
  \BibitemOpen
  \bibfield  {author} {\bibinfo {author} {\bibfnamefont {S.}~\bibnamefont
  {Moudgalya}}, \bibinfo {author} {\bibfnamefont {S.}~\bibnamefont {Rachel}},
  \bibinfo {author} {\bibfnamefont {B.~A.}\ \bibnamefont {Bernevig}},\ and\
  \bibinfo {author} {\bibfnamefont {N.}~\bibnamefont {Regnault}},\ }\bibfield
  {title} {\bibinfo {title} {Exact excited states of nonintegrable models},\
  }\href {https://doi.org/10.1103/PhysRevB.98.235155} {\bibfield  {journal}
  {\bibinfo  {journal} {Phys. Rev. B}\ }\textbf {\bibinfo {volume} {98}},\
  \bibinfo {pages} {235155} (\bibinfo {year} {2018})}\BibitemShut {NoStop}%
\bibitem [{\citenamefont {van Voorden}\ \emph {et~al.}(2020)\citenamefont {van
  Voorden}, \citenamefont {Min\'a\ifmmode~\check{r}\else \v{r}\fi{}},\ and\
  \citenamefont {Schoutens}}]{Kareljan2020}%
  \BibitemOpen
  \bibfield  {author} {\bibinfo {author} {\bibfnamefont {B.}~\bibnamefont {van
  Voorden}}, \bibinfo {author} {\bibfnamefont {J.~c.~v.}\ \bibnamefont
  {Min\'a\ifmmode~\check{r}\else \v{r}\fi{}}},\ and\ \bibinfo {author}
  {\bibfnamefont {K.}~\bibnamefont {Schoutens}},\ }\bibfield  {title} {\bibinfo
  {title} {Quantum many-body scars in transverse field {I}sing ladders and
  beyond},\ }\href {https://doi.org/10.1103/PhysRevB.101.220305} {\bibfield
  {journal} {\bibinfo  {journal} {Phys. Rev. B}\ }\textbf {\bibinfo {volume}
  {101}},\ \bibinfo {pages} {220305} (\bibinfo {year} {2020})}\BibitemShut
  {NoStop}%
\bibitem [{\citenamefont {Michailidis}\ \emph {et~al.}(2020)\citenamefont
  {Michailidis}, \citenamefont {Turner}, \citenamefont
  {Papi\ifmmode~\acute{c}\else \'{c}\fi{}}, \citenamefont {Abanin},\ and\
  \citenamefont {Serbyn}}]{Serbyn2020}%
  \BibitemOpen
  \bibfield  {author} {\bibinfo {author} {\bibfnamefont {A.~A.}\ \bibnamefont
  {Michailidis}}, \bibinfo {author} {\bibfnamefont {C.~J.}\ \bibnamefont
  {Turner}}, \bibinfo {author} {\bibfnamefont {Z.}~\bibnamefont
  {Papi\ifmmode~\acute{c}\else \'{c}\fi{}}}, \bibinfo {author} {\bibfnamefont
  {D.~A.}\ \bibnamefont {Abanin}},\ and\ \bibinfo {author} {\bibfnamefont
  {M.}~\bibnamefont {Serbyn}},\ }\bibfield  {title} {\bibinfo {title}
  {Stabilizing two-dimensional quantum scars by deformation and
  synchronization},\ }\href {https://doi.org/10.1103/PhysRevResearch.2.022065}
  {\bibfield  {journal} {\bibinfo  {journal} {Phys. Rev. Res.}\ }\textbf
  {\bibinfo {volume} {2}},\ \bibinfo {pages} {022065} (\bibinfo {year}
  {2020})}\BibitemShut {NoStop}%
\bibitem [{\citenamefont {Lin}\ \emph {et~al.}(2020)\citenamefont {Lin},
  \citenamefont {Calvera},\ and\ \citenamefont {Hsieh}}]{Lin2020}%
  \BibitemOpen
  \bibfield  {author} {\bibinfo {author} {\bibfnamefont {C.-J.}\ \bibnamefont
  {Lin}}, \bibinfo {author} {\bibfnamefont {V.}~\bibnamefont {Calvera}},\ and\
  \bibinfo {author} {\bibfnamefont {T.~H.}\ \bibnamefont {Hsieh}},\ }\bibfield
  {title} {\bibinfo {title} {Quantum many-body scar states in two-dimensional
  {R}ydberg atom arrays},\ }\href {https://doi.org/10.1103/PhysRevB.101.220304}
  {\bibfield  {journal} {\bibinfo  {journal} {Phys. Rev. B}\ }\textbf {\bibinfo
  {volume} {101}},\ \bibinfo {pages} {220304} (\bibinfo {year}
  {2020})}\BibitemShut {NoStop}%
\bibitem [{\citenamefont {Schecter}\ and\ \citenamefont
  {Iadecola}(2019)}]{Schecter2019}%
  \BibitemOpen
  \bibfield  {author} {\bibinfo {author} {\bibfnamefont {M.}~\bibnamefont
  {Schecter}}\ and\ \bibinfo {author} {\bibfnamefont {T.}~\bibnamefont
  {Iadecola}},\ }\bibfield  {title} {\bibinfo {title} {Weak ergodicity breaking
  and quantum many-body scars in spin-1 {XY} magnets},\ }\href
  {https://doi.org/10.1103/PhysRevLett.123.147201} {\bibfield  {journal}
  {\bibinfo  {journal} {Phys. Rev. Lett.}\ }\textbf {\bibinfo {volume} {123}},\
  \bibinfo {pages} {147201} (\bibinfo {year} {2019})}\BibitemShut {NoStop}%
\bibitem [{\citenamefont {Chattopadhyay}\ \emph {et~al.}(2020)\citenamefont
  {Chattopadhyay}, \citenamefont {Pichler}, \citenamefont {Lukin},\ and\
  \citenamefont {Ho}}]{Chattopadhyay2020}%
  \BibitemOpen
  \bibfield  {author} {\bibinfo {author} {\bibfnamefont {S.}~\bibnamefont
  {Chattopadhyay}}, \bibinfo {author} {\bibfnamefont {H.}~\bibnamefont
  {Pichler}}, \bibinfo {author} {\bibfnamefont {M.~D.}\ \bibnamefont {Lukin}},\
  and\ \bibinfo {author} {\bibfnamefont {W.~W.}\ \bibnamefont {Ho}},\
  }\bibfield  {title} {\bibinfo {title} {Quantum many-body scars from virtual
  entangled pairs},\ }\href {https://doi.org/10.1103/PhysRevB.101.174308}
  {\bibfield  {journal} {\bibinfo  {journal} {Phys. Rev. B}\ }\textbf {\bibinfo
  {volume} {101}},\ \bibinfo {pages} {174308} (\bibinfo {year}
  {2020})}\BibitemShut {NoStop}%
\bibitem [{\citenamefont {Gotta}\ \emph {et~al.}(2023)\citenamefont {Gotta},
  \citenamefont {Moudgalya},\ and\ \citenamefont {Mazza}}]{Gotta2023}%
  \BibitemOpen
  \bibfield  {author} {\bibinfo {author} {\bibfnamefont {L.}~\bibnamefont
  {Gotta}}, \bibinfo {author} {\bibfnamefont {S.}~\bibnamefont {Moudgalya}},\
  and\ \bibinfo {author} {\bibfnamefont {L.}~\bibnamefont {Mazza}},\ }\bibfield
   {title} {\bibinfo {title} {Asymptotic quantum many-body scars},\ }\href
  {https://doi.org/10.1103/PhysRevLett.131.190401} {\bibfield  {journal}
  {\bibinfo  {journal} {Phys. Rev. Lett.}\ }\textbf {\bibinfo {volume} {131}},\
  \bibinfo {pages} {190401} (\bibinfo {year} {2023})}\BibitemShut {NoStop}%
\bibitem [{\citenamefont {Chen}\ \emph {et~al.}(2023)\citenamefont {Chen},
  \citenamefont {Chen},\ and\ \citenamefont {Zhu}}]{Chen2023}%
  \BibitemOpen
  \bibfield  {author} {\bibinfo {author} {\bibfnamefont {Q.}~\bibnamefont
  {Chen}}, \bibinfo {author} {\bibfnamefont {S.~A.}\ \bibnamefont {Chen}},\
  and\ \bibinfo {author} {\bibfnamefont {Z.}~\bibnamefont {Zhu}},\ }\bibfield
  {title} {\bibinfo {title} {{Weak ergodicity breaking in non-Hermitian
  many-body systems}},\ }\href {https://doi.org/10.21468/SciPostPhys.15.2.052}
  {\bibfield  {journal} {\bibinfo  {journal} {SciPost Phys.}\ }\textbf
  {\bibinfo {volume} {15}},\ \bibinfo {pages} {052} (\bibinfo {year}
  {2023})}\BibitemShut {NoStop}%
\bibitem [{\citenamefont {Omiya}\ and\ \citenamefont
  {M\"uller}(2023{\natexlab{a}})}]{Omiya2023}%
  \BibitemOpen
  \bibfield  {author} {\bibinfo {author} {\bibfnamefont {K.}~\bibnamefont
  {Omiya}}\ and\ \bibinfo {author} {\bibfnamefont {M.}~\bibnamefont
  {M\"uller}},\ }\bibfield  {title} {\bibinfo {title} {Fractionalization paves
  the way to local projector embeddings of quantum many-body scars},\ }\href
  {https://doi.org/10.1103/PhysRevB.108.054412} {\bibfield  {journal} {\bibinfo
   {journal} {Phys. Rev. B}\ }\textbf {\bibinfo {volume} {108}},\ \bibinfo
  {pages} {054412} (\bibinfo {year} {2023}{\natexlab{a}})}\BibitemShut
  {NoStop}%
\bibitem [{\citenamefont {Logari\ifmmode~\acute{c}\else \'{c}\fi{}}\ \emph
  {et~al.}(2024)\citenamefont {Logari\ifmmode~\acute{c}\else \'{c}\fi{}},
  \citenamefont {Dooley}, \citenamefont {Pappalardi},\ and\ \citenamefont
  {Goold}}]{John2024}%
  \BibitemOpen
  \bibfield  {author} {\bibinfo {author} {\bibfnamefont {L.}~\bibnamefont
  {Logari\ifmmode~\acute{c}\else \'{c}\fi{}}}, \bibinfo {author} {\bibfnamefont
  {S.}~\bibnamefont {Dooley}}, \bibinfo {author} {\bibfnamefont
  {S.}~\bibnamefont {Pappalardi}},\ and\ \bibinfo {author} {\bibfnamefont
  {J.}~\bibnamefont {Goold}},\ }\bibfield  {title} {\bibinfo {title} {Quantum
  many-body scars in dual-unitary circuits},\ }\href
  {https://doi.org/10.1103/PhysRevLett.132.010401} {\bibfield  {journal}
  {\bibinfo  {journal} {Phys. Rev. Lett.}\ }\textbf {\bibinfo {volume} {132}},\
  \bibinfo {pages} {010401} (\bibinfo {year} {2024})}\BibitemShut {NoStop}%
\bibitem [{\citenamefont {Wang}\ \emph {et~al.}(2024)\citenamefont {Wang},
  \citenamefont {Yuan}, \citenamefont {Zhang}, \citenamefont {Wang},
  \citenamefont {Deng},\ and\ \citenamefont {Duan}}]{Wang2024}%
  \BibitemOpen
  \bibfield  {author} {\bibinfo {author} {\bibfnamefont {H.-R.}\ \bibnamefont
  {Wang}}, \bibinfo {author} {\bibfnamefont {D.}~\bibnamefont {Yuan}}, \bibinfo
  {author} {\bibfnamefont {S.-Y.}\ \bibnamefont {Zhang}}, \bibinfo {author}
  {\bibfnamefont {Z.}~\bibnamefont {Wang}}, \bibinfo {author} {\bibfnamefont
  {D.-L.}\ \bibnamefont {Deng}},\ and\ \bibinfo {author} {\bibfnamefont
  {L.-M.}\ \bibnamefont {Duan}},\ }\bibfield  {title} {\bibinfo {title}
  {Embedding quantum many-body scars into decoherence-free subspaces},\ }\href
  {https://doi.org/10.1103/PhysRevLett.132.150401} {\bibfield  {journal}
  {\bibinfo  {journal} {Phys. Rev. Lett.}\ }\textbf {\bibinfo {volume} {132}},\
  \bibinfo {pages} {150401} (\bibinfo {year} {2024})}\BibitemShut {NoStop}%
\bibitem [{\citenamefont {Moudgalya}\ \emph {et~al.}(2020)\citenamefont
  {Moudgalya}, \citenamefont {Regnault},\ and\ \citenamefont
  {Bernevig}}]{Andrei2020}%
  \BibitemOpen
  \bibfield  {author} {\bibinfo {author} {\bibfnamefont {S.}~\bibnamefont
  {Moudgalya}}, \bibinfo {author} {\bibfnamefont {N.}~\bibnamefont
  {Regnault}},\ and\ \bibinfo {author} {\bibfnamefont {B.~A.}\ \bibnamefont
  {Bernevig}},\ }\bibfield  {title} {\bibinfo {title}
  {$\ensuremath{\eta}$-pairing in hubbard models: From spectrum generating
  algebras to quantum many-body scars},\ }\href
  {https://doi.org/10.1103/PhysRevB.102.085140} {\bibfield  {journal} {\bibinfo
   {journal} {Phys. Rev. B}\ }\textbf {\bibinfo {volume} {102}},\ \bibinfo
  {pages} {085140} (\bibinfo {year} {2020})}\BibitemShut {NoStop}%
\bibitem [{\citenamefont {Sanada}\ \emph {et~al.}(2023)\citenamefont {Sanada},
  \citenamefont {Miao},\ and\ \citenamefont {Katsura}}]{Sanada2023}%
  \BibitemOpen
  \bibfield  {author} {\bibinfo {author} {\bibfnamefont {K.}~\bibnamefont
  {Sanada}}, \bibinfo {author} {\bibfnamefont {Y.}~\bibnamefont {Miao}},\ and\
  \bibinfo {author} {\bibfnamefont {H.}~\bibnamefont {Katsura}},\ }\bibfield
  {title} {\bibinfo {title} {Quantum many-body scars in spin models with
  multibody interactions},\ }\href
  {https://doi.org/10.1103/PhysRevB.108.155102} {\bibfield  {journal} {\bibinfo
   {journal} {Phys. Rev. B}\ }\textbf {\bibinfo {volume} {108}},\ \bibinfo
  {pages} {155102} (\bibinfo {year} {2023})}\BibitemShut {NoStop}%
\bibitem [{\citenamefont {Deng}\ and\ \citenamefont {Yang}(2023)}]{Deng2023}%
  \BibitemOpen
  \bibfield  {author} {\bibinfo {author} {\bibfnamefont {W.}~\bibnamefont
  {Deng}}\ and\ \bibinfo {author} {\bibfnamefont {Z.-C.}\ \bibnamefont
  {Yang}},\ }\bibfield  {title} {\bibinfo {title} {Using models with static
  quantum many-body scars to generate time-crystalline behavior under periodic
  driving},\ }\href {https://doi.org/10.1103/PhysRevB.108.205129} {\bibfield
  {journal} {\bibinfo  {journal} {Phys. Rev. B}\ }\textbf {\bibinfo {volume}
  {108}},\ \bibinfo {pages} {205129} (\bibinfo {year} {2023})}\BibitemShut
  {NoStop}%
\bibitem [{\citenamefont {Pakrouski}\ \emph {et~al.}(2020)\citenamefont
  {Pakrouski}, \citenamefont {Pallegar}, \citenamefont {Popov},\ and\
  \citenamefont {Klebanov}}]{Pakrouski2020}%
  \BibitemOpen
  \bibfield  {author} {\bibinfo {author} {\bibfnamefont {K.}~\bibnamefont
  {Pakrouski}}, \bibinfo {author} {\bibfnamefont {P.~N.}\ \bibnamefont
  {Pallegar}}, \bibinfo {author} {\bibfnamefont {F.~K.}\ \bibnamefont
  {Popov}},\ and\ \bibinfo {author} {\bibfnamefont {I.~R.}\ \bibnamefont
  {Klebanov}},\ }\bibfield  {title} {\bibinfo {title} {Many-body scars as a
  group invariant sector of hilbert space},\ }\href
  {https://doi.org/10.1103/PhysRevLett.125.230602} {\bibfield  {journal}
  {\bibinfo  {journal} {Phys. Rev. Lett.}\ }\textbf {\bibinfo {volume} {125}},\
  \bibinfo {pages} {230602} (\bibinfo {year} {2020})}\BibitemShut {NoStop}%
\bibitem [{\citenamefont {O'Dea}\ \emph {et~al.}(2020)\citenamefont {O'Dea},
  \citenamefont {Burnell}, \citenamefont {Chandran},\ and\ \citenamefont
  {Khemani}}]{ODea2020}%
  \BibitemOpen
  \bibfield  {author} {\bibinfo {author} {\bibfnamefont {N.}~\bibnamefont
  {O'Dea}}, \bibinfo {author} {\bibfnamefont {F.}~\bibnamefont {Burnell}},
  \bibinfo {author} {\bibfnamefont {A.}~\bibnamefont {Chandran}},\ and\
  \bibinfo {author} {\bibfnamefont {V.}~\bibnamefont {Khemani}},\ }\bibfield
  {title} {\bibinfo {title} {From tunnels to towers: Quantum scars from lie
  algebras and $q$-deformed lie algebras},\ }\href
  {https://doi.org/10.1103/PhysRevResearch.2.043305} {\bibfield  {journal}
  {\bibinfo  {journal} {Phys. Rev. Res.}\ }\textbf {\bibinfo {volume} {2}},\
  \bibinfo {pages} {043305} (\bibinfo {year} {2020})}\BibitemShut {NoStop}%
\bibitem [{\citenamefont {Pakrouski}\ \emph {et~al.}(2021)\citenamefont
  {Pakrouski}, \citenamefont {Pallegar}, \citenamefont {Popov},\ and\
  \citenamefont {Klebanov}}]{Pakrouski2021}%
  \BibitemOpen
  \bibfield  {author} {\bibinfo {author} {\bibfnamefont {K.}~\bibnamefont
  {Pakrouski}}, \bibinfo {author} {\bibfnamefont {P.~N.}\ \bibnamefont
  {Pallegar}}, \bibinfo {author} {\bibfnamefont {F.~K.}\ \bibnamefont
  {Popov}},\ and\ \bibinfo {author} {\bibfnamefont {I.~R.}\ \bibnamefont
  {Klebanov}},\ }\bibfield  {title} {\bibinfo {title} {Group theoretic approach
  to many-body scar states in fermionic lattice models},\ }\href
  {https://doi.org/10.1103/PhysRevResearch.3.043156} {\bibfield  {journal}
  {\bibinfo  {journal} {Phys. Rev. Res.}\ }\textbf {\bibinfo {volume} {3}},\
  \bibinfo {pages} {043156} (\bibinfo {year} {2021})}\BibitemShut {NoStop}%
\bibitem [{\citenamefont {Shibata}\ \emph {et~al.}(2020)\citenamefont
  {Shibata}, \citenamefont {Yoshioka},\ and\ \citenamefont
  {Katsura}}]{Shibata2020}%
  \BibitemOpen
  \bibfield  {author} {\bibinfo {author} {\bibfnamefont {N.}~\bibnamefont
  {Shibata}}, \bibinfo {author} {\bibfnamefont {N.}~\bibnamefont {Yoshioka}},\
  and\ \bibinfo {author} {\bibfnamefont {H.}~\bibnamefont {Katsura}},\
  }\bibfield  {title} {\bibinfo {title} {Onsager's scars in disordered spin
  chains},\ }\href {https://doi.org/10.1103/PhysRevLett.124.180604} {\bibfield
  {journal} {\bibinfo  {journal} {Phys. Rev. Lett.}\ }\textbf {\bibinfo
  {volume} {124}},\ \bibinfo {pages} {180604} (\bibinfo {year}
  {2020})}\BibitemShut {NoStop}%
\bibitem [{\citenamefont {Moudgalya}\ and\ \citenamefont
  {Motrunich}(2022)}]{Moudgalyaprx2022}%
  \BibitemOpen
  \bibfield  {author} {\bibinfo {author} {\bibfnamefont {S.}~\bibnamefont
  {Moudgalya}}\ and\ \bibinfo {author} {\bibfnamefont {O.~I.}\ \bibnamefont
  {Motrunich}},\ }\bibfield  {title} {\bibinfo {title} {Hilbert space
  fragmentation and commutant algebras},\ }\href
  {https://doi.org/10.1103/PhysRevX.12.011050} {\bibfield  {journal} {\bibinfo
  {journal} {Phys. Rev. X}\ }\textbf {\bibinfo {volume} {12}},\ \bibinfo
  {pages} {011050} (\bibinfo {year} {2022})}\BibitemShut {NoStop}%
\bibitem [{\citenamefont {Moudgalya}\ \emph {et~al.}(2022)\citenamefont
  {Moudgalya}, \citenamefont {Bernevig},\ and\ \citenamefont
  {Regnault}}]{Sanjay2022}%
  \BibitemOpen
  \bibfield  {author} {\bibinfo {author} {\bibfnamefont {S.}~\bibnamefont
  {Moudgalya}}, \bibinfo {author} {\bibfnamefont {B.~A.}\ \bibnamefont
  {Bernevig}},\ and\ \bibinfo {author} {\bibfnamefont {N.}~\bibnamefont
  {Regnault}},\ }\bibfield  {title} {\bibinfo {title} {Quantum many-body scars
  and {H}ilbert space fragmentation: a review of exact results},\ }\href
  {https://doi.org/10.1088/1361-6633/ac73a0} {\bibfield  {journal} {\bibinfo
  {journal} {Reports on Progress in Physics}\ }\textbf {\bibinfo {volume}
  {85}},\ \bibinfo {pages} {086501} (\bibinfo {year} {2022})}\BibitemShut
  {NoStop}%
\bibitem [{\citenamefont {Francica}\ and\ \citenamefont
  {Dell'Anna}(2023)}]{Francica2023}%
  \BibitemOpen
  \bibfield  {author} {\bibinfo {author} {\bibfnamefont {G.}~\bibnamefont
  {Francica}}\ and\ \bibinfo {author} {\bibfnamefont {L.}~\bibnamefont
  {Dell'Anna}},\ }\bibfield  {title} {\bibinfo {title} {Hilbert space
  fragmentation in a long-range system},\ }\href
  {https://doi.org/10.1103/PhysRevB.108.045127} {\bibfield  {journal} {\bibinfo
   {journal} {Phys. Rev. B}\ }\textbf {\bibinfo {volume} {108}},\ \bibinfo
  {pages} {045127} (\bibinfo {year} {2023})}\BibitemShut {NoStop}%
\bibitem [{\citenamefont {Andreadakis}\ and\ \citenamefont
  {Zanardi}(2023)}]{Andreadakis2023}%
  \BibitemOpen
  \bibfield  {author} {\bibinfo {author} {\bibfnamefont {F.}~\bibnamefont
  {Andreadakis}}\ and\ \bibinfo {author} {\bibfnamefont {P.}~\bibnamefont
  {Zanardi}},\ }\bibfield  {title} {\bibinfo {title} {Coherence generation,
  symmetry algebras, and {H}ilbert space fragmentation},\ }\href
  {https://doi.org/10.1103/PhysRevA.107.062402} {\bibfield  {journal} {\bibinfo
   {journal} {Phys. Rev. A}\ }\textbf {\bibinfo {volume} {107}},\ \bibinfo
  {pages} {062402} (\bibinfo {year} {2023})}\BibitemShut {NoStop}%
\bibitem [{\citenamefont {Mizuta}\ \emph {et~al.}(2020)\citenamefont {Mizuta},
  \citenamefont {Takasan},\ and\ \citenamefont {Kawakami}}]{Mizuta2020}%
  \BibitemOpen
  \bibfield  {author} {\bibinfo {author} {\bibfnamefont {K.}~\bibnamefont
  {Mizuta}}, \bibinfo {author} {\bibfnamefont {K.}~\bibnamefont {Takasan}},\
  and\ \bibinfo {author} {\bibfnamefont {N.}~\bibnamefont {Kawakami}},\
  }\bibfield  {title} {\bibinfo {title} {Exact floquet quantum many-body scars
  under rydberg blockade},\ }\href
  {https://doi.org/10.1103/PhysRevResearch.2.033284} {\bibfield  {journal}
  {\bibinfo  {journal} {Phys. Rev. Res.}\ }\textbf {\bibinfo {volume} {2}},\
  \bibinfo {pages} {033284} (\bibinfo {year} {2020})}\BibitemShut {NoStop}%
\bibitem [{\citenamefont {Machado}\ \emph {et~al.}(2020)\citenamefont
  {Machado}, \citenamefont {Else}, \citenamefont {Kahanamoku-Meyer},
  \citenamefont {Nayak},\ and\ \citenamefont {Yao}}]{Machado2020}%
  \BibitemOpen
  \bibfield  {author} {\bibinfo {author} {\bibfnamefont {F.}~\bibnamefont
  {Machado}}, \bibinfo {author} {\bibfnamefont {D.~V.}\ \bibnamefont {Else}},
  \bibinfo {author} {\bibfnamefont {G.~D.}\ \bibnamefont {Kahanamoku-Meyer}},
  \bibinfo {author} {\bibfnamefont {C.}~\bibnamefont {Nayak}},\ and\ \bibinfo
  {author} {\bibfnamefont {N.~Y.}\ \bibnamefont {Yao}},\ }\bibfield  {title}
  {\bibinfo {title} {Long-range prethermal phases of nonequilibrium matter},\
  }\href {https://doi.org/10.1103/PhysRevX.10.011043} {\bibfield  {journal}
  {\bibinfo  {journal} {Phys. Rev. X}\ }\textbf {\bibinfo {volume} {10}},\
  \bibinfo {pages} {011043} (\bibinfo {year} {2020})}\BibitemShut {NoStop}%
\bibitem [{\citenamefont {Choi}\ \emph {et~al.}(2019)\citenamefont {Choi},
  \citenamefont {Turner}, \citenamefont {Pichler}, \citenamefont {Ho},
  \citenamefont {Michailidis}, \citenamefont {Papi\ifmmode~\acute{c}\else
  \'{c}\fi{}}, \citenamefont {Serbyn}, \citenamefont {Lukin},\ and\
  \citenamefont {Abanin}}]{Dmitry2019}%
  \BibitemOpen
  \bibfield  {author} {\bibinfo {author} {\bibfnamefont {S.}~\bibnamefont
  {Choi}}, \bibinfo {author} {\bibfnamefont {C.~J.}\ \bibnamefont {Turner}},
  \bibinfo {author} {\bibfnamefont {H.}~\bibnamefont {Pichler}}, \bibinfo
  {author} {\bibfnamefont {W.~W.}\ \bibnamefont {Ho}}, \bibinfo {author}
  {\bibfnamefont {A.~A.}\ \bibnamefont {Michailidis}}, \bibinfo {author}
  {\bibfnamefont {Z.}~\bibnamefont {Papi\ifmmode~\acute{c}\else \'{c}\fi{}}},
  \bibinfo {author} {\bibfnamefont {M.}~\bibnamefont {Serbyn}}, \bibinfo
  {author} {\bibfnamefont {M.~D.}\ \bibnamefont {Lukin}},\ and\ \bibinfo
  {author} {\bibfnamefont {D.~A.}\ \bibnamefont {Abanin}},\ }\bibfield  {title}
  {\bibinfo {title} {Emergent {SU(2)} dynamics and perfect quantum many-body
  scars},\ }\href {https://doi.org/10.1103/PhysRevLett.122.220603} {\bibfield
  {journal} {\bibinfo  {journal} {Phys. Rev. Lett.}\ }\textbf {\bibinfo
  {volume} {122}},\ \bibinfo {pages} {220603} (\bibinfo {year}
  {2019})}\BibitemShut {NoStop}%
\bibitem [{\citenamefont {Bull}\ \emph {et~al.}(2020)\citenamefont {Bull},
  \citenamefont {Desaules},\ and\ \citenamefont {Papi\ifmmode~\acute{c}\else
  \'{c}\fi{}}}]{Bull2020}%
  \BibitemOpen
  \bibfield  {author} {\bibinfo {author} {\bibfnamefont {K.}~\bibnamefont
  {Bull}}, \bibinfo {author} {\bibfnamefont {J.-Y.}\ \bibnamefont {Desaules}},\
  and\ \bibinfo {author} {\bibfnamefont {Z.}~\bibnamefont
  {Papi\ifmmode~\acute{c}\else \'{c}\fi{}}},\ }\bibfield  {title} {\bibinfo
  {title} {Quantum scars as embeddings of weakly broken {L}ie algebra
  representations},\ }\href {https://doi.org/10.1103/PhysRevB.101.165139}
  {\bibfield  {journal} {\bibinfo  {journal} {Phys. Rev. B}\ }\textbf {\bibinfo
  {volume} {101}},\ \bibinfo {pages} {165139} (\bibinfo {year}
  {2020})}\BibitemShut {NoStop}%
\bibitem [{\citenamefont {Omiya}\ and\ \citenamefont
  {M\"uller}(2023{\natexlab{b}})}]{Markus2023}%
  \BibitemOpen
  \bibfield  {author} {\bibinfo {author} {\bibfnamefont {K.}~\bibnamefont
  {Omiya}}\ and\ \bibinfo {author} {\bibfnamefont {M.}~\bibnamefont
  {M\"uller}},\ }\bibfield  {title} {\bibinfo {title} {Quantum many-body scars
  in bipartite {R}ydberg arrays originating from hidden projector embedding},\
  }\href {https://doi.org/10.1103/PhysRevA.107.023318} {\bibfield  {journal}
  {\bibinfo  {journal} {Phys. Rev. A}\ }\textbf {\bibinfo {volume} {107}},\
  \bibinfo {pages} {023318} (\bibinfo {year} {2023}{\natexlab{b}})}\BibitemShut
  {NoStop}%
\bibitem [{\citenamefont {Lin}\ and\ \citenamefont
  {Motrunich}(2019)}]{Lin2019}%
  \BibitemOpen
  \bibfield  {author} {\bibinfo {author} {\bibfnamefont {C.-J.}\ \bibnamefont
  {Lin}}\ and\ \bibinfo {author} {\bibfnamefont {O.~I.}\ \bibnamefont
  {Motrunich}},\ }\bibfield  {title} {\bibinfo {title} {Exact quantum many-body
  scar states in the {R}ydberg-blockaded atom chain},\ }\href
  {https://doi.org/10.1103/PhysRevLett.122.173401} {\bibfield  {journal}
  {\bibinfo  {journal} {Phys. Rev. Lett.}\ }\textbf {\bibinfo {volume} {122}},\
  \bibinfo {pages} {173401} (\bibinfo {year} {2019})}\BibitemShut {NoStop}%
\bibitem [{\citenamefont {Mark}\ \emph {et~al.}(2020)\citenamefont {Mark},
  \citenamefont {Lin},\ and\ \citenamefont {Motrunich}}]{Mark2020}%
  \BibitemOpen
  \bibfield  {author} {\bibinfo {author} {\bibfnamefont {D.~K.}\ \bibnamefont
  {Mark}}, \bibinfo {author} {\bibfnamefont {C.-J.}\ \bibnamefont {Lin}},\ and\
  \bibinfo {author} {\bibfnamefont {O.~I.}\ \bibnamefont {Motrunich}},\
  }\bibfield  {title} {\bibinfo {title} {Exact eigenstates in the lesanovsky
  model, proximity to integrability and the {PXP} model, and approximate scar
  states},\ }\href {https://doi.org/10.1103/PhysRevB.101.094308} {\bibfield
  {journal} {\bibinfo  {journal} {Phys. Rev. B}\ }\textbf {\bibinfo {volume}
  {101}},\ \bibinfo {pages} {094308} (\bibinfo {year} {2020})}\BibitemShut
  {NoStop}%
\bibitem [{\citenamefont {Khemani}\ \emph {et~al.}(2019)\citenamefont
  {Khemani}, \citenamefont {Laumann},\ and\ \citenamefont
  {Chandran}}]{Anushya2019}%
  \BibitemOpen
  \bibfield  {author} {\bibinfo {author} {\bibfnamefont {V.}~\bibnamefont
  {Khemani}}, \bibinfo {author} {\bibfnamefont {C.~R.}\ \bibnamefont
  {Laumann}},\ and\ \bibinfo {author} {\bibfnamefont {A.}~\bibnamefont
  {Chandran}},\ }\bibfield  {title} {\bibinfo {title} {Signatures of
  integrability in the dynamics of {R}ydberg-blockaded chains},\ }\href
  {https://doi.org/10.1103/PhysRevB.99.161101} {\bibfield  {journal} {\bibinfo
  {journal} {Phys. Rev. B}\ }\textbf {\bibinfo {volume} {99}},\ \bibinfo
  {pages} {161101} (\bibinfo {year} {2019})}\BibitemShut {NoStop}%
\bibitem [{\citenamefont {You}\ \emph {et~al.}(2020)\citenamefont {You},
  \citenamefont {Sun}, \citenamefont {Ren}, \citenamefont {Yu},\ and\
  \citenamefont {Ole\ifmmode~\acute{s}\else \'{s}\fi{}}}]{You2020}%
  \BibitemOpen
  \bibfield  {author} {\bibinfo {author} {\bibfnamefont {W.-L.}\ \bibnamefont
  {You}}, \bibinfo {author} {\bibfnamefont {G.}~\bibnamefont {Sun}}, \bibinfo
  {author} {\bibfnamefont {J.}~\bibnamefont {Ren}}, \bibinfo {author}
  {\bibfnamefont {W.~C.}\ \bibnamefont {Yu}},\ and\ \bibinfo {author}
  {\bibfnamefont {A.~M.}\ \bibnamefont {Ole\ifmmode~\acute{s}\else
  \'{s}\fi{}}},\ }\bibfield  {title} {\bibinfo {title} {Quantum phase
  transitions in the spin-1 {K}itaev-{H}eisenberg chain},\ }\href
  {https://doi.org/10.1103/PhysRevB.102.144437} {\bibfield  {journal} {\bibinfo
   {journal} {Phys. Rev. B}\ }\textbf {\bibinfo {volume} {102}},\ \bibinfo
  {pages} {144437} (\bibinfo {year} {2020})}\BibitemShut {NoStop}%
\bibitem [{\citenamefont {You}\ \emph {et~al.}(2022)\citenamefont {You},
  \citenamefont {Zhao}, \citenamefont {Ren}, \citenamefont {Sun}, \citenamefont
  {Li},\ and\ \citenamefont {Ole\ifmmode~\acute{s}\else \'{s}\fi{}}}]{You2022}%
  \BibitemOpen
  \bibfield  {author} {\bibinfo {author} {\bibfnamefont {W.-L.}\ \bibnamefont
  {You}}, \bibinfo {author} {\bibfnamefont {Z.}~\bibnamefont {Zhao}}, \bibinfo
  {author} {\bibfnamefont {J.}~\bibnamefont {Ren}}, \bibinfo {author}
  {\bibfnamefont {G.}~\bibnamefont {Sun}}, \bibinfo {author} {\bibfnamefont
  {L.}~\bibnamefont {Li}},\ and\ \bibinfo {author} {\bibfnamefont {A.~M.}\
  \bibnamefont {Ole\ifmmode~\acute{s}\else \'{s}\fi{}}},\ }\bibfield  {title}
  {\bibinfo {title} {Quantum many-body scars in spin-1 {K}itaev chains},\
  }\href {https://doi.org/10.1103/PhysRevResearch.4.013103} {\bibfield
  {journal} {\bibinfo  {journal} {Phys. Rev. Res.}\ }\textbf {\bibinfo {volume}
  {4}},\ \bibinfo {pages} {013103} (\bibinfo {year} {2022})}\BibitemShut
  {NoStop}%
\bibitem [{\citenamefont {Zhang}\ \emph
  {et~al.}(2023{\natexlab{b}})\citenamefont {Zhang}, \citenamefont {Wang},
  \citenamefont {Liu}, \citenamefont {Ren}, \citenamefont {Li}, \citenamefont
  {Wu}, \citenamefont {Ole\ifmmode~\acute{s}\else \'{s}\fi{}},\ and\
  \citenamefont {You}}]{Zhangwy2023}%
  \BibitemOpen
  \bibfield  {author} {\bibinfo {author} {\bibfnamefont {W.-Y.}\ \bibnamefont
  {Zhang}}, \bibinfo {author} {\bibfnamefont {Y.-N.}\ \bibnamefont {Wang}},
  \bibinfo {author} {\bibfnamefont {D.}~\bibnamefont {Liu}}, \bibinfo {author}
  {\bibfnamefont {J.}~\bibnamefont {Ren}}, \bibinfo {author} {\bibfnamefont
  {J.}~\bibnamefont {Li}}, \bibinfo {author} {\bibfnamefont {N.}~\bibnamefont
  {Wu}}, \bibinfo {author} {\bibfnamefont {A.~M.}\ \bibnamefont
  {Ole\ifmmode~\acute{s}\else \'{s}\fi{}}},\ and\ \bibinfo {author}
  {\bibfnamefont {W.-L.}\ \bibnamefont {You}},\ }\bibfield  {title} {\bibinfo
  {title} {Quantum many-body scars in spin-1 {K}itaev chain with uniaxial
  single-ion anisotropy},\ }\href {https://doi.org/10.1103/PhysRevB.108.104411}
  {\bibfield  {journal} {\bibinfo  {journal} {Phys. Rev. B}\ }\textbf {\bibinfo
  {volume} {108}},\ \bibinfo {pages} {104411} (\bibinfo {year}
  {2023}{\natexlab{b}})}\BibitemShut {NoStop}%
\bibitem [{\citenamefont {Ebadi}\ \emph {et~al.}(2021)\citenamefont {Ebadi},
  \citenamefont {Wang}, \citenamefont {Levine}, \citenamefont {Keesling},
  \citenamefont {Semeghini}, \citenamefont {Omran}, \citenamefont {Bluvstein},
  \citenamefont {Samajdar}, \citenamefont {Pichler}, \citenamefont {Ho},
  \citenamefont {Choi}, \citenamefont {Sachdev}, \citenamefont {Greiner},
  \citenamefont {Vuleti{\'{c}}},\ and\ \citenamefont {Lukin}}]{Ebadi2021}%
  \BibitemOpen
  \bibfield  {author} {\bibinfo {author} {\bibfnamefont {S.}~\bibnamefont
  {Ebadi}}, \bibinfo {author} {\bibfnamefont {T.~T.}\ \bibnamefont {Wang}},
  \bibinfo {author} {\bibfnamefont {H.}~\bibnamefont {Levine}}, \bibinfo
  {author} {\bibfnamefont {A.}~\bibnamefont {Keesling}}, \bibinfo {author}
  {\bibfnamefont {G.}~\bibnamefont {Semeghini}}, \bibinfo {author}
  {\bibfnamefont {A.}~\bibnamefont {Omran}}, \bibinfo {author} {\bibfnamefont
  {D.}~\bibnamefont {Bluvstein}}, \bibinfo {author} {\bibfnamefont
  {R.}~\bibnamefont {Samajdar}}, \bibinfo {author} {\bibfnamefont
  {H.}~\bibnamefont {Pichler}}, \bibinfo {author} {\bibfnamefont {W.~W.}\
  \bibnamefont {Ho}}, \bibinfo {author} {\bibfnamefont {S.}~\bibnamefont
  {Choi}}, \bibinfo {author} {\bibfnamefont {S.}~\bibnamefont {Sachdev}},
  \bibinfo {author} {\bibfnamefont {M.}~\bibnamefont {Greiner}}, \bibinfo
  {author} {\bibfnamefont {V.}~\bibnamefont {Vuleti{\'{c}}}},\ and\ \bibinfo
  {author} {\bibfnamefont {M.~D.}\ \bibnamefont {Lukin}},\ }\bibfield  {title}
  {\bibinfo {title} {Quantum phases of matter on a 256-atom programmable
  quantum simulator},\ }\href {https://doi.org/10.1038/s41586-021-03582-4}
  {\bibfield  {journal} {\bibinfo  {journal} {Nature}\ }\textbf {\bibinfo
  {volume} {595}},\ \bibinfo {pages} {227} (\bibinfo {year}
  {2021})}\BibitemShut {NoStop}%
\bibitem [{\citenamefont {Bluvstein}\ \emph {et~al.}(2021)\citenamefont
  {Bluvstein}, \citenamefont {Omran}, \citenamefont {Levine}, \citenamefont
  {Keesling}, \citenamefont {Semeghini}, \citenamefont {Ebadi}, \citenamefont
  {Wang}, \citenamefont {Michailidis}, \citenamefont {Maskara}, \citenamefont
  {Ho}, \citenamefont {Choi}, \citenamefont {Serbyn}, \citenamefont {Greiner},
  \citenamefont {Vuletić},\ and\ \citenamefont {Lukin}}]{Lukin2021_2}%
  \BibitemOpen
  \bibfield  {author} {\bibinfo {author} {\bibfnamefont {D.}~\bibnamefont
  {Bluvstein}}, \bibinfo {author} {\bibfnamefont {A.}~\bibnamefont {Omran}},
  \bibinfo {author} {\bibfnamefont {H.}~\bibnamefont {Levine}}, \bibinfo
  {author} {\bibfnamefont {A.}~\bibnamefont {Keesling}}, \bibinfo {author}
  {\bibfnamefont {G.}~\bibnamefont {Semeghini}}, \bibinfo {author}
  {\bibfnamefont {S.}~\bibnamefont {Ebadi}}, \bibinfo {author} {\bibfnamefont
  {T.~T.}\ \bibnamefont {Wang}}, \bibinfo {author} {\bibfnamefont {A.~A.}\
  \bibnamefont {Michailidis}}, \bibinfo {author} {\bibfnamefont
  {N.}~\bibnamefont {Maskara}}, \bibinfo {author} {\bibfnamefont {W.~W.}\
  \bibnamefont {Ho}}, \bibinfo {author} {\bibfnamefont {S.}~\bibnamefont
  {Choi}}, \bibinfo {author} {\bibfnamefont {M.}~\bibnamefont {Serbyn}},
  \bibinfo {author} {\bibfnamefont {M.}~\bibnamefont {Greiner}}, \bibinfo
  {author} {\bibfnamefont {V.}~\bibnamefont {Vuletić}},\ and\ \bibinfo
  {author} {\bibfnamefont {M.~D.}\ \bibnamefont {Lukin}},\ }\bibfield  {title}
  {\bibinfo {title} {Controlling quantum many-body dynamics in driven {R}ydberg
  atom arrays},\ }\href {https://doi.org/10.1126/science.abg2530} {\bibfield
  {journal} {\bibinfo  {journal} {Science}\ }\textbf {\bibinfo {volume}
  {371}},\ \bibinfo {pages} {1355} (\bibinfo {year} {2021})}\BibitemShut
  {NoStop}%
\bibitem [{\citenamefont {Ghosh}\ \emph {et~al.}(2018)\citenamefont {Ghosh},
  \citenamefont {Sen},\ and\ \citenamefont {Sengupta}}]{Ghosh2018}%
  \BibitemOpen
  \bibfield  {author} {\bibinfo {author} {\bibfnamefont {R.}~\bibnamefont
  {Ghosh}}, \bibinfo {author} {\bibfnamefont {A.}~\bibnamefont {Sen}},\ and\
  \bibinfo {author} {\bibfnamefont {K.}~\bibnamefont {Sengupta}},\ }\bibfield
  {title} {\bibinfo {title} {Ramp and periodic dynamics across non-{I}sing
  critical points},\ }\href {https://doi.org/10.1103/PhysRevB.97.014309}
  {\bibfield  {journal} {\bibinfo  {journal} {Phys. Rev. B}\ }\textbf {\bibinfo
  {volume} {97}},\ \bibinfo {pages} {014309} (\bibinfo {year}
  {2018})}\BibitemShut {NoStop}%
\bibitem [{\citenamefont {Whitsitt}\ \emph {et~al.}(2018)\citenamefont
  {Whitsitt}, \citenamefont {Samajdar},\ and\ \citenamefont
  {Sachdev}}]{Whitsitt2018}%
  \BibitemOpen
  \bibfield  {author} {\bibinfo {author} {\bibfnamefont {S.}~\bibnamefont
  {Whitsitt}}, \bibinfo {author} {\bibfnamefont {R.}~\bibnamefont {Samajdar}},\
  and\ \bibinfo {author} {\bibfnamefont {S.}~\bibnamefont {Sachdev}},\
  }\bibfield  {title} {\bibinfo {title} {Quantum field theory for the chiral
  clock transition in one spatial dimension},\ }\href
  {https://doi.org/10.1103/PhysRevB.98.205118} {\bibfield  {journal} {\bibinfo
  {journal} {Phys. Rev. B}\ }\textbf {\bibinfo {volume} {98}},\ \bibinfo
  {pages} {205118} (\bibinfo {year} {2018})}\BibitemShut {NoStop}%
\bibitem [{\citenamefont {Keesling}\ \emph {et~al.}(2019)\citenamefont
  {Keesling}, \citenamefont {Omran}, \citenamefont {Levine}, \citenamefont
  {Bernien}, \citenamefont {Pichler}, \citenamefont {Choi}, \citenamefont
  {Samajdar}, \citenamefont {Schwartz}, \citenamefont {Silvi}, \citenamefont
  {Sachdev}, \citenamefont {Zoller}, \citenamefont {Endres}, \citenamefont
  {Greiner}, \citenamefont {Vuleti{\'{c}}},\ and\ \citenamefont
  {Lukin}}]{Keesling2019}%
  \BibitemOpen
  \bibfield  {author} {\bibinfo {author} {\bibfnamefont {A.}~\bibnamefont
  {Keesling}}, \bibinfo {author} {\bibfnamefont {A.}~\bibnamefont {Omran}},
  \bibinfo {author} {\bibfnamefont {H.}~\bibnamefont {Levine}}, \bibinfo
  {author} {\bibfnamefont {H.}~\bibnamefont {Bernien}}, \bibinfo {author}
  {\bibfnamefont {H.}~\bibnamefont {Pichler}}, \bibinfo {author} {\bibfnamefont
  {S.}~\bibnamefont {Choi}}, \bibinfo {author} {\bibfnamefont {R.}~\bibnamefont
  {Samajdar}}, \bibinfo {author} {\bibfnamefont {S.}~\bibnamefont {Schwartz}},
  \bibinfo {author} {\bibfnamefont {P.}~\bibnamefont {Silvi}}, \bibinfo
  {author} {\bibfnamefont {S.}~\bibnamefont {Sachdev}}, \bibinfo {author}
  {\bibfnamefont {P.}~\bibnamefont {Zoller}}, \bibinfo {author} {\bibfnamefont
  {M.}~\bibnamefont {Endres}}, \bibinfo {author} {\bibfnamefont
  {M.}~\bibnamefont {Greiner}}, \bibinfo {author} {\bibfnamefont
  {V.}~\bibnamefont {Vuleti{\'{c}}}},\ and\ \bibinfo {author} {\bibfnamefont
  {M.~D.}\ \bibnamefont {Lukin}},\ }\bibfield  {title} {\bibinfo {title}
  {Quantum {K}ibble--{Z}urek mechanism and critical dynamics on a programmable
  {R}ydberg simulator},\ }\href {https://doi.org/10.1038/s41586-019-1070-1}
  {\bibfield  {journal} {\bibinfo  {journal} {Nature}\ }\textbf {\bibinfo
  {volume} {568}},\ \bibinfo {pages} {207} (\bibinfo {year}
  {2019})}\BibitemShut {NoStop}%
\bibitem [{\citenamefont {Bull}\ \emph {et~al.}(2019)\citenamefont {Bull},
  \citenamefont {Martin},\ and\ \citenamefont {Papi\ifmmode~\acute{c}\else
  \'{c}\fi{}}}]{Bull2019}%
  \BibitemOpen
  \bibfield  {author} {\bibinfo {author} {\bibfnamefont {K.}~\bibnamefont
  {Bull}}, \bibinfo {author} {\bibfnamefont {I.}~\bibnamefont {Martin}},\ and\
  \bibinfo {author} {\bibfnamefont {Z.}~\bibnamefont
  {Papi\ifmmode~\acute{c}\else \'{c}\fi{}}},\ }\bibfield  {title} {\bibinfo
  {title} {Systematic construction of scarred many-body dynamics in 1{D}
  lattice models},\ }\href {https://doi.org/10.1103/PhysRevLett.123.030601}
  {\bibfield  {journal} {\bibinfo  {journal} {Phys. Rev. Lett.}\ }\textbf
  {\bibinfo {volume} {123}},\ \bibinfo {pages} {030601} (\bibinfo {year}
  {2019})}\BibitemShut {NoStop}%
\bibitem [{\citenamefont {Mukherjee}\ \emph {et~al.}(2020)\citenamefont
  {Mukherjee}, \citenamefont {Sen}, \citenamefont {Sen},\ and\ \citenamefont
  {Sengupta}}]{Mukherjee2020_2}%
  \BibitemOpen
  \bibfield  {author} {\bibinfo {author} {\bibfnamefont {B.}~\bibnamefont
  {Mukherjee}}, \bibinfo {author} {\bibfnamefont {A.}~\bibnamefont {Sen}},
  \bibinfo {author} {\bibfnamefont {D.}~\bibnamefont {Sen}},\ and\ \bibinfo
  {author} {\bibfnamefont {K.}~\bibnamefont {Sengupta}},\ }\bibfield  {title}
  {\bibinfo {title} {Dynamics of the vacuum state in a periodically driven
  {R}ydberg chain},\ }\href {https://doi.org/10.1103/PhysRevB.102.075123}
  {\bibfield  {journal} {\bibinfo  {journal} {Phys. Rev. B}\ }\textbf {\bibinfo
  {volume} {102}},\ \bibinfo {pages} {075123} (\bibinfo {year}
  {2020})}\BibitemShut {NoStop}%
\bibitem [{\citenamefont {Surace}\ \emph {et~al.}(2020)\citenamefont {Surace},
  \citenamefont {Mazza}, \citenamefont {Giudici}, \citenamefont {Lerose},
  \citenamefont {Gambassi},\ and\ \citenamefont {Dalmonte}}]{Surace2020_2}%
  \BibitemOpen
  \bibfield  {author} {\bibinfo {author} {\bibfnamefont {F.~M.}\ \bibnamefont
  {Surace}}, \bibinfo {author} {\bibfnamefont {P.~P.}\ \bibnamefont {Mazza}},
  \bibinfo {author} {\bibfnamefont {G.}~\bibnamefont {Giudici}}, \bibinfo
  {author} {\bibfnamefont {A.}~\bibnamefont {Lerose}}, \bibinfo {author}
  {\bibfnamefont {A.}~\bibnamefont {Gambassi}},\ and\ \bibinfo {author}
  {\bibfnamefont {M.}~\bibnamefont {Dalmonte}},\ }\bibfield  {title} {\bibinfo
  {title} {Lattice gauge theories and string dynamics in {R}ydberg atom quantum
  simulators},\ }\href {https://doi.org/10.1103/PhysRevX.10.021041} {\bibfield
  {journal} {\bibinfo  {journal} {Phys. Rev. X}\ }\textbf {\bibinfo {volume}
  {10}},\ \bibinfo {pages} {021041} (\bibinfo {year} {2020})}\BibitemShut
  {NoStop}%
\bibitem [{\citenamefont {Zache}\ \emph {et~al.}(2019)\citenamefont {Zache},
  \citenamefont {Mueller}, \citenamefont {Schneider}, \citenamefont
  {Jendrzejewski}, \citenamefont {Berges},\ and\ \citenamefont
  {Hauke}}]{Zache2019}%
  \BibitemOpen
  \bibfield  {author} {\bibinfo {author} {\bibfnamefont {T.~V.}\ \bibnamefont
  {Zache}}, \bibinfo {author} {\bibfnamefont {N.}~\bibnamefont {Mueller}},
  \bibinfo {author} {\bibfnamefont {J.~T.}\ \bibnamefont {Schneider}}, \bibinfo
  {author} {\bibfnamefont {F.}~\bibnamefont {Jendrzejewski}}, \bibinfo {author}
  {\bibfnamefont {J.}~\bibnamefont {Berges}},\ and\ \bibinfo {author}
  {\bibfnamefont {P.}~\bibnamefont {Hauke}},\ }\bibfield  {title} {\bibinfo
  {title} {Dynamical topological transitions in the massive schwinger model
  with a $\ensuremath{\theta}$ term},\ }\href
  {https://doi.org/10.1103/PhysRevLett.122.050403} {\bibfield  {journal}
  {\bibinfo  {journal} {Phys. Rev. Lett.}\ }\textbf {\bibinfo {volume} {122}},\
  \bibinfo {pages} {050403} (\bibinfo {year} {2019})}\BibitemShut {NoStop}%
\bibitem [{\citenamefont {Zache}\ \emph {et~al.}(2022)\citenamefont {Zache},
  \citenamefont {Van~Damme}, \citenamefont {Halimeh}, \citenamefont {Hauke},\
  and\ \citenamefont {Banerjee}}]{Zache2022}%
  \BibitemOpen
  \bibfield  {author} {\bibinfo {author} {\bibfnamefont {T.~V.}\ \bibnamefont
  {Zache}}, \bibinfo {author} {\bibfnamefont {M.}~\bibnamefont {Van~Damme}},
  \bibinfo {author} {\bibfnamefont {J.~C.}\ \bibnamefont {Halimeh}}, \bibinfo
  {author} {\bibfnamefont {P.}~\bibnamefont {Hauke}},\ and\ \bibinfo {author}
  {\bibfnamefont {D.}~\bibnamefont {Banerjee}},\ }\bibfield  {title} {\bibinfo
  {title} {Toward the continuum limit of a $(1+1)\mathrm{D}$ quantum link
  schwinger model},\ }\href {https://doi.org/10.1103/PhysRevD.106.L091502}
  {\bibfield  {journal} {\bibinfo  {journal} {Phys. Rev. D}\ }\textbf {\bibinfo
  {volume} {106}},\ \bibinfo {pages} {L091502} (\bibinfo {year}
  {2022})}\BibitemShut {NoStop}%
\bibitem [{\citenamefont {Desaules}\ \emph {et~al.}(2023)\citenamefont
  {Desaules}, \citenamefont {Hudomal}, \citenamefont {Banerjee}, \citenamefont
  {Sen}, \citenamefont {Papi\ifmmode~\acute{c}\else \'{c}\fi{}},\ and\
  \citenamefont {Halimeh}}]{Desaules2023}%
  \BibitemOpen
  \bibfield  {author} {\bibinfo {author} {\bibfnamefont {J.-Y.}\ \bibnamefont
  {Desaules}}, \bibinfo {author} {\bibfnamefont {A.}~\bibnamefont {Hudomal}},
  \bibinfo {author} {\bibfnamefont {D.}~\bibnamefont {Banerjee}}, \bibinfo
  {author} {\bibfnamefont {A.}~\bibnamefont {Sen}}, \bibinfo {author}
  {\bibfnamefont {Z.}~\bibnamefont {Papi\ifmmode~\acute{c}\else \'{c}\fi{}}},\
  and\ \bibinfo {author} {\bibfnamefont {J.~C.}\ \bibnamefont {Halimeh}},\
  }\bibfield  {title} {\bibinfo {title} {Prominent quantum many-body scars in a
  truncated {S}chwinger model},\ }\href
  {https://doi.org/10.1103/PhysRevB.107.205112} {\bibfield  {journal} {\bibinfo
   {journal} {Phys. Rev. B}\ }\textbf {\bibinfo {volume} {107}},\ \bibinfo
  {pages} {205112} (\bibinfo {year} {2023})}\BibitemShut {NoStop}%
\bibitem [{\citenamefont {Wang}\ \emph {et~al.}(2023)\citenamefont {Wang},
  \citenamefont {Zhang}, \citenamefont {Yao}, \citenamefont {Liu},
  \citenamefont {Zhu}, \citenamefont {Zheng}, \citenamefont {Wang},
  \citenamefont {Zhai}, \citenamefont {Yuan},\ and\ \citenamefont
  {Pan}}]{Pan2023_2}%
  \BibitemOpen
  \bibfield  {author} {\bibinfo {author} {\bibfnamefont {H.-Y.}\ \bibnamefont
  {Wang}}, \bibinfo {author} {\bibfnamefont {W.-Y.}\ \bibnamefont {Zhang}},
  \bibinfo {author} {\bibfnamefont {Z.}~\bibnamefont {Yao}}, \bibinfo {author}
  {\bibfnamefont {Y.}~\bibnamefont {Liu}}, \bibinfo {author} {\bibfnamefont
  {Z.-H.}\ \bibnamefont {Zhu}}, \bibinfo {author} {\bibfnamefont {Y.-G.}\
  \bibnamefont {Zheng}}, \bibinfo {author} {\bibfnamefont {X.-K.}\ \bibnamefont
  {Wang}}, \bibinfo {author} {\bibfnamefont {H.}~\bibnamefont {Zhai}}, \bibinfo
  {author} {\bibfnamefont {Z.-S.}\ \bibnamefont {Yuan}},\ and\ \bibinfo
  {author} {\bibfnamefont {J.-W.}\ \bibnamefont {Pan}},\ }\bibfield  {title}
  {\bibinfo {title} {Interrelated thermalization and quantum criticality in a
  lattice gauge simulator},\ }\href
  {https://doi.org/10.1103/PhysRevLett.131.050401} {\bibfield  {journal}
  {\bibinfo  {journal} {Phys. Rev. Lett.}\ }\textbf {\bibinfo {volume} {131}},\
  \bibinfo {pages} {050401} (\bibinfo {year} {2023})}\BibitemShut {NoStop}%
\bibitem [{\citenamefont {Zhang}\ \emph {et~al.}()\citenamefont {Zhang},
  \citenamefont {Liu}, \citenamefont {Cheng}, \citenamefont {He}, \citenamefont
  {Wang}, \citenamefont {Wang}, \citenamefont {Zhu}, \citenamefont {Su},
  \citenamefont {Zhou}, \citenamefont {Zheng}, \citenamefont {Sun},
  \citenamefont {Yang}, \citenamefont {Hauke}, \citenamefont {Zheng},
  \citenamefont {Halimeh}, \citenamefont {Yuan},\ and\ \citenamefont
  {Pan}}]{zhang2023observation}%
  \BibitemOpen
  \bibfield  {author} {\bibinfo {author} {\bibfnamefont {W.-Y.}\ \bibnamefont
  {Zhang}}, \bibinfo {author} {\bibfnamefont {Y.}~\bibnamefont {Liu}}, \bibinfo
  {author} {\bibfnamefont {Y.}~\bibnamefont {Cheng}}, \bibinfo {author}
  {\bibfnamefont {M.-G.}\ \bibnamefont {He}}, \bibinfo {author} {\bibfnamefont
  {H.-Y.}\ \bibnamefont {Wang}}, \bibinfo {author} {\bibfnamefont {T.-Y.}\
  \bibnamefont {Wang}}, \bibinfo {author} {\bibfnamefont {Z.-H.}\ \bibnamefont
  {Zhu}}, \bibinfo {author} {\bibfnamefont {G.-X.}\ \bibnamefont {Su}},
  \bibinfo {author} {\bibfnamefont {Z.-Y.}\ \bibnamefont {Zhou}}, \bibinfo
  {author} {\bibfnamefont {Y.-G.}\ \bibnamefont {Zheng}}, \bibinfo {author}
  {\bibfnamefont {H.}~\bibnamefont {Sun}}, \bibinfo {author} {\bibfnamefont
  {B.}~\bibnamefont {Yang}}, \bibinfo {author} {\bibfnamefont {P.}~\bibnamefont
  {Hauke}}, \bibinfo {author} {\bibfnamefont {W.}~\bibnamefont {Zheng}},
  \bibinfo {author} {\bibfnamefont {J.~C.}\ \bibnamefont {Halimeh}}, \bibinfo
  {author} {\bibfnamefont {Z.-S.}\ \bibnamefont {Yuan}},\ and\ \bibinfo
  {author} {\bibfnamefont {J.-W.}\ \bibnamefont {Pan}},\ }\href@noop {}
  {\bibinfo {title} {Observation of microscopic confinement dynamics by a
  tunable topological $\theta$-angle}},\ \Eprint
  {https://arxiv.org/abs/2306.11794 (2023)} {arXiv:2306.11794 (2023)}
  \BibitemShut {NoStop}%
\bibitem [{\citenamefont {Vidal}(2007)}]{Vidal2007}%
  \BibitemOpen
  \bibfield  {author} {\bibinfo {author} {\bibfnamefont {G.}~\bibnamefont
  {Vidal}},\ }\bibfield  {title} {\bibinfo {title} {Classical simulation of
  infinite-size quantum lattice systems in one spatial dimension},\ }\href
  {https://doi.org/10.1103/PhysRevLett.98.070201} {\bibfield  {journal}
  {\bibinfo  {journal} {Phys. Rev. Lett.}\ }\textbf {\bibinfo {volume} {98}},\
  \bibinfo {pages} {070201} (\bibinfo {year} {2007})}\BibitemShut {NoStop}%
\bibitem [{\citenamefont {Halimeh}\ \emph {et~al.}(2022)\citenamefont
  {Halimeh}, \citenamefont {McCulloch}, \citenamefont {Yang},\ and\
  \citenamefont {Hauke}}]{Halimeh2022}%
  \BibitemOpen
  \bibfield  {author} {\bibinfo {author} {\bibfnamefont {J.~C.}\ \bibnamefont
  {Halimeh}}, \bibinfo {author} {\bibfnamefont {I.~P.}\ \bibnamefont
  {McCulloch}}, \bibinfo {author} {\bibfnamefont {B.}~\bibnamefont {Yang}},\
  and\ \bibinfo {author} {\bibfnamefont {P.}~\bibnamefont {Hauke}},\ }\bibfield
   {title} {\bibinfo {title} {Tuning the topological
  $\ensuremath{\theta}$-angle in cold-atom quantum simulators of gauge
  theories},\ }\href {https://doi.org/10.1103/PRXQuantum.3.040316} {\bibfield
  {journal} {\bibinfo  {journal} {PRX Quantum}\ }\textbf {\bibinfo {volume}
  {3}},\ \bibinfo {pages} {040316} (\bibinfo {year} {2022})}\BibitemShut
  {NoStop}%
\bibitem [{\citenamefont {Buyens}\ \emph {et~al.}(2016)\citenamefont {Buyens},
  \citenamefont {Haegeman}, \citenamefont {Verschelde}, \citenamefont
  {Verstraete},\ and\ \citenamefont {Van~Acoleyen}}]{Buyens2016}%
  \BibitemOpen
  \bibfield  {author} {\bibinfo {author} {\bibfnamefont {B.}~\bibnamefont
  {Buyens}}, \bibinfo {author} {\bibfnamefont {J.}~\bibnamefont {Haegeman}},
  \bibinfo {author} {\bibfnamefont {H.}~\bibnamefont {Verschelde}}, \bibinfo
  {author} {\bibfnamefont {F.}~\bibnamefont {Verstraete}},\ and\ \bibinfo
  {author} {\bibfnamefont {K.}~\bibnamefont {Van~Acoleyen}},\ }\bibfield
  {title} {\bibinfo {title} {Confinement and string breaking for
  {${\mathrm{QED}}_{2}$} in the hamiltonian picture},\ }\href
  {https://doi.org/10.1103/PhysRevX.6.041040} {\bibfield  {journal} {\bibinfo
  {journal} {Phys. Rev. X}\ }\textbf {\bibinfo {volume} {6}},\ \bibinfo {pages}
  {041040} (\bibinfo {year} {2016})}\BibitemShut {NoStop}%
\bibitem [{\citenamefont {Coleman}(1976)}]{COLEMAN1976}%
  \BibitemOpen
  \bibfield  {author} {\bibinfo {author} {\bibfnamefont {S.}~\bibnamefont
  {Coleman}},\ }\bibfield  {title} {\bibinfo {title} {More about the massive
  schwinger model},\ }\href
  {https://doi.org/https://doi.org/10.1016/0003-4916(76)90280-3} {\bibfield
  {journal} {\bibinfo  {journal} {Annals of Physics}\ }\textbf {\bibinfo
  {volume} {101}},\ \bibinfo {pages} {239} (\bibinfo {year}
  {1976})}\BibitemShut {NoStop}%
\bibitem [{\citenamefont {Bi}\ and\ \citenamefont
  {Senthil}(2019)}]{Senthil2019}%
  \BibitemOpen
  \bibfield  {author} {\bibinfo {author} {\bibfnamefont {Z.}~\bibnamefont
  {Bi}}\ and\ \bibinfo {author} {\bibfnamefont {T.}~\bibnamefont {Senthil}},\
  }\bibfield  {title} {\bibinfo {title} {Adventure in topological phase
  transitions in $3+1$-{D}: Non-{A}belian deconfined quantum criticalities and
  a possible duality},\ }\href {https://doi.org/10.1103/PhysRevX.9.021034}
  {\bibfield  {journal} {\bibinfo  {journal} {Phys. Rev. X}\ }\textbf {\bibinfo
  {volume} {9}},\ \bibinfo {pages} {021034} (\bibinfo {year}
  {2019})}\BibitemShut {NoStop}%
\bibitem [{\citenamefont {Prakash}\ \emph {et~al.}(2023)\citenamefont
  {Prakash}, \citenamefont {Fava},\ and\ \citenamefont
  {Parameswaran}}]{Prakash2023}%
  \BibitemOpen
  \bibfield  {author} {\bibinfo {author} {\bibfnamefont {A.}~\bibnamefont
  {Prakash}}, \bibinfo {author} {\bibfnamefont {M.}~\bibnamefont {Fava}},\ and\
  \bibinfo {author} {\bibfnamefont {S.~A.}\ \bibnamefont {Parameswaran}},\
  }\bibfield  {title} {\bibinfo {title} {Multiversality and unnecessary
  criticality in one dimension},\ }\href
  {https://doi.org/10.1103/PhysRevLett.130.256401} {\bibfield  {journal}
  {\bibinfo  {journal} {Phys. Rev. Lett.}\ }\textbf {\bibinfo {volume} {130}},\
  \bibinfo {pages} {256401} (\bibinfo {year} {2023})}\BibitemShut {NoStop}%
\bibitem [{\citenamefont {Sen}\ \emph {et~al.}(2010)\citenamefont {Sen},
  \citenamefont {Shankar}, \citenamefont {Dhar},\ and\ \citenamefont
  {Ramola}}]{Sen2010}%
  \BibitemOpen
  \bibfield  {author} {\bibinfo {author} {\bibfnamefont {D.}~\bibnamefont
  {Sen}}, \bibinfo {author} {\bibfnamefont {R.}~\bibnamefont {Shankar}},
  \bibinfo {author} {\bibfnamefont {D.}~\bibnamefont {Dhar}},\ and\ \bibinfo
  {author} {\bibfnamefont {K.}~\bibnamefont {Ramola}},\ }\bibfield  {title}
  {\bibinfo {title} {Spin-1 kitaev model in one dimension},\ }\href
  {https://doi.org/10.1103/PhysRevB.82.195435} {\bibfield  {journal} {\bibinfo
  {journal} {Phys. Rev. B}\ }\textbf {\bibinfo {volume} {82}},\ \bibinfo
  {pages} {195435} (\bibinfo {year} {2010})}\BibitemShut {NoStop}%
\bibitem [{\citenamefont {Schrieffer}\ and\ \citenamefont
  {Wolff}(1966)}]{Schrieffer1966}%
  \BibitemOpen
  \bibfield  {author} {\bibinfo {author} {\bibfnamefont {J.~R.}\ \bibnamefont
  {Schrieffer}}\ and\ \bibinfo {author} {\bibfnamefont {P.~A.}\ \bibnamefont
  {Wolff}},\ }\bibfield  {title} {\bibinfo {title} {Relation between the
  {A}nderson and {K}ondo {H}amiltonians},\ }\href
  {https://doi.org/10.1103/PhysRev.149.491} {\bibfield  {journal} {\bibinfo
  {journal} {Phys. Rev.}\ }\textbf {\bibinfo {volume} {149}},\ \bibinfo {pages}
  {491} (\bibinfo {year} {1966})}\BibitemShut {NoStop}%
\bibitem [{\citenamefont {Peng}\ and\ \citenamefont {Cui}(2022)}]{cui2022}%
  \BibitemOpen
  \bibfield  {author} {\bibinfo {author} {\bibfnamefont {C.}~\bibnamefont
  {Peng}}\ and\ \bibinfo {author} {\bibfnamefont {X.}~\bibnamefont {Cui}},\
  }\bibfield  {title} {\bibinfo {title} {Bridging quantum many-body scars and
  quantum integrability in ising chains with transverse and longitudinal
  fields},\ }\href {https://doi.org/10.1103/PhysRevB.106.214311} {\bibfield
  {journal} {\bibinfo  {journal} {Phys. Rev. B}\ }\textbf {\bibinfo {volume}
  {106}},\ \bibinfo {pages} {214311} (\bibinfo {year} {2022})}\BibitemShut
  {NoStop}%
\bibitem [{\citenamefont {Yao}\ \emph {et~al.}(2022{\natexlab{b}})\citenamefont
  {Yao}, \citenamefont {Pan}, \citenamefont {Liu},\ and\ \citenamefont
  {Zhai}}]{zhaih2022}%
  \BibitemOpen
  \bibfield  {author} {\bibinfo {author} {\bibfnamefont {Z.}~\bibnamefont
  {Yao}}, \bibinfo {author} {\bibfnamefont {L.}~\bibnamefont {Pan}}, \bibinfo
  {author} {\bibfnamefont {S.}~\bibnamefont {Liu}},\ and\ \bibinfo {author}
  {\bibfnamefont {H.}~\bibnamefont {Zhai}},\ }\bibfield  {title} {\bibinfo
  {title} {Quantum many-body scars and quantum criticality},\ }\href
  {https://doi.org/10.1103/PhysRevB.105.125123} {\bibfield  {journal} {\bibinfo
   {journal} {Phys. Rev. B}\ }\textbf {\bibinfo {volume} {105}},\ \bibinfo
  {pages} {125123} (\bibinfo {year} {2022}{\natexlab{b}})}\BibitemShut
  {NoStop}%
\bibitem [{\citenamefont {Mukherjee}\ \emph {et~al.}(2022)\citenamefont
  {Mukherjee}, \citenamefont {Sen},\ and\ \citenamefont
  {Sengupta}}]{Mukherjee2022}%
  \BibitemOpen
  \bibfield  {author} {\bibinfo {author} {\bibfnamefont {B.}~\bibnamefont
  {Mukherjee}}, \bibinfo {author} {\bibfnamefont {A.}~\bibnamefont {Sen}},\
  and\ \bibinfo {author} {\bibfnamefont {K.}~\bibnamefont {Sengupta}},\
  }\bibfield  {title} {\bibinfo {title} {Periodically driven {R}ydberg chains
  with staggered detuning},\ }\href
  {https://doi.org/10.1103/PhysRevB.106.064305} {\bibfield  {journal} {\bibinfo
   {journal} {Phys. Rev. B}\ }\textbf {\bibinfo {volume} {106}},\ \bibinfo
  {pages} {064305} (\bibinfo {year} {2022})}\BibitemShut {NoStop}%
\bibitem [{\citenamefont {Shen}\ \emph {et~al.}()\citenamefont {Shen},
  \citenamefont {Qin}, \citenamefont {Desaules}, \citenamefont {Papić},\ and\
  \citenamefont {Lee}}]{shen2024}%
  \BibitemOpen
  \bibfield  {author} {\bibinfo {author} {\bibfnamefont {R.}~\bibnamefont
  {Shen}}, \bibinfo {author} {\bibfnamefont {F.}~\bibnamefont {Qin}}, \bibinfo
  {author} {\bibfnamefont {J.-Y.}\ \bibnamefont {Desaules}}, \bibinfo {author}
  {\bibfnamefont {Z.}~\bibnamefont {Papić}},\ and\ \bibinfo {author}
  {\bibfnamefont {C.~H.}\ \bibnamefont {Lee}},\ }\href@noop {} {\bibinfo
  {title} {Enhanced many-body quantum scars from the non-{H}ermitian {F}ock
  skin effect}},\ \Eprint {https://arxiv.org/abs/2403.02395 (2024)}
  {arXiv:2403.02395 (2024)} \BibitemShut {NoStop}%
\bibitem [{\citenamefont {Iadecola}\ \emph {et~al.}(2019)\citenamefont
  {Iadecola}, \citenamefont {Schecter},\ and\ \citenamefont
  {Xu}}]{Iadecola2019}%
  \BibitemOpen
  \bibfield  {author} {\bibinfo {author} {\bibfnamefont {T.}~\bibnamefont
  {Iadecola}}, \bibinfo {author} {\bibfnamefont {M.}~\bibnamefont {Schecter}},\
  and\ \bibinfo {author} {\bibfnamefont {S.}~\bibnamefont {Xu}},\ }\bibfield
  {title} {\bibinfo {title} {Quantum many-body scars from magnon
  condensation},\ }\href {https://doi.org/10.1103/PhysRevB.100.184312}
  {\bibfield  {journal} {\bibinfo  {journal} {Phys. Rev. B}\ }\textbf {\bibinfo
  {volume} {100}},\ \bibinfo {pages} {184312} (\bibinfo {year}
  {2019})}\BibitemShut {NoStop}%
\bibitem [{\citenamefont {Turner}\ \emph
  {et~al.}(2018{\natexlab{b}})\citenamefont {Turner}, \citenamefont
  {Michailidis}, \citenamefont {Abanin}, \citenamefont {Serbyn},\ and\
  \citenamefont {Papi\ifmmode~\acute{c}\else \'{c}\fi{}}}]{TurnerPRB2018}%
  \BibitemOpen
  \bibfield  {author} {\bibinfo {author} {\bibfnamefont {C.~J.}\ \bibnamefont
  {Turner}}, \bibinfo {author} {\bibfnamefont {A.~A.}\ \bibnamefont
  {Michailidis}}, \bibinfo {author} {\bibfnamefont {D.~A.}\ \bibnamefont
  {Abanin}}, \bibinfo {author} {\bibfnamefont {M.}~\bibnamefont {Serbyn}},\
  and\ \bibinfo {author} {\bibfnamefont {Z.}~\bibnamefont
  {Papi\ifmmode~\acute{c}\else \'{c}\fi{}}},\ }\bibfield  {title} {\bibinfo
  {title} {Quantum scarred eigenstates in a rydberg atom chain: Entanglement,
  breakdown of thermalization, and stability to perturbations},\ }\href
  {https://doi.org/10.1103/PhysRevB.98.155134} {\bibfield  {journal} {\bibinfo
  {journal} {Phys. Rev. B}\ }\textbf {\bibinfo {volume} {98}},\ \bibinfo
  {pages} {155134} (\bibinfo {year} {2018}{\natexlab{b}})}\BibitemShut
  {NoStop}%
\bibitem [{\citenamefont {Chen}\ and\ \citenamefont
  {Iadecola}(2021)}]{Chen2021}%
  \BibitemOpen
  \bibfield  {author} {\bibinfo {author} {\bibfnamefont {I.-C.}\ \bibnamefont
  {Chen}}\ and\ \bibinfo {author} {\bibfnamefont {T.}~\bibnamefont
  {Iadecola}},\ }\bibfield  {title} {\bibinfo {title} {Emergent symmetries and
  slow quantum dynamics in a {R}ydberg-atom chain with confinement},\ }\href
  {https://doi.org/10.1103/PhysRevB.103.214304} {\bibfield  {journal} {\bibinfo
   {journal} {Phys. Rev. B}\ }\textbf {\bibinfo {volume} {103}},\ \bibinfo
  {pages} {214304} (\bibinfo {year} {2021})}\BibitemShut {NoStop}%
\bibitem [{\citenamefont {Sala}\ \emph {et~al.}(2020)\citenamefont {Sala},
  \citenamefont {Rakovszky}, \citenamefont {Verresen}, \citenamefont {Knap},\
  and\ \citenamefont {Pollmann}}]{Sala2020}%
  \BibitemOpen
  \bibfield  {author} {\bibinfo {author} {\bibfnamefont {P.}~\bibnamefont
  {Sala}}, \bibinfo {author} {\bibfnamefont {T.}~\bibnamefont {Rakovszky}},
  \bibinfo {author} {\bibfnamefont {R.}~\bibnamefont {Verresen}}, \bibinfo
  {author} {\bibfnamefont {M.}~\bibnamefont {Knap}},\ and\ \bibinfo {author}
  {\bibfnamefont {F.}~\bibnamefont {Pollmann}},\ }\bibfield  {title} {\bibinfo
  {title} {Ergodicity breaking arising from hilbert space fragmentation in
  dipole-conserving hamiltonians},\ }\href
  {https://doi.org/10.1103/PhysRevX.10.011047} {\bibfield  {journal} {\bibinfo
  {journal} {Phys. Rev. X}\ }\textbf {\bibinfo {volume} {10}},\ \bibinfo
  {pages} {011047} (\bibinfo {year} {2020})}\BibitemShut {NoStop}%
\bibitem [{\citenamefont {Yang}\ \emph {et~al.}(2020)\citenamefont {Yang},
  \citenamefont {Liu}, \citenamefont {Gorshkov},\ and\ \citenamefont
  {Iadecola}}]{Yang2020}%
  \BibitemOpen
  \bibfield  {author} {\bibinfo {author} {\bibfnamefont {Z.-C.}\ \bibnamefont
  {Yang}}, \bibinfo {author} {\bibfnamefont {F.}~\bibnamefont {Liu}}, \bibinfo
  {author} {\bibfnamefont {A.~V.}\ \bibnamefont {Gorshkov}},\ and\ \bibinfo
  {author} {\bibfnamefont {T.}~\bibnamefont {Iadecola}},\ }\bibfield  {title}
  {\bibinfo {title} {Hilbert-space fragmentation from strict confinement},\
  }\href {https://doi.org/10.1103/PhysRevLett.124.207602} {\bibfield  {journal}
  {\bibinfo  {journal} {Phys. Rev. Lett.}\ }\textbf {\bibinfo {volume} {124}},\
  \bibinfo {pages} {207602} (\bibinfo {year} {2020})}\BibitemShut {NoStop}%
\bibitem [{\citenamefont {Zhang}\ \emph
  {et~al.}(2023{\natexlab{c}})\citenamefont {Zhang}, \citenamefont {Yuan},
  \citenamefont {Iadecola}, \citenamefont {Xu},\ and\ \citenamefont
  {Deng}}]{Zhangprl2023}%
  \BibitemOpen
  \bibfield  {author} {\bibinfo {author} {\bibfnamefont {S.-Y.}\ \bibnamefont
  {Zhang}}, \bibinfo {author} {\bibfnamefont {D.}~\bibnamefont {Yuan}},
  \bibinfo {author} {\bibfnamefont {T.}~\bibnamefont {Iadecola}}, \bibinfo
  {author} {\bibfnamefont {S.}~\bibnamefont {Xu}},\ and\ \bibinfo {author}
  {\bibfnamefont {D.-L.}\ \bibnamefont {Deng}},\ }\bibfield  {title} {\bibinfo
  {title} {Extracting quantum many-body scarred eigenstates with matrix product
  states},\ }\href {https://doi.org/10.1103/PhysRevLett.131.020402} {\bibfield
  {journal} {\bibinfo  {journal} {Phys. Rev. Lett.}\ }\textbf {\bibinfo
  {volume} {131}},\ \bibinfo {pages} {020402} (\bibinfo {year}
  {2023}{\natexlab{c}})}\BibitemShut {NoStop}%
\bibitem [{\citenamefont {Windt}\ and\ \citenamefont
  {Pichler}(2022)}]{Windt2022}%
  \BibitemOpen
  \bibfield  {author} {\bibinfo {author} {\bibfnamefont {B.}~\bibnamefont
  {Windt}}\ and\ \bibinfo {author} {\bibfnamefont {H.}~\bibnamefont
  {Pichler}},\ }\bibfield  {title} {\bibinfo {title} {Squeezing quantum
  many-body scars},\ }\href {https://doi.org/10.1103/PhysRevLett.128.090606}
  {\bibfield  {journal} {\bibinfo  {journal} {Phys. Rev. Lett.}\ }\textbf
  {\bibinfo {volume} {128}},\ \bibinfo {pages} {090606} (\bibinfo {year}
  {2022})}\BibitemShut {NoStop}%
\end{thebibliography}%

\end{document}